\documentclass[10pt]{iopart}
\usepackage{graphicx}
\usepackage{amsfonts}
\usepackage{amsbsy}
\usepackage{bbm}
\usepackage{dsfont}
\usepackage{color}

\newcommand{\argmax}{\mathrm{argmax}}
\newcommand{\argmin}{\mathrm{argmin}}

\newcommand{\extr}{\mathrm{extr}}

\newcommand{\x}{\mathbf{x}}

\newcommand{\m}{\mathbf{m}}

\newcommand{\bc}{\mathbf{c}}

\newcommand{\Lmatrix}{   \ensuremath{\boldsymbol{\Lambda}}         }
\newcommand{\cmatrix}{  \mathbf{C}  }
\newcommand{\Cov}{   \ensuremath{\boldsymbol{\Sigma}}  }

\newcommand{\A}{  \ensuremath{\mathbf{A}}   }
\newcommand{\I}{\mathbf{I}} 
\newcommand{\Ind}{\mathds{1}}

\newcommand{\thetav}{  \ensuremath{\boldsymbol{\theta}}   }

\newcommand{\xmatrix}{ \ensuremath{ \mathbf{X}}}
\newcommand{\av}{\ensuremath{\mathbf{a}}}
\newcommand{\nullv}{\ensuremath{\mathbf{0}}}
\newcommand{\Q}{  \ensuremath{\mathbf{Q}}   }
\newcommand{\cv}{\ensuremath{ \mathbf{c} }}

\newcommand{\muv}{  \ensuremath{\boldsymbol{\mu}}   }

\begin{document}

\title[Replica analysis of  Bayesian data clustering]{Replica analysis of Bayesian data clustering} 

\author{Alexander Mozeika$^{\dag}$ and Anthony CC Coolen$^{\dag\ddag\S}$}
\address{$\dag$Institute for Mathematical and Molecular Biomedicine, King's College London, Hodgkin Building, London SE1 1UL, UK\\[1mm]
$\ddag$ Department of Mathematics, King's College London, The Strand, London WC2R 2LS, UK 
\\[1mm]
$\S$ London Institute for Mathematical Sciences, 35a South St, Mayfair, London, W1K 2XF, UK}

%check pacs numbers!!!
% 05.50.+q	Lattice theory and statistics (Ising, Potts, etc.) 
%05.70.Fh	Phase transitions: general studies
%02.50.-r	Probability theory, stochastic processes, and statistics
%02.50.Tt	 Inference methods
%75.10.Nr	Spin-glass and other random models (for spin glasses and other random magnets, see 75.50.Lk)

\ead{alexander.mozeika@kcl.ac.uk, ton.coolen@kcl.ac.uk}

\begin{abstract}
We use statistical mechanics to study model-based Bayesian data clustering.  In this approach, each partition of the data into clusters is regarded as  a microscopic system state, the negative data log-likelihood  gives the energy of each state, and  the data set realisation acts as disorder.  Optimal clustering  corresponds to the  ground state of the system, and is hence obtained from the free energy via a low `temperature' limit. We assume that  for large sample sizes the free energy density is self-averaging, and we use the replica method to compute the asymptotic free energy  density.   The main order parameter  in the resulting  (replica symmetric) theory,  the distribution of the data over the clusters, satisfies a self-consistent equation which can be solved by a population dynamics  algorithm. From this order parameter one computes the average free energy,  and all relevant macroscopic characteristics of the problem.  The theory describes numerical experiments perfectly, and gives a significant  improvement over the mean-field theory that was used to study this model  in past.  
\end{abstract}

\section{Introduction\label{section:Intro}}
%Statistical mechanics, and  statistical inference
Analytical tools of statistical mechanics  are nowadays applied widely   to statistical inference problems (see e.g. \cite{Advani2016} and references therein).  The central object of study in parameter inference is an expression for the likelihood of the data, which encodes information about the  model that generated the data  and the sampling process. The traditional maximum likelihood (ML) method infers model parameters from the data, but is often intractable (see e.g. \cite{Mozeika2014}) or can lead to overfitting \cite{Coolen2017}.  The Bayesian framework represents a more rigorous approach to parameter inference. It requires assumptions about the `prior probability' of model parameters,  and expresses the `posterior probability'  of the parameters, given the data, in terms of the data likelihood.  In the so-called maximum a posteriori probability (MAP)  method, one computes the most probable parameters, according to the posterior probability.  MAP cures overfitting in ML partially by providing  a `regulariser'  \cite{Advani2016}.  Both ML and MAP methods can be seen as optimisation problems, in which the data likelihood and  posterior parameter probability, respectively,  play the role of the objective function.  With a trivial sign change this objective function can be mapped into an `energy' function to be minimised, so that ML and MAP parameter inference can both equivalently be seen as computing a ground state in statistical mechanics \cite{Nishimori2001, Mezard2009}.  

%Bayesian clustering, etc 
Clustering is a popular type of inference where one seeks to allocate statistically similar data points to the same category (or cluster),  in an unsupervised way.  It is  used  in astrophysics~\cite{Souza2017}, biology \cite{Hanage2009}, and many other areas.  The assumed data likelihood  in  ML and Bayesian model-based clustering methods is usually a Gaussian Mixture Model (GMM) \cite{Souza2017, Bishop2006}.   The  GMM likelihood,  however,   is  analytically intractable,  and  one hence tends to resort to variational  approximations \cite{Bishop2006} or computationally intensive  Monte Carlo methods \cite{Nobile2007}.  Furthermore,  the number of model parameters,  in particular the number of partitions of the data, is \emph{extensive}, even if we fix the dimension of the data to be finite, which leads to additional difficulties \cite{Guihenneuc2005}.  

%difficulties in Bayesian clustering 
For this reason, not many analytical  results are  available  for model-based clustering (MBC),  leaving  mostly (many) numerical studies.  Here,  even when the number of parameters $d$ is kept finite, the matrix  of  `allocation'  variables $\cmatrix$~\cite{Bishop2006} which we ultimately want to infer is  growing  with the sample size $N$.  The situation is complicated further if, in addition to $\cmatrix$,  we are also inferring the number of true clusters $K$. In the GMM approach, the number of clusters  is usually  found by  adding a `penalty'  term to the log-likelihood function, such as for the Bayesian information criterion (BIC) or the integrated complete-data likelihood (ICL) \cite{Scrucca2016}. These penalty based approaches  sometimes lead to conflicting results~\cite{Souza2017}.

%Statistical Mechanics 
  A direct solution to the above problems is to follow the approach of  statistical mechanics and compute the partition function \cite{Nishimori2001, Mezard2009}.  This  approach is usually  not  pursued  by statisticians, and in this case  has not yet been pursued fully by physicists either (in spite of their familiarity with such calculations).  Popular Machine Learning textbooks written by physicists, such as~\cite{Bishop2006}  or  the more recent~\cite{Barber2012}, cover only the (algorithmic)  variational mean-field approach for the case when  $K$ is unknown, and the (non-Bayesian) expectation-maximisation  algorithm for the case  when  $K$  is known.  Most statistical mechanics approaches to data clustering \cite{Rose1990, Blatt1996, Luksza2010} use some heuristic measure of data dissimilarity as an energy function, rather than an actual statistical model of the data, or limit themselves to the simple case of assuming only two clusters \cite{Watkin1994,Barkai1994,Biehl1994} in the high dimensional  regime where $d\rightarrow\infty$ and $N\rightarrow\infty$, with $d/N$ finite.   
  
The work of  \cite{Barkai1994} and \cite{Biehl1994}  is mainly  concerned with the inference of parameters of two isotropic  Gaussians from a balanced sample, i.e. a very restricted  model of the data  which does not take into account correlations, different cluster sizes, data with more than two clusters, etc.  The former is concerned with the inference of the centres   of the assumed Gaussians, and the latter with finding a single `direction' in the data. Hence both studies  do not formally address the  MBC problem.  Furthermore, in  \cite{Watkin1994}  the  Bayesian approach is  used to infer `prototype vectors', such as centres  of Gaussians, etc., of the same dimension as the data, so also this work is not addressing the  MBC problem systematically either.  Finally, we note that none of the above papers refer to previous work on  MBC in the low-dimensional regime of finite $d$ and  $N\rightarrow\infty$. To our knowledge, only one study considers the  high-dimensional regime of a specific Bayesian GMM clustering problem, namely \cite{Lesieur2016}.  A systematic statistical mechanical  treatment of the Bayesian clustering  problem is still lacking.

  %What we do in this paper
In this paper we consider a more general model-based  Bayesian clustering protocol,  which allows for simultaneous inference of the number of clusters in the data and their components, based on stochastic partitions of the data (SPD) \cite{Corander2009}.  SPD assumes  priors on the partitions to compute the MAP estimate of data partitions.  The mean-field (MF) theory of Bayesian SPD inference  was developed recently in \cite{Mozeika2018}.  That study used the negative log-likelihood as the energy function, and computed its average over the data and the partitions. It led to a simple and intuitive analytical framework,  which  makes non-trivial predictions about  low energy states and the corresponding (MAP) data partitions.  However, these predictions are only correct in the  regime of  `weak'  correlations \cite{Mozeika2018}.   In this paper we pursue a full statistical mechanical treatment of the Bayesian clustering problem covering \emph{all} correlation regimes.  To this end we analyse the free energy, and we use the replica method \cite{Mezard1987} to compute its average over the data. This, unlike MF, allows us to compute  the average energy of the \emph{optimal} partitions.   Furthermore, the present analysis  produces a simple  algorithmic framework,  with the  population dynamics \cite{Mezard1987} clustering algorithm at its heart,  for the  \emph{simultaneous} inference of the number of clusters in the data and their components.  This can be seen as  a first \emph{non-variational}  result for this type of problems \cite{Bishop2006}.

 \section{Model of the data and Bayesian cluster inference\label{section:model}}
  Let us assume that we  observe a data  sample   $\xmatrix=\{\x_1, \ldots, \x_N\}$, where $\x_i \in \mathbb{R}^d$ for all $i$. Each vector $\x_i$ are assumed to have been  generated independently from one of $K$ distributions, which are members of a parametrized family $P(\x|\thetav)$. $M_1$ data-points are sampled from $P(\x\vert\thetav_1)$, with parameter $\thetav_1$, $M_2$ data-points are sampled from $P(\x\vert\thetav_2)$, etc. We clearly have the constraint  $\sum_{\mu=1}^K M_\mu=N$,  and we assume that $M_\mu\geq1$ for all $\mu$.  We will say that $\x_i$  (or its index $i$) belongs to `cluster'  $\mu$ if $\x_i$ was sampled from $P(\x\vert\thetav_\mu)$.  The above  sampling scenario can be described by the following distribution: 
\begin{eqnarray}
P(\xmatrix\vert\cmatrix,K, \thetav_1,\ldots,\thetav_K)=\prod_{\mu=1}^{K}\prod_{i=1}^N P^{\,c_{i\mu}}(\x_{i}\vert\thetav_{\mu}) \label{eq:P(X|..)}
\end{eqnarray}
which is parametrised by the  the partition  matrix,  or `allocation' matrix~\cite{Bishop2006},  $\cmatrix$.   Each element  of this matrix $\left[\cmatrix\right]_{i\mu}=c_{i\mu}$ computes an indicator function $\Ind\left[\x_i\sim P(\x\vert\thetav_\mu)\right]$, i.e. is nonzero if and only if $\x_i$ is sampled from $P(\x\vert\thetav_\mu)$.  Furthermore, we have $\sum_{\mu\leq K}c_{i\mu}=1$ for all $i\in\left[N\right] $\footnote{Throughout this paper the notation   $\left[N\right]$ will be used to represent  the set  $\{1,\ldots,N\}$.}, i.e. $\x_i$ belongs to only one cluster, and $M_\mu(\cmatrix)=\sum_{i\leq N} c_{i\mu}\geq1$   for all   $\mu\in\left[K\right]$, i.e. empty clusters are not allowed\footnote{We note that the distribution (\ref{eq:P(X|..)}) could  be also defined  by using  set notation, see e.g.   \cite{Mozeika2018}.   }.  

Suppose we now want to infer the partition  matrix $\cmatrix$ and the number of clusters $K$. The Bayesian approach  to this problem (see e.g.~\cite{Bishop2006}) would be to assume prior distributions for parameters and  partitions, $P(\thetav_{\mu})$ and $P(\cmatrix,K)=P(\cmatrix\vert K)P(K)$\footnote{The simplest route,  following the `Principle of Insufficient Reason', is to choose uniform  $P(\cmatrix\vert K)$ and $P(K)$. The former is then given by   $P(\cmatrix\vert K)=1/K!\,\mathcal{S}(N,K)$, where $\mathcal{S}(N,K)$ is the Stirling number of the second kind ($\mathcal{S}(N,K)\simeq K^N/K!$ as $N\rightarrow\infty$~\cite{Rennie1969}), and the latter is given by  $P(K)=1/N$.}, and to consider subsequently the  posterior distribution 
\begin{eqnarray}
P(\cmatrix,K|\xmatrix)&=& \frac{P(\xmatrix|\cmatrix,K)P(\cmatrix|K)P(K)}{\sum_{\tilde{K}=1}^N P(\tilde{K}) \sum_{\tilde\cmatrix}P(\xmatrix|\tilde\cmatrix,\tilde K)P(\tilde{\cmatrix}|\tilde{K})}
\nonumber
\\
&=&\frac{\rme^{-N\hat{F}_N(\cmatrix,\, \xmatrix)} P(\cmatrix\vert K) P(K) }{\sum_{\tilde{K}=1}^N\!P(\tilde{K})\!\sum_{  \tilde{\cmatrix}    }\rme^{-N\hat{F}_N(\tilde{\cmatrix},\, \xmatrix)} P( \tilde{\cmatrix} \vert \tilde{K})}
\label{eq:P(C|X)},
\end{eqnarray}
where we have defined the  log-likelihood density
\begin{eqnarray}
\hat{F}_N(\cmatrix,\, \xmatrix)&=& -\frac{1}{N}\sum_{\mu=1}^K\log \left\langle\rme^{\sum_{i=1}^N c_{i\mu}\log P(\x_{i}\vert\thetav_{\mu})}\right\rangle_{\thetav_\mu} \label{def:F-hat}
\end{eqnarray}
and the short-hand $\left\langle  f(\thetav_\mu) \right\rangle_{\thetav_\mu}\!=\int\! \rmd\thetav_{\mu}~  P(\thetav_{\mu})  f(\thetav_\mu) $.  Expression (\ref{eq:P(C|X)}) can be used  to infer the most probable partition $\cmatrix$ \cite{Mozeika2018}.  For  each $K\leq N$ we can compute  
\begin{eqnarray}
\hat{\cmatrix}\,\vert K&=& \argmax_{\cmatrix}\, P(\cmatrix\vert \xmatrix,K)\nonumber\\
&=&\argmax_{\cmatrix}\big[\rme^{-N\hat{F}_N(\cmatrix,\, \xmatrix)} P(\cmatrix\vert K) \big]
\label{eq:c|k}
\end{eqnarray}
and the MAP estimator 
\begin{eqnarray}
(\hat{\cmatrix},\hat{K})&=&  \argmax_{\cmatrix,K}\, P(\cmatrix,K|\xmatrix)\nonumber
\\
&=&  \argmax_{\cmatrix,K}\big[\rme^{-N\hat{F}_N(\cmatrix,\, \xmatrix)} P(\cmatrix\vert K) P(K) \big]
\label{eq:C-hat}
\end{eqnarray}
Furthermore, we can use (\ref{eq:P(C|X)}) to compute the distribution of cluster sizes
\begin{eqnarray}
P(K\vert\xmatrix)&=&\frac{\rme^{-N    f_N(K, \xmatrix)  }  P(K) }{\sum_{\tilde{K}=1}^N  P(\tilde{K})\, \rme^{-N  f_N(\tilde{K}, \xmatrix)} }\label{eq:P(K|X)},
\end{eqnarray}
where
\begin{eqnarray}
 f_N(K, \xmatrix)&=&-\frac{1}{N}\log\Big[\sum_{\cmatrix}   \rme^{-N\hat{F}_N(\cmatrix,\, \xmatrix)} P(\cmatrix\vert K)\Big]
 \label{def:f-beta-1}.
\end{eqnarray}
%
% \Red{The distribution $P(K\vert\xmatrix)$ simplifies in the asymptotic regime if  $\min_{K}  f_N(K, \xmatrix)$   is unique and exists for $N\rightarrow\infty$. In this case, averages of the form  $\sum_{K=1}^NP(K\vert\xmatrix)a_K$, where  $a_K$ is  some `test function', are given by 
%%
%\begin{eqnarray}
%%
%\sum_{K=1}^NP(K\vert\xmatrix)a_K
%&=&   \frac{\sum_{K=1}^N  P(K)   \rme^{-N  \Delta_N(K,\xmatrix) }    a_K}{\sum_{K^{\prime}=1}^N  P(K^{\prime})\, \rme^{-N    \Delta_N(K^\prime,\xmatrix)}     }  \label{eq:P(K|X)-large-N-comp-1},
%%
%\end{eqnarray}
%%
%where $\Delta_N(K,\xmatrix)= f_N(K, \xmatrix)  -\min_{\tilde{K}}  f_N (\tilde{K}, \xmatrix  )$.  We note that 
%%
%\begin{eqnarray}
%\sum_{K=1}^N  P(K)   \rme^{-N  \Delta_N(K,\xmatrix) }    a_K &=& \sum_{K=1}^N  \Ind\left[ \Delta_N(K,\xmatrix)\!=\!0 \right] P(K) a_K 
%\label{eq:P(K|X)-large-N-comp-2}
%\\
%&&\hspace*{-10mm} +\sum_{K=1}^N  \Ind\left[ \Delta_N(K,\xmatrix)\!>\!0 \right]  P(K)   \rme^{-N  \Delta_N(K,\xmatrix) }    a_K 
%\nonumber
%%
%\end{eqnarray}
%%
%and hence 
%%
%\begin{eqnarray}
%%
%\sum_{K=1}^NP(K\vert\xmatrix)a_K=\sum_{K=1}^N  \delta_{K,\argmin_{\tilde{K}}  f_N (\tilde{K}, \xmatrix  )} a_K \mbox{ as } N\rightarrow\infty.\label{eq:P(K|X)-large-N-comp-3}
%%
%\end{eqnarray}
%%
%Thus $\lim_{N\to\infty}P(K\vert\xmatrix)=\delta_{K,\argmin_{\tilde{K}} \lim_{N\to\infty} f_N (\tilde{K}, \xmatrix  )}$. }

%
\section{Statistical mechanics and replica approach\label{ssection:replica}}

\subsection{Size independent identities}

When the prior $P(\cmatrix,K)=P(\cmatrix\vert K)P(K)$ is chosen to be uniform\footnote{For non-uniform $P(\cmatrix\vert K)$ we have to minimise  $\hat{F}_N(\cmatrix,\xmatrix) -N^{-1}\log P(\cmatrix\vert K)$ instead of $\hat{F}_N(\cmatrix,\xmatrix)$.}, MAP inference of clusters and cluster numbers according to (\ref{eq:c|k},\ref{eq:C-hat}) requires 
finding the minimum  $\min_{\cmatrix}\hat{F}_N(\cmatrix,\xmatrix)$ of the negative log-likelihood (\ref{def:F-hat}), which is a function of the data  $\xmatrix=(\x_1,\ldots,\x_N)$. Here  we assume that  $\xmatrix$  is sampled from the distribution  % and $\cmatrix$ is the allocation matrix.  
\begin{eqnarray}
q(\xmatrix\vert L)&=&\sum_{\cmatrix}q(\cmatrix\vert L) \left\{\prod_{\nu=1}^L \prod_{i=1}^N     q_{\nu}^{c_{i\nu}}(\x_{i})    \right\}  \label{eq:q(X)}, 
\end{eqnarray}
%which is equivalent to (\ref{eq:q(X)-1}). 
where $q(\cmatrix\vert L)$   and  $q_{\nu}(\x)$ are, respectively, the `true' distribution of partitions, of size $L$,  and the true distribution of data in these partitions.  We note that the above expression will generally differ from the form (\ref{eq:P(C|X)}), which  allows to study various scenarios describing `mismatch'  between the assumed model and the actual data.

The minimum of $\hat{F}_N(\cmatrix,\xmatrix)$ can be computed within the statistical mechanics framework (see e.g.~\cite{Mezard2009}),  via the zero `temperature'  limit  of  the `free energy' (density), using 
$\min_{\cmatrix}\hat{F}_N(\cmatrix,\xmatrix)=\lim_{\beta\rightarrow\infty}     f_N(\beta,\xmatrix) $, with 
\begin{eqnarray}
 f_N(\beta,\xmatrix)&=&-\frac{1}{\beta N}\log \sum_{\cmatrix}\rme^{-\beta N  \hat{F}_N(\cmatrix, \xmatrix)}\label{def:f-emp}
\end{eqnarray}
Although the free energy $ f_N(\beta,\xmatrix)$  is a function of the randomly generated data $\xmatrix$, we expect that  in the thermodynamic limit  $N\rightarrow\infty$, i.e. for  inference with an  infinite amount of data, it will be self-averaging, i.e. $\lim_{N\to\infty}\big\{
\langle  f_N^2(\beta,\xmatrix)\rangle_{\xmatrix} -\langle f_N(\beta,\xmatrix)\rangle_{\xmatrix}^2\big\}=0$. This implies that   instead of (\ref{def:f-emp}) we can work with the average free energy density
\begin{eqnarray}
f_N(\beta)&=&-\frac{1}{\beta N}\Big\langle\log \sum_{\cmatrix}\rme^{-\beta N  \hat{F}_N(\cmatrix, \xmatrix)}\Big\rangle_{\xmatrix}\label{def:f-aver-1},  
\end{eqnarray}
where the average $\left\langle\cdots\right\rangle_{\xmatrix}$ is generated by the distribution $q(\xmatrix\vert L)$. We note that if the prior  $P(\cmatrix\vert K)$ is uniform,  i.e.  $P(\cmatrix\vert K)=1/K!\,\mathcal{S}(N,K)$, then  $f_N(\beta)$ is equivalent to  
\begin{eqnarray}
f_N(\beta)&=&-\frac{1}{\beta N}\Big\langle\log \sum_{\cmatrix} P(\cmatrix\vert K)\rme^{-\beta N  \hat{F}_N(\cmatrix, \xmatrix)}\Big\rangle_{\xmatrix}+\phi_N(\beta)
.\label{def:f-aver-2}
%
%&&~~~~~~~~~~~~~~~~~~~~~~~~~-\frac{1}{\beta N} \log (K!\,\mathcal{S}(N,K))\nonumber
%
\end{eqnarray}
with $\phi_N(\beta)\!=\! -\frac{1}{\beta N}\log [K! \mathcal{S}(N,K)]$. 
The replica identity $\langle\log z\rangle=\lim_{n\rightarrow0}n^{-1}\log\langle z^n\rangle$ allows us to write the relevant part of the average free energy density as
\begin{eqnarray}
\hspace*{-10mm}
f_N(\beta)-\phi_N(\beta)&=&-\lim_{n\rightarrow0}\frac{1}{\beta N n}\log \Big\langle\Big[
\sum_{\cmatrix} P(\cmatrix\vert K)\rme^{-\beta N  \hat{F}_N(\cmatrix, \xmatrix)}\Big]^n\Big\rangle_{\xmatrix}.
\label{eq:f-aver-repl}
\end{eqnarray}
The standard route for  computing  averages  via the replica method~\cite{Mezard1987} is to evaluate the above for integer $n$, following by taking $n\to 0$ via analytical continuation. So
\begin{eqnarray}
\hspace*{-25mm}
 \Big\langle\Big[\sum_{\cmatrix} P(\cmatrix\vert K)\rme^{-\beta N  \hat{F}_N(\cmatrix, \xmatrix)}\Big]^n\Big\rangle_{\!\xmatrix}
\!
&=&  \sum_{\cmatrix^1} \cdots  \sum_{\cmatrix^n} \Big[ \prod_{\alpha=1}^n P(\cmatrix^\alpha\vert K)\Big] \Big\langle  \rme^{-\beta N   \sum_{\alpha=1}^n\hat{F}_N(\cmatrix^\alpha, \xmatrix)}    \Big\rangle_{\!\xmatrix}\nonumber
\hspace*{-2mm}\\
\hspace*{-25mm}
&=& \Big\langle \Big\langle  \rme^{-\beta N   \sum_{\alpha=1}^n\hat{F}_N(\cmatrix^\alpha, \xmatrix)}   \Big\rangle_{\left\{\cmatrix^\alpha\right\}}  \Big\rangle_{\xmatrix}  \label{eq:disorder-aver-comp-1}, 
\end{eqnarray}
where the average $\left\langle  \cdots  \right  \rangle_{\left  \{\cmatrix^\alpha  \right\}}$  refers to the replicated distribution $\prod_{\alpha=1}^n P(\cmatrix^\alpha\vert K)$.   
We next compute the average over $\xmatrix$  (see \ref{app:disorder-average} for details) which leads  us to  the following integral
\begin{eqnarray}
\hspace*{-10mm}
\Big\langle\!\Big\langle  \rme^{-\beta N   \sum_{\alpha=1}^n\hat{F}_N(\cmatrix^\alpha, \xmatrix)}     
\Big\rangle\!\Big\rangle_{\!\!\left\{\cmatrix^\alpha\right\},\xmatrix}
&=& \int\!\{\rmd \Q\,\rmd\hat{\Q}\,\rmd A\,\rmd\hat{A}\}\, \rme^{N    \Psi[\{\Q, \hat{\Q}\};\{A,\hat{A}\}]  }     \label{eq:disorder-aver-comp-8}, 
\end{eqnarray}
with\\[-4mm]
\begin{eqnarray}
\hspace*{-10mm}
\Psi[\{\Q, \hat{\Q}\};\{A,\hat{A}\}]&=&\rmi   \sum_{\alpha=1}^n\sum_{\mu=1}^K  \int\! \rmd \x~ \hat{Q}_{\mu}^\alpha(\x )  Q_{\mu}^\alpha(\x )
+\rmi \sum_{\nu, \muv  }\hat{A}(\nu,\muv)  A(\nu,\muv)
\nonumber\\
\hspace*{-10mm}
&&+ \beta \sum_{\alpha=1}^n\sum_{\mu=1}^K\frac{1}{N}\log \langle \rme^{N\int\rmd\x~Q_{\mu}^\alpha(\x )  \log P(\x\vert\thetav_{\mu})   } \rangle_{\thetav_{\mu}}  \nonumber\\
\hspace*{-10mm}
&&+\sum_{\nu, \muv } A(\nu,\muv ) \log\!     \int \!  \rmd\x~    q_{\nu}(\x)\,    \rme^{-\rmi \sum_{\alpha=1}^n \hat{Q}_{\mu_{\alpha}}^\alpha(\x ) }   \nonumber\\
\hspace*{-10mm}
&&+\frac{1}{N}\log\left\langle\rme^{-\rmi N\sum_{\nu, \muv  }\hat{A}(\nu,\muv)  A(\nu,\muv\vert\cmatrix,\left\{\cmatrix^\alpha\right\})} \right \rangle_{\left\{\cmatrix^\alpha\right\};\cmatrix}
\label{def:Psi-repl},
\end{eqnarray}
where the average $\langle\cdots  \rangle_{\left\{\cmatrix^\alpha\right\};\cmatrix} $ refers to  the distribution  $q(\cmatrix\vert L)\prod_{\alpha=1}^n P(\cmatrix^\alpha\vert K)$. Finally,  using the above result in our formula for the average free energy  (\ref{eq:f-aver-repl}) gives us
\begin{eqnarray}
\hspace*{-10mm}
f_N(\beta)&=&-\lim_{n\rightarrow0}\frac{1}{\beta N n}\log \int\{\rmd \Q\,\rmd\hat{\Q}\,\rmd A\,\rmd\hat{A}\}\, \rme^{N    \Psi[\{\Q, \hat{\Q}\};\{A,\hat{A}\}]  } +\phi_N(\beta)\label{eq:f-aver-repl-int}.
\end{eqnarray}

\subsection{Inference for large $N$\label{subsection:Large-N-inference}}
For finite $N$, equation  (\ref{eq:f-aver-repl-int}) is as complicated as its predecessor (\ref{def:f-aver-2}). The former can, however,  be computed via saddle-point integration when $N\rightarrow\infty$, provided we are allowed to take this limit first and the replica limit  $n\rightarrow0$ later.  Now we obtain
\begin{eqnarray}
f(\beta)&=&-\frac{1}{\beta }\lim_{n\rightarrow0}\frac{1}{n}\extr_{\{\Q, \hat{\Q}, A,\hat{A}\}} \Psi[\{\Q, \hat{\Q}\};\{A,\hat{A}\}]  +\phi(\beta)
\label{eq:f-aver-repl-large-N},
\end{eqnarray}
where $\phi(\beta)=\lim_{N\rightarrow\infty}\phi_N(\beta)$.  The further calculation requires knowledge of   the average in the last term of the functional  (\ref{def:Psi-repl}), which 
 can be written in the form 
\begin{eqnarray}
\hspace*{-19mm}
\left\langle\rme^{-\rmi N\sum_{\nu, \muv  }\hat{A}(\nu,\muv)  A(\nu,\muv\vert\cmatrix,\left\{\cmatrix^\alpha\right\})} \right \rangle_{\!\!\left\{\cmatrix^\alpha\right\};\cmatrix} 
&=&\!\! \!\sum_{\left\{N(\nu,\muv)\right\}}  \! \!\! P_N\!\left[ \left\{N(\nu,\muv)\right\}\right]  \rme^{-\rmi\sum_{\nu, \muv  }\hat{A}(\nu,\muv) N(\nu,\muv)    }\!,
\nonumber
\\[-2mm]
&&\label{eq:mgf}
\end{eqnarray}
 where the set of  variables $\left\{N(\nu,\muv)\right\}$, which are governed by the distribution 
\begin{eqnarray}
\hspace*{-15mm}
P_N\!\left[ \left\{N(\nu,\muv)\right\}\right]
&&=\sum_\cmatrix \sum_{\left\{\cmatrix^\alpha\right\}}  q(\cmatrix\vert L) \left\{\prod_{\alpha=1}^n p(\cmatrix^\alpha\vert K)\right\} \prod_{\nu, \muv  } \delta_{N(\nu,\muv);   N A(\nu,\muv\vert\cmatrix,\left\{\cmatrix^\alpha\right\}) },
\nonumber
\\[-2mm]&&
\label{def:Prob-N(nu,mu)}
\end{eqnarray}
are  subject to the hard constraints  $\sum_{\nu,\muv} N(\nu,\muv)=N$ (the sample size), $\sum_{\muv} N(\nu,\muv)=N(\nu)$ (the sample size  of a data generated from $q_\nu(\x)$), and $\sum_{\nu,\muv\setminus\mu_\alpha} N(\nu,\muv)=N(\mu_\alpha)>0$ (the size of  the cluster $\mu_\alpha$ in replica $\alpha$). To compute the average (\ref{eq:mgf}) we will  assume that  for $N\to\infty$ the distribution $P_N\!\left[ \left\{N(\nu,\muv)\right\}\right]$ approaches  the associated (soft constrained) multinomial distribution 
\begin{eqnarray}
\tilde{P}_N\!\left[ \left\{N(\nu,\muv)\right\}\right]&=& 
\frac{N!}{\prod_{\nu,\muv}N(\nu,\muv)!}\prod_{\nu,\muv} \tilde{A}(\nu,\muv)^{N(\nu,\muv)}\label{def:mult},
\end{eqnarray}
where $\sum_{\nu,\muv}\tilde{A}(\nu,\muv)=1$ and  $\tilde{A}(\nu,\muv)>0$.  In this case   we would find simply
\begin{eqnarray}
\hspace*{-10mm}
\left\langle\rme^{-\rmi N\sum_{\nu, \muv  }\hat{A}(\nu,\muv)  A(\nu,\muv\vert\cmatrix,\left\{\cmatrix^\alpha\right\})} \right \rangle_{\!\!\left\{\cmatrix^\alpha\right\};\cmatrix}  &=&   \Big\{\sum_{\nu, \muv  } \tilde{A}(\nu,\muv) \,\rme^{-\rmi\hat{A}(\nu,\muv)    }\Big\}^N\label{eq:mgf-mult}.
\end{eqnarray}
The 
above assumption can by  justified  by the following  large deviations argument.  

\subsection{Particle gas representation of replicated partitions\label{sssection:urn}}
The multinomial distribution  (\ref{def:mult}) describes $n$ copies, i.e. replicas,  of $N$ `particles'  distributed over $K$  reservoirs. For  $\A=(\av_1,\ldots,\av_N)$ this distribution is given by 
\begin{eqnarray}
P(\A)=\prod_{i=1}^N P(\av_i)  \label{def:P(A)},
\end{eqnarray}
 where $P(\av_i)=\tilde{A}(\nu,\muv)= {\rm Prob}(a_i(1)\!=\!\nu,a_i(2)\!=\!\mu_1,\ldots,a_i(n\!+\!1)\!=\!\mu_n)$  denotes the probability that a particle $i$ has `colour' $\nu\in[L]$ and is in `reservoir' $\mu_1\in[K]$ of replica $n=1$,  reservoir $\mu_2\in[K]$ of replica $n=2$, etc.  The state $\A$ of this `gas'  of particles is a `partition' if the reservoirs are not empty,  i.e.  if $N_{\mu_\alpha}^\alpha(\A)= \sum_{i\leq N}\delta_{\mu_\alpha;\, a_i(\alpha+1)}>0$ for all $\alpha$ and $\mu_\alpha$. If $\A$ is sampled from the distribution $P(\A)$, this will happen with high probability as $N\rightarrow\infty$ if the marginal $\tilde{A}(\mu_\alpha)=\sum_{\nu, \muv\setminus\mu_\alpha}\tilde{A}(\nu,\muv)>0$. 
To show this we  first compute  the average  $\left\langle N_{\mu_\alpha}^\alpha(\A)\right\rangle_\A=\sum_{\A}P(\A)\,N_{\mu_\alpha}^\alpha(\A)$: 
\begin{eqnarray}
\hspace*{-10mm}
\left\langle N_{\mu_\alpha}^\alpha(\A)\right\rangle_\A&=&\sum_{i=1}^N \sum_{\av_i} P(\av_i)\,\delta_{\mu_\alpha;\, a_i(\alpha+1)}\nonumber\\
\hspace*{-10mm}
&=&N \sum_{\av_i} P(\av_i)\delta_{\mu_\alpha;\, a_i(\alpha+1)}=N\!\sum_{\nu, \muv\setminus\mu_\alpha}\tilde{A}(\nu,\muv)
=N \tilde{A}(\mu_\alpha)\label{eq:<N>}.~~~
\end{eqnarray}
Thus the average $\left\langle N_{\mu_\alpha}^\alpha(\A)\right\rangle_\A>0$.  Secondly,  for $\epsilon>0$ we consider the  probability of observing the event $N_{\mu_\alpha}^\alpha(\A)\notin(N (\tilde{A}(\mu_\alpha)-\epsilon ),N (\tilde{A}(\mu_\alpha)+\epsilon ))$. Clearly, 
\begin{eqnarray}
&&\hspace*{-15mm}
{\rm Prob}\left (N_{\mu_\alpha}^\alpha\!(\A)\notin(N (\tilde{A}(\mu_\alpha)\!-\!\epsilon ),N (\tilde{A}(\mu_\alpha)\!+\!\epsilon )) \right ) 
\label{eq:tails} \\
&=&{\rm Prob}\left  (  N_{\mu_\alpha}^\alpha\!(\A)/N\leq \tilde{A}(\mu_\alpha)\!-\!\epsilon \right  ) +{\rm Prob}\left  (  N_{\mu_\alpha}^\alpha\!(\A)/N \geq  \tilde{A}(\mu_\alpha)\!+\!\epsilon \right  ).
\nonumber
\end{eqnarray}
For any $\lambda>0$, the second term can be bounded using Markov's inequality, as follows 
\begin{eqnarray}
\hspace*{-10mm}
{\rm Prob}\left(  N_{\mu_\alpha}^\alpha\!(\A)/N \geq  \tilde{A}(\mu_\alpha)\!+\!\epsilon \right) 
&=&  {\rm Prob}\left  (  \rme^{\lambda N_{\mu_\alpha}^\alpha(\A)} \geq  \rme^{\lambda N (\tilde{A}(\mu_\alpha)+\epsilon )}\right  )  \nonumber\\
\hspace*{-10mm}
&\leq&    \langle\rme^{\lambda N_{\mu_\alpha}^\alpha(\A)} \rangle_{\A} ~ \rme^{-\lambda N (\tilde{A}(\mu_\alpha)+\epsilon )}\label{eq:r-tail-comp-1},
\end{eqnarray}
with the average
\begin{eqnarray}
\langle \rme^{\lambda N_{\mu_\alpha}^\alpha(\A)}  \rangle_{\A}&=&\sum_{\A}P(\A)\, \rme^{\lambda N_{\mu_\alpha}^\alpha(\A)}= \prod_{i=1}^N\left\{\sum_{\av_i} P(\av_i) \,  \rme^{\lambda\delta_{\mu_\alpha;\, a_i(\alpha+1)}    }\right\}
\nonumber
\\
&=& \big[1+\tilde{A}(\mu_\alpha) (\rme^\lambda -1)   \big]^N\label{eq:r-tail-comp-2}.
\end{eqnarray}
Hence
\begin{eqnarray}
&&{\rm Prob}\left  (  N_{\mu_\alpha}^\alpha(\A) \geq  N (\tilde{A}(\mu_\alpha)\!+\!\epsilon )\right  )  \leq \rme^{-N \mathrm{I}(\lambda,\epsilon)}\label{eq:r-tail-comp-3},
\end{eqnarray}
where $\mathrm{I}(\lambda,\epsilon)=-\log(1+\tilde{A}(\mu_\alpha) (\rme^\lambda -1)   )+  \lambda(\tilde{A}(\mu_\alpha)+\epsilon)$ is a \emph{rate function}. The latter  has its maximum at $\lambda^*= \log [ ( \tilde{A}(\mu_\alpha)^{2}\!+\!\tilde{A}(\mu_\alpha)\epsilon\!-\!\tilde{A}(\mu_\alpha)\!-\!\epsilon)/(\tilde{A}(\mu_\alpha) ( \tilde{A}(\mu_\alpha)\!-\!1\!+\!
\epsilon )  ]
$, and  $\mathrm{I}(\lambda^*,\epsilon)=D( \tilde{A}(\mu_\alpha)\!+\!\epsilon \, \vert  \vert \tilde{A}(\mu_\alpha)    )$, where $D( p \, \vert  \vert q   )=p\log(\frac{p}{q})+(1-p)\log(\frac{1-p}{1-q})\geq0$ is the Kullback-Leibler divergence~\cite{Cover2012} of binary distributions with probabilities $p,q\in[0,1]$. We may now write
\begin{eqnarray}
{\rm Prob}\left  (  N_{\mu_\alpha}^\alpha(\A) \geq  N (\tilde{A}(\mu_\alpha)\!+\!\epsilon )\right  )  &\leq& \rme^{-N      D( \tilde{A}(\mu_\alpha)+\epsilon \, \vert  \vert \tilde{A}(\mu_\alpha)    ) }\label{eq:r-tail-bound}.
\end{eqnarray}
Following similar steps to bound the first term of (\ref{eq:tails}) gives us also the inequality
\begin{eqnarray}
{\rm Prob} \left ( N_{\mu_\alpha}^\alpha(\A) \leq  N (\tilde{A}(\mu_\alpha)\!-\!\epsilon )\right  ) 
&\leq& \rme^{-N      D( \tilde{A}(\mu_\alpha)-\epsilon \, \vert  \vert \tilde{A}(\mu_\alpha)    ) }\label{eq:l-tail-bound}.
\end{eqnarray}
In combination, our two bounds directly lead to
\begin{eqnarray}
&&\hspace*{-15mm} {\rm Prob}\left  (  N_{\mu_\alpha}^\alpha(\A)\notin(N (\tilde{A}(\mu_\alpha)\!-\!\epsilon ),N (\tilde{A}(\mu_\alpha)\!+\!\epsilon )) \right )  \nonumber\\
&\leq&  2\,\rme^{-N     \min_{\sigma\in\{-1,1\}} D( \tilde{A}(\mu_\alpha)+\sigma\epsilon \, \vert  \vert \tilde{A}(\mu_\alpha)    ) } \label{eq:tails-bound-1}
\end{eqnarray}
The probability for one or more of the events $N_{\mu_\alpha}^\alpha\!(\A)\notin(N (\tilde{A}(\mu_\alpha)\!-\!\epsilon ),N (\tilde{A}(\mu_\alpha)\!+\!\epsilon ))$ to occur (of which there are $nK$ )   can be bounded  using Boole's inequality in combination with (\ref{eq:tails-bound-1}), as follows
\begin{eqnarray}
&&\hspace*{-15mm} {\rm Prob}\left  (  \cup_{\alpha,\mu_\alpha} \left\{N_{\mu_\alpha}^\alpha(\A)\notin(N (\tilde{A}(\mu_\alpha)\!-\!\epsilon ),N (\tilde{A}(\mu_\alpha)\!+\!\epsilon )) \right\}\right  ) \nonumber\\
&\leq &   \sum_{\alpha=1}^n  \sum_{\mu_\alpha=1}^K {\rm Prob}\left  (  N_{\mu_\alpha}^\alpha(\A)\notin(N (\tilde{A}(\mu_\alpha)\!-\!\epsilon )),N (\tilde{A}(\mu_\alpha)\!+\!\epsilon )) \right )\nonumber\\  
&\leq &   2nK\,\rme^{-N     \min_{\alpha,\mu_\alpha} \min_{\sigma\in\{-1,1\}} D( \tilde{A}(\mu_\alpha)+\sigma\epsilon \, \vert  \vert \tilde{A}(\mu_\alpha)    ) } \label{eq:tails-bound-2}.
\end{eqnarray}
We conclude that for $N\rightarrow\infty$ the deviations of the  random variables $N_{\mu_\alpha}^\alpha(\A)$  from  their averages $N \tilde{A}(\mu_\alpha)$  decay exponentially with $N$. 

Let us next consider the  entropy density 
\begin{eqnarray}
H(\A)/N&=&-\sum_{\av_i} P(\av_i)   \log P(\av_i)= - \sum_{\nu,\muv} \tilde{A}(\nu,\muv)\log \tilde{A}(\nu,\muv)\nonumber\\
&=&-\sum_{\nu} \tilde{A}(\nu)\log \tilde{A}(\nu)- \sum_{\nu,\muv} \tilde{A}(\nu)\tilde{A}(\muv\vert\nu)\log \tilde{A}(\muv\vert\nu)\label{eq:H(A)}.
\end{eqnarray}
If we assume that 
\begin{eqnarray}
\tilde{A}(\muv\vert\nu)&=&\prod_{\alpha=1}^n \tilde{A}(\mu_\alpha\vert\nu)\label{eq:A-prior},
\end{eqnarray}
then 
\begin{eqnarray}
H(\A)/N&=&-\sum_{\nu} \tilde{A}(\nu)\log \tilde{A}(\nu)- n\sum_{\nu,\mu} \tilde{A}(\nu)\tilde{A}(\mu\vert\nu)\log \tilde{A}(\mu\vert\nu)\label{eq:H(A)-RS}.
\end{eqnarray}
The entropy of the distribution  $q(\cmatrix\vert L) \left\{\prod_{\alpha=1}^n p(\cmatrix^\alpha\vert K)\right\}$, used in (\ref{def:Prob-N(nu,mu)}), is given by 
\begin{eqnarray}
\hspace*{-18mm}
H(p,q)/N&=&-\frac{1}{N}\sum_\cmatrix \sum_{\left\{\cmatrix^\alpha\right\}}  q(\cmatrix\vert L) \Big[\prod_{\alpha=1}^n p(\cmatrix^\alpha\vert K)\Big]\log \Big\{q(\cmatrix\vert L) \Big[\prod_{\alpha=1}^n p(\cmatrix^\alpha\vert K)\Big]\Big\} \nonumber
\\
\hspace*{-18mm}
&=&H(q)/N+nH(p)/N\label{eq:H(p,q)},
\end{eqnarray}
with $H(q)=-\sum_\cmatrix  q(\cmatrix\vert L)\log q(\cmatrix\vert L) $ and $H(p)=-\sum_\cmatrix  p(\cmatrix\vert K)\log p(\cmatrix\vert K)$.  For the case of uniform distributions $q(\cmatrix\vert L)=1/L!\mathcal{S}(N,L)$ and $p(\cmatrix\vert K)=1/K!\mathcal{S}(N,K)$ the latter entropies are, respectively,  $\log (L!\mathcal{S}(N,L))$ and $\log (K!\mathcal{S}(N,K))$. This gives us $H(p,q)/N=\log(L)+n\log(K)$ in the limit $N\rightarrow\infty$.  Comparing this asymptotic result for $H(p,q)/N$ with $H(\A)/N$ in (\ref{eq:H(A)-RS}), we see that the two expressions are equal for large $N$ when $\tilde{A}(\nu)=1/L$ and $\tilde{A}(\mu\vert\nu)=1/K$. In this case, the distribution (\ref{def:Prob-N(nu,mu)}) apparently approaches the multinomial distribution (\ref{def:mult}).  We expect this also to be true when the distribution $q(\cmatrix\vert L)$ is uniform,  but subject to the constraints $\sum_{i=1}^Nc_{i\nu}=N\tilde{A}(\nu)$. 

\section{ Replica Symmetric theory\label{section:RS}}

\subsection{Simplification of the saddle-point problem}
Using the assumptions  (\ref{eq:mgf-mult}) and (\ref{eq:A-prior}), we obtain  a simplified expression for (\ref{def:Psi-repl}):
\begin{eqnarray}
\hspace*{-15mm}
\Psi[\{\Q, \hat{\Q}\};\{A,\hat{A}\}]&=&\rmi   \sum_{\alpha=1}^n\sum_{\mu=1}^K  \int\!\rmd \x~ \hat{Q}_{\mu}^\alpha(\x )  Q_{\mu}^\alpha(\x ) 
\nonumber
\\
\hspace*{-15mm}
&&
+\sum_{\nu, \muv  }A(\nu,\muv)\Big[\rmi\hat{A}(\nu,\muv)  
+\log\!  \int \! \rmd\x ~     q_{\nu}(\x)\,    \rme^{-\rmi \sum_{\alpha=1}^n \hat{Q}_{\mu_{\alpha}}^\alpha(\x ) }  \Big]
\nonumber\\
\hspace*{-15mm}
&&+ \beta \sum_{\alpha=1}^n\sum_{\mu=1}^K\frac{1}{N}\log \Big\langle \rme^{N\int\!\rmd\x  ~Q_{\mu}^\alpha(\x )  \log P(\x\vert\thetav_{\mu})   } \Big\rangle_{\thetav_{\mu}}  \nonumber\\
\hspace*{-15mm}
&&+\log  \Big[\sum_{\nu, \muv  } \tilde{A}(\nu)\rme^{-\rmi\hat{A}(\nu,\muv)    } \prod_{\alpha=1}^n \tilde{A}(\mu_\alpha\vert\nu)\Big]
\label{eq:Psi-comp-1} 
\end{eqnarray}
 The extrema of this functional are seen to be the solutions of the following equations:
\begin{eqnarray}
\hat{A}(\nu,\muv)&=&\rmi  \log  \int \! \rmd\x~    q_{\nu}(\x)\,    \rme^{-\rmi \sum_{\alpha=1}^n \hat{Q}_{\mu_{\alpha}}^\alpha(\x ) }   \\
A(\nu,\muv)&=&  \frac{\tilde{A}(\nu) \rme^{-\rmi \hat{A}(\nu,\muv)}\prod_{\alpha=1}^n\tilde{A}(\mu_\alpha\vert\nu)}{ \sum_{\tilde{\nu}, \tilde{\muv}  } \tilde{A}(\tilde{\nu}) \rme^{-\rmi \hat{A}(\tilde{\nu},\tilde{\muv})}\prod_{\alpha=1}^n\tilde{A}(\tilde{\mu}_\alpha\vert\tilde{\nu})}\\
% \sum_{\alpha=1}^n\sum_{\mu=1}^K  \int \hat{Q}_{\mu}^\alpha(\x )  Q_{\mu}^\alpha(\x ) \rmd \x
%
Q_{\mu}^\alpha(\x ) &=&\sum_{\nu, \muv }\delta_{\mu;\mu_\alpha}   A(\nu,\muv)    \frac{     q_{\nu}(\x)\,    \rme^{-\rmi \sum_{\gamma=1}^n \hat{Q}_{\mu_{\gamma}}^\gamma(\x ) }  }{   \int \! \rmd\tilde{\x} ~    q_{\nu}(\tilde{\x})\,    \rme^{-\rmi \sum_{\gamma=1}^n \hat{Q}_{\mu_{\gamma}}^\gamma(\tilde{\x} ) }     } \\
\hat{Q}_{\mu}^\alpha(\x ) &=&\rmi\beta \frac{ \langle \rme^{N\int \!\rmd\tilde{\x}  ~Q_{\mu}^\alpha(\tilde{\x} )  \log P(\tilde{\x}\vert\thetav)  } \log P(\x\vert\thetav)\rangle_{\thetav} }{ \langle \rme^{N\int\! \rmd\tilde{\x} ~ Q_{\mu}^\alpha(\tilde{\x} )  \log P(\tilde{\x}\vert\thetav)   } \rangle_{\thetav}} \label{eq:SP-comp-1}. 
\end{eqnarray}
For $N\to\infty$ we can evaluate 
the integrals in the  last equation with the Laplace method~\cite{DeBruijn1981},  giving
\begin{eqnarray}
\hat{Q}_{\mu}^\alpha(\x )&=&\rmi\beta \log P(\x\vert\thetav_{\mu}^\alpha)\nonumber\\
\thetav_{\mu}^\alpha&=&\argmax_{\thetav}  \int\!\rmd\x~ Q_{\mu}^\alpha(\x )  \log P(\x\vert\thetav)\label{eq:SP-comp-2}.
\end{eqnarray}
Upon eliminating the conjugate order parameters $\{\hat{\Q},\hat{A}\}$ from our coupled equations  and considering large $N$, we obtain after some straightforward manipulations the following expression for the nontrivial part of the average free energy (\ref{eq:f-aver-repl-large-N}), 
\begin{eqnarray}
\hspace*{-20mm}
f(\beta)-\phi(\beta) &=&-\lim_{n\rightarrow0}\frac{1}{\beta n}
\log  \Bigg\{
\sum_{\nu} \tilde{A}(\nu) \int \! \rmd\x~    q_{\nu}(\x)  \prod_{\alpha=1}^n\Bigg[\sum_{\mu=1}^K 
 \tilde{A}(\mu\vert\nu)
 \rme^{\beta\log P(\x\vert\thetav_{\mu}^\alpha) } \Bigg]
\Bigg\}
\nonumber
\\[-0mm] \hspace*{-20mm}&& \label{eq:f-R}
\end{eqnarray}
and the following closed equations for the remaining order parameters $\{\Q,A\}$:
\begin{eqnarray}
\hspace*{-10mm} 
Q_{\mu}^\alpha(\x ) &=&\sum_{\nu, \muv }\delta_{\mu;\mu_\alpha}  A(\nu,\muv)   \frac{     q_{\nu}(\x)\,  
    \rme^{\sum_{\gamma=1}^n \beta \log P(\x\vert\thetav_{\mu_\gamma}^\gamma) }
 }{   \int \!  \rmd\tilde{\x}~    q_{\nu}(\tilde{\x})\,   
      \rme^{\sum_{\gamma=1}^n \beta \log P( \tilde{\x}\vert\thetav_{\mu_\gamma}^\gamma) }
    }  \label{eq:Q-R}, 
\\
\hspace*{-10mm} 
A(\nu,\muv)&=&  \frac{\tilde{A}(\nu)  
 \int \!  \rmd\x   ~   q_{\nu}(\x)     \Big[ \prod_{\alpha=1}^n\tilde{A}(\mu_\alpha\vert\nu)\, \rme^{ \beta \log P(\x\vert\thetav_{\mu_\alpha}^\alpha) }  \Big]
    }{ \sum_{\tilde{\nu}} \tilde{A}(\tilde{\nu})
    \int \!  \rmd\x  ~   q_{\tilde{\nu}}(\x)   \Big[
    \prod_{\alpha=1}^n\sum_{\tilde{\mu}_\alpha}\tilde{A}(\tilde{\mu}_\alpha\vert\tilde{\nu})\, \rme^{\beta \log P(\x\vert\thetav_{\tilde{\mu}_\alpha}^\alpha) }   \Big]
    }\label{eq:A-R}
\end{eqnarray}
In order to take the replica limit $n\rightarrow0$  in  (\ref{eq:f-R},\ref{eq:Q-R},\ref{eq:A-R}) we will make the the `replica symmetry'  (RS) assumption~\cite{Mezard1987}, which here translates into  $Q_{\mu_\alpha}^\alpha\!(\x )=Q_{\mu_\alpha}\!(\x )$. It then follows from (\ref{eq:SP-comp-2}), in turn,  that $\thetav_{\!\mu}^\alpha=\thetav_{\!\mu_\alpha}$.  The RS structure allows us to take the replica limit (see \ref{app:replica-limit} for details) and find the following equations:
\begin{eqnarray}
Q_{\mu}(\x ) &=&\sum_{\nu}\tilde{A}(\nu)\, q_{\nu}(\x)  \frac{
   \tilde{A}(\mu\vert\nu) \,\rme^{    \beta \log P(   \x\vert\thetav_{\mu}      )      }     
    }{   \sum_{\tilde{\mu}}\tilde{A}(\tilde{\mu}\vert\nu)\,\rme^{    \beta \log P(   \x\vert\thetav_{\tilde{\mu}}      )      }       }\nonumber\\
    \thetav_{\mu}&=&\argmax_{\thetav}  \int\!\rmd\x~ Q_{\mu}(\x )  \log P(\x\vert\thetav)
    \label{eq:Q-RS}
 \\
A(\mu\vert\nu)&=&\int \!  \rmd\x~    q_{\nu}(\x)   \frac{\tilde{A}(\mu\vert\nu)\,\rme^{ \beta \log P(\x\vert\thetav_{\mu}) }  }{ \sum_{\tilde{\mu}}\tilde{A}(\tilde{\mu}\vert\nu)\, \rme^{ \beta \log P(\x\vert\thetav_{\tilde{\mu}}) } } 
\nonumber\\
A(\nu)&=&  \tilde{A}(\nu)\label{eq:A-RS}
\end{eqnarray}
and the asymptotic form of the average free energy
\begin{eqnarray}
\hspace*{-10mm}
f(\beta)&=&-\frac{1}{\beta } \!\int\!\rmd\x \sum_{\nu=1}^L \tilde{A}(\nu)     q_{\nu}(\x) \log\Big[    \sum_{\mu=1  }^K   \tilde{A}(\mu\vert\nu)\,   \rme^{ \beta \log P(\x\vert\thetav_{\mu}) } \Big] +\phi(\beta)
 \label{eq:f-RS}.
\end{eqnarray}
The physical meaning of  the order parameters $Q_{\mu}(\x )$ and $A(\mu\vert\nu)$  becomes clear if we define the following two densities
\begin{eqnarray}
Q_\mu(\x\vert\cmatrix,\xmatrix)&=&\frac{1}{N}\sum_{i=1}^N c_{i\mu}\,\delta(\x-\x_i)   \label{def:Q-order-parameter}
\\
A(\nu,\mu\vert\cmatrix,\xmatrix)&=& \frac{1}{N}\!\sum_{i=1}^N  c_{i\mu}  \Ind\left[ \x_i \sim q_\nu(\x)\right] \label{def:A-order-parameter}.
\end{eqnarray}
If we sample  $\cmatrix$  from the Gibbs-Boltzmann distribution 
\begin{eqnarray}
P_\beta(\cmatrix\vert\xmatrix) &=&\frac{1}{Z_\beta(\xmatrix)   }P(\cmatrix\vert K)\rme^{-\beta N  \hat{F}_N(\cmatrix, \xmatrix)}\label{def:P(C|X)},
\end{eqnarray}
where $Z_\beta(\xmatrix)=\sum_{\cmatrix} P(\cmatrix\vert K)\rme^{-\beta N  \hat{F}_N(\cmatrix, \xmatrix)}$ is the associated partition function,  and with the conditional averages $\left\langle G(\cmatrix)\right\rangle_{\cmatrix\vert\xmatrix}=\sum_{\cmatrix} P(\cmatrix\vert K)G(\cmatrix)$, then one finds that 
\begin{eqnarray}
 Q_\mu(\x) &=&\lim_{N\rightarrow\infty}\left\langle \left\langle  Q_\mu(\x\vert\cmatrix,\xmatrix)\right\rangle_{\cmatrix\vert\xmatrix}  \right\rangle_{\xmatrix} \label{eq:Q-phys-meaning},
\\
 A(\nu,\mu) &=&\lim_{N\rightarrow\infty}  \left\langle \left\langle  A(\nu,\mu\vert\cmatrix,\xmatrix)\right\rangle_{\cmatrix\vert\xmatrix}  \right\rangle_{\xmatrix} \label{eq:A-phys-meaning},
\end{eqnarray}
(see \ref{app:physical-meaning} for details). So, asymptotically,  $ Q_\mu(\x) $ 
 is the average distribution of data in cluster $\mu$, and $A(\nu,\mu) $
is the average fraction of data originating from the distribution $q_\nu(\x)$ that are allocated by the clustering process to cluster $\mu$. 

\subsection{RS theory for $\beta\rightarrow\infty$ \label{sssection:RS-T-0}}

Let us study the behaviour of the RS order parameter equations (\ref{eq:Q-RS}), (\ref{eq:A-RS}) and (\ref{eq:f-RS}) in the zero temperature limit  $\beta\rightarrow\infty$.  First, for the order parameter $Q_{\mu}(\x )$, governed by the equation  (\ref{eq:Q-RS}), and any  test function $a_\mu$ we consider the sum 
\begin{eqnarray}
\hspace*{-15mm}
\sum_{\mu}Q_{\mu}(\x ) a_\mu 
&=& \sum_{\nu} \tilde{A}(\nu)\,  q_{\nu}(\x)\,   \frac{ \sum_{\mu} \tilde{A}(\mu\vert\nu)\,
    \rme^{    \beta \log P(   \x\vert\thetav_{\mu}      )     }     
    a_\mu  }{   \sum_{\mu^{\prime\prime}}\tilde{A}(\mu^{\prime\prime}\vert\nu)\,\rme^{    \beta \log P(   \x\vert\thetav_{\mu^{\prime\prime}}      )      }      } \nonumber\\
    \hspace*{-15mm}
&=& \sum_{\nu} \tilde{A}(\nu)\, q_{\nu}(\x)\,\frac{ \sum_{\mu^\prime} \tilde{A}(\mu^\prime\vert\nu)\,
    \rme^{   - \beta (  \max_{\tilde{\mu}}\log P(   \x\vert\thetav_{\tilde{\mu}}      ) -\log P(   \x\vert\thetav_{\mu^\prime}      ) )   }     
    a_{\mu^\prime}  }{   \sum_{\mu^{\prime\prime}}\tilde{A}(\mu^{\prime\prime}\vert\nu)\,   \rme^{   - \beta( \max_{\tilde{\mu}}\log P(   \x\vert\thetav_{\tilde{\mu}}      )-\log P(   \x\vert\thetav_{\mu^{\prime\prime}}     )  )   }          } \nonumber\\ 
    \hspace*{-15mm}
    &=& \sum_{\nu} \tilde{A}(\nu)\, q_{\nu}(\x)\,\frac{ \sum_{\mu^\prime} \tilde{A}(\mu^\prime\vert\nu)\,
    \rme^{   - \beta \Delta_{\mu^\prime}(     \x  )   }     
    a_{\mu^\prime}  }{   \sum_{\mu^{\prime\prime}}\tilde{A}(\mu^{\prime\prime}\vert\nu)\,   \rme^{   - \beta    \Delta_{\mu^{\prime\prime}}(     \x  )   }          }, \label{eq:Q-RS-T-0-comp-1}
    \end{eqnarray}
where $ \Delta_{\mu}(     \x  )= \max_{\tilde{\mu}}\log P(   \x\vert\thetav_{\tilde{\mu}}      ) -\log P(   \x\vert\thetav_{\mu}      ) $. For $\beta\to\infty$ the average will tend to
\begin{eqnarray}
\hspace*{-10mm}
\lim_{\beta\to\infty}
\frac{ \sum_{\mu^\prime} \tilde{A}(\mu^\prime\vert\nu)\,
    \rme^{   - \beta \Delta_{\mu^\prime}(     \x  )   }    \, 
    a_{\mu^\prime}  }{   \sum_{\mu^{\prime\prime}}\tilde{A}(\mu^{\prime\prime}\vert\nu)\,   \rme^{   - \beta    \Delta_{\mu^{\prime\prime}}(     \x  )   }          } 
&=& \frac{\sum_{\mu^\prime}\Ind\left[    \Delta_{\mu^\prime}(     \x  )=0         \right] \tilde{A}(\mu^\prime\vert\nu)\,  a_{\mu^\prime}  
     }{   \sum_{\mu^{\prime\prime}} \Ind\left[     \Delta_{\mu^{\prime\prime}}(     \x  )=0             \right]  \tilde{A}(\mu^{\prime\prime}\vert\nu)         }.
      \label{eq:Q-RS-T-0-comp-2}
    \end{eqnarray}
Hence for $\beta\to\infty$ we may write
\begin{eqnarray}
Q_{\mu}(\x ) &=&\sum_{\nu} \tilde{A}(\nu)\, q_{\nu}(\x)\,
    \frac{    \Ind\left[    \Delta_{\mu}(     \x  )=0         \right] \tilde{A}(\mu \vert\nu) 
     }{   \sum_{\mu^{\prime}} \Ind\left[     \Delta_{\mu^{\prime}}(     \x  )=0             \right]  \tilde{A}(\mu^{\prime}\vert\nu)         }
    \label{eq:Q-RS-T-0}.
     %
    %  \thetav_{\mu}&=&\argmax_{\thetav}  \int\!\rmd\x~ Q_{\mu}(\x )  \log P(\x\vert\thetav)
    \end{eqnarray}
Similarly, equation  (\ref{eq:A-RS}) for the order parameter $A(\mu\vert\nu)$ gives us
\begin{eqnarray}
\sum_\mu A(\mu\vert\nu) a_\mu &=&
\int \!  \rmd\x~    q_{\nu}(\x)   \frac{\sum_\mu \tilde{A}(\mu\vert\nu)\,\rme^{- \beta \Delta_{\mu}(\x ) }  a_\mu}{ \sum_{\tilde{\mu}}\tilde{A}(\tilde{\mu}\vert\nu)\, \rme^{- \beta \Delta_{\tilde{\mu}}(\x  ) } } ,
\end{eqnarray}
so for $\beta\to\infty$ we may write, assuming the expectation and limit operators commute, 
\begin{eqnarray}
\sum_\mu A(\mu\vert\nu) a_\mu &=&
\int \!   \rmd\x~   q_{\nu}(\x) \frac{\sum_\mu   \Ind\left[     \Delta_{\mu }(     \x  )=0             \right]  \tilde{A}(\mu\vert\nu)  \,   a_\mu}{ \sum_{\tilde{\mu}}    \Ind\left[     \Delta_{\tilde{\mu} }(     \x  )=0             \right]  \tilde{A}(\tilde{\mu}\vert\nu) }.
\label{eq:A-RS-T-0-comp-1}
\end{eqnarray}
We note that $A(\mu)=\int\!\rmd\x~ Q_{\mu}(\x )$, as a  consequence  of  the  (\ref{def:Q-order-parameter})  and (\ref{def:A-order-parameter}).
Finally, taking $\beta\rightarrow\infty$  in the average free energy density (\ref{eq:f-RS}) gives us 
\begin{eqnarray}
\hspace*{-20mm}
&&
\hspace*{-10mm}
\lim_{\beta\to\infty}\Big[f(\beta)-\phi(\beta)\Big]
\nonumber\\
\hspace*{-20mm}
&=&-\lim_{\beta\to\infty}\frac{1}{\beta }\! \sum_{\nu} \tilde{A}(\nu) \!\int\! \rmd\x~    q_{\nu}(\x)\log\Big[  \rme^{ \beta\max_{\tilde{\mu} }\log P(\x\vert\thetav_{\tilde{\mu}}) }   \sum_{\mu=1  }^K   \tilde{A}(\mu\vert\nu)   \rme^{ -\beta \Delta_{\mu}(\x   ) } \Big] \nonumber\\
\hspace*{-20mm}
&=&- \sum_{\nu} \tilde{A}(\nu) \int\!    \rmd\x ~q_{\nu}(\x)  \max_{\mu}\log P(   \x\vert\thetav_{\mu}      )  \nonumber\\
\hspace*{-20mm}
&&-\lim_{\beta\to\infty}\frac{1}{\beta }\sum_{\nu} \tilde{A}(\nu)\int\! \rmd\x~  q_{\nu}(\x)
\nonumber
\\
\hspace*{-20mm}&&
\hspace*{5mm}\times\log\Bigg[     \sum_{\mu=1  }^K   \tilde{A}(\mu\vert\nu)\Big(   
\Ind\left[   \Delta_{\mu}(     \x  )\!>\!0  \right]\rme^{ -\beta \Delta_\mu (\x  ) } 
 +\Ind\left[     \Delta_{\mu }(     \x  )\!=\!0  \right] \Big) \Bigg ] \nonumber\\
\hspace*{-20mm}
&=&- \!\int\!  \rmd\x \sum_{\nu} \tilde{A}(\nu)     q_{\nu}(\x)  \max_{\mu}\log P(   \x\vert\thetav_{\mu}      ) \nonumber\\
\hspace*{-20mm}
&&-\lim_{\beta\to\infty}\frac{1}{\beta }\sum_{\nu} \tilde{A}(\nu)\int\! \rmd\x ~q_{\nu}(\x) \log\Big[     
\sum_{\mu=1  }^K     \tilde{A}(\mu\vert\nu)\Ind\left[     \Delta_{\mu }(     \x  )\!=\!0             \right] \Big] \nonumber\\
\hspace*{-20mm}
&&-\lim_{\beta\to\infty}\frac{1}{\beta }\sum_{\nu} \tilde{A}(\nu)\! \!\int\!  \rmd\x~  q_{\nu}(\x)\log\left[1\!+\!  \frac{\sum_{\mu=1  }^K \Ind\left[     \Delta_{\mu}(     \x  )>0             \right]\tilde{A}(\mu\vert\nu)\,   \rme^{ -\beta \Delta_\mu (\x  ) } }{\sum_{\mu=1  }^K   \Ind\left[     \Delta_{\mu }(     \x  )=0             \right]\tilde{A}(\mu\vert\nu)} \right ]\nonumber\\
\hspace*{-16mm}
&=&-  \sum_{\nu} \tilde{A}(\nu)\!\int\!  \rmd\x~ q_{\nu}(\x)  \max_{\mu}\log P(   \x\vert\thetav_{\mu}      )  \label{eq:f-RS-T-0-comp-1},
\end{eqnarray}
The average energy $e(\beta)= \lim_{N\rightarrow\infty}\langle \langle  \hat{F}_N(\cmatrix,\, \xmatrix)\rangle_{\cmatrix\vert\xmatrix} \rangle_{\xmatrix}$ is given by (see \ref{app:energy})
\begin{eqnarray}
e(\beta)&=&  -\sum_{\mu=1}^K\int\! \rmd\x~ Q_\mu(\x) \log P(\x\vert\thetav_\mu)\label{eq:e-RS},
\end{eqnarray}
where $Q_\mu(\x)$ is a solution of the equation (\ref{eq:Q-RS}).  The latter reduces to  (\ref{eq:Q-RS-T-0}) when $\beta\rightarrow\infty$, and hence in this limit we find
\begin{eqnarray}
\hspace*{-15mm}
e(\infty)
&=&-\!\sum_{\mu=1}^K\sum_{\nu} \tilde{A}(\nu) \! \int\!\rmd\x ~  q_{\nu}(\x)\log P(\x\vert\thetav_\mu) 
    \frac{ \Ind\left[    \Delta_{\mu}(     \x  )\!=\!0         \right] \tilde{A}(\mu \vert\nu) 
     }{   \sum_{\mu^{\prime}} \Ind\left[     \Delta_{\mu^{\prime}}(     \x  )\!=\!0             \right]  \tilde{A}(\mu^{\prime}\vert\nu)         }
\label{eq:e-RS-T-0}.
\end{eqnarray}
It is trivial to show (and intuitive) that $e(\infty)=f(\infty)$. 
For finite $\beta$, the average free energy $f(\beta)-\phi_N$ and the energy $e(\beta)$,  given by equations (\ref{eq:f-RS},\ref{eq:e-RS}),  can be used to compute the average entropy density  of the Gibbs-Boltzmann distribution (\ref{def:P(C|X)}) via the  Helmholtz free energy  $f(\beta)=e(\beta)-\frac{1}{\beta}s(\beta)$, 
\begin{eqnarray}
 s(\beta)&=& -\lim_{N\rightarrow\infty}\frac{1}{N}\Big\langle\sum_{\cmatrix} P_\beta(\cmatrix\vert\xmatrix)\log P_\beta(\cmatrix\vert\xmatrix)\Big\rangle_{\xmatrix}\label{def:aver-entropy}
\end{eqnarray}
 From the  Helmholtz free energy we immediately infer that $ \lim_{\beta\rightarrow\infty}s(\beta)/\beta=0$.

\subsection{RS theory for $\beta\rightarrow0$ \label{sssection:RS-T-inf}}
The RS  theory simplifies considerably  in the high temperature limit  $\beta\rightarrow0$.  Here  the order parameter $Q_{\mu}(\x )$, which is governed by the equation (\ref{eq:Q-RS}), is given by 
\begin{eqnarray}
Q_{\mu}(\x ) &=&\sum_{\nu}  \tilde{A}(\nu)  \tilde{A}(\mu\vert\nu) \, q_{\nu}(\x).  \label{eq:Q-RS-T-inf}
    \end{eqnarray}
The fraction of  data points originating from the distribution $q_{\nu}(\x)$ assigned to cluster $\mu$, $A(\mu, \nu)$, is $\tilde{A}(\nu) \tilde{A}(\mu\vert\nu)$ due to (\ref{eq:A-RS}). Using this in  (\ref{eq:e-RS}) gives the average energy
\begin{eqnarray}
e(0)&=&  - \sum_{\nu=1}^L  \tilde{A}(\nu)  \sum_{\mu=1}^K\tilde{A}(\mu\vert\nu) \int\! \rmd\x  ~q_{\nu}(\x)\log P(\x\vert\thetav_\mu) \label{eq:e-RS-T-inf}
\end{eqnarray}
where $\thetav_{\mu}=\argmax_{\thetav}  \int\!\rmd\x~ Q_{\mu}(\x )  \log P(\x\vert\thetav)$. We note that (\ref{eq:e-RS-T-inf}) is equal  to 
\begin{eqnarray}
F  (\tilde{A}  )&=&  \sum_{\nu=1}^L  \tilde{A}(\nu)  \sum_{\mu=1}^K \tilde{A}(\mu\vert\nu) D( q_\nu\vert\vert P_\mu) +\sum_{\nu=1}^L \tilde{A}(\nu) H(q_\nu), \label{eq:F-MF}
\end{eqnarray}
where $H(q_\nu)$ is  the differential entropy of $q_{\nu}(\x)$,  which is also the  entropy  function of  the mean-field theory~\cite{Mozeika2018}.  
For finite $N$, the average energy $e_N(\beta)=\langle \langle  \hat{F}_N(\cmatrix,\, \xmatrix)\rangle_{\cmatrix} \rangle_{\xmatrix}$ is a monotonic non-increasing function of $\beta$. Also the limits $ \lim_{\beta\rightarrow\infty} e_N(\beta)$ and $ \lim_{\beta\rightarrow0} e_N(\beta)$ exist. Thus $e_N(\infty) \leq  e_N(0)$ for $N$ finite and hence the average energy $e(\infty)$ is bounded from above by the mean-field entropy $F  (\tilde{A} )$, i. e. $e(\infty)\leq F (\tilde{A}  )$. For model distributions $P(\x\vert\thetav_\mu)$ with non-overlapping supports for different $\theta_\mu$, this upper bound can be optimised by replacing $F (\tilde{A}  )$ with  $\min_{ \tilde{A}} F (\tilde{A}  )$ and hence in this case
\begin{eqnarray}
e(\infty)&\leq& \min_{ \tilde{A}} F (\tilde{A}  )\label{eq:e-F-ineq}.%true for non-overlapping supports but for overlapping?
\end{eqnarray}
The minimum is computed over all  prior parameters  $\tilde{A}(\mu\vert\nu)$ satisfying the  constraints $\tilde{A}(\mu\vert\nu)>0$ and  $\sum_{\mu\leq K}\tilde{A}(\mu\vert\nu)=1$.  Finally, we note that for $K=1$, as a consequence  of $Q_{\mu}(\x ) =\sum_{\nu\leq L}  \tilde{A}(\nu)   \, q_{\nu}(\x)$,  we will have  $e(\infty)= F (\tilde{A}  )$.

\subsection{Recovery of true partitions\label{sssection:true-part}}

Equation (\ref{eq:Q-RS-T-0}) for  $Q_{\mu}(\x )$ can be used to derive 
the following expression  for the distribution  
$\tilde{Q}_{\mu}(\x ) =Q_{\mu}(\x )/\int\!\rmd\tilde{\x}~ Q_{\mu}(\tilde{\x} ) $ of data that are assigned to cluster $\mu$: 
\begin{eqnarray}
\tilde{Q}_{\mu}(\x ) %&=&\frac{Q_{\mu}(\x )}{\int Q_{\mu}(\tilde{\x} ) \rmd\tilde{\x}}  \nonumber\\
&=&\frac{\sum_{\nu} \tilde{A}(\nu)\, q_{\nu}(\x)\,
    \frac{    \Ind\left[    \Delta_{\mu}(     \x  )=0         \right] \tilde{A}(\mu \vert\nu) 
     }{  Z_\nu(\x)        }}{ \int \! \rmd\tilde{\x}~    \sum_{\nu} \tilde{A}(\nu)\, q_{\nu}(\tilde{\x})\,
    \frac{    \Ind\left[    \Delta_{\mu}(     \tilde{\x} )=0         \right] \tilde{A}(\mu \vert\nu) 
     }{  Z_\nu(\tilde{\x})        }   }    \label{eq:Q-RS-T-0-norm}\\
    \Delta_{\mu}(     \x  )&=& \max_{\tilde{\mu}}\log P(   \x\vert\thetav_{\tilde{\mu}}      ) -\log P(   \x\vert\thetav_{\mu}      )\nonumber\\
     \thetav_{\mu}&=&\argmax_{\thetav}  \int\!\rmd\x~ \tilde{Q}_{\mu}(\x )  \log P(\x\vert\thetav)\nonumber,
    \end{eqnarray}
 where $Z_\nu(\x)= \sum_{\mu} \Ind\left[     \Delta_{\mu}(     \x  )=0             \right]  \tilde{A}(\mu\vert\nu) $.
Suppose we knew the number of true clusters, i.e.  $K=L$.  If our clustering procedure  was  perfect  we would then expect that each cluster holds data from  at most one distribution, i.e. we expect  $\tilde{Q}_{\mu}(\x )=q_\mu(   \x )$  to be a solution of the following equation 
\begin{eqnarray}
q_\mu(   \x ) &=&   \frac{\sum_{\nu} \tilde{A}(\nu)\, q_{\nu}(\x)\,
    \frac{    \Ind\left[    \Delta_{\mu}(     \x  )=0         \right] \tilde{A}(\mu \vert\nu) 
     }{  Z_\nu(\x)        }   }  { \int  \! \rmd\tilde{\x}  ~ \sum_{\nu} \tilde{A}(\nu)\, q_{\nu}(\tilde{\x})\,
    \frac{    \Ind\left[    \Delta_{\mu}(     \tilde{\x} )=0         \right] \tilde{A}(\mu \vert\nu) 
     }{  Z_\nu(\tilde{\x})        }   }\label{eq:Q-RS-T-0-norm-sol} .%\label{eq:Q-RS-T-0-norm},
    \end{eqnarray}
This  is certainly true if $ \Ind\left[    \Delta_{\mu}(     \x  )\!=\!0         \right] \tilde{A}(\mu \vert\nu)  = \delta_{\nu; \mu} Z_\nu(\x)  $ for all $\x$ in the domain of $q_\mu(   \x )$.  The latter  condition implies that   $\int \! \rmd\x ~q_{\nu}(\x)~ \Ind\left[    \Delta_{\mu}(     \x  )\!=\!0         \right] \tilde{A}(\mu \vert\nu)  Z^{-1}_\nu(\x) =\delta_{\nu;\mu}$ which, by the definition  of order parameter $A(\mu\vert\nu)$, is equivalent to $A(\mu\vert\nu)=\delta_{\nu;\mu}$, i.e. all data from the distribution $q_\nu(   \x )$ are in cluster $\mu$. Thus  if 
\begin{eqnarray}
\int \!\rmd\x~ q_{\nu}(\x)\, \frac{    \Ind\left[    \Delta_{\mu}(     \x  )=0         \right] \tilde{A}(\mu \vert\nu)  }{  Z_\nu(\x)        }&=&\delta_{\nu;\mu}\label{eq:clust-recovery-cond-1}
    \end{eqnarray}
holds for all pairs  $(\nu,\mu)$ in a bijective mapping of the set $[K]$ to itself,  then $\tilde{Q}_{\mu}(\x )=q_\mu(   \x )$  is a solution of equation (\ref{eq:Q-RS-T-0-norm-sol}).
Let us define the set $S_P(\x)=\left\{\mu\,\vert \,\Delta_{\mu}(     \x  )=0 \right\}$ and consider the average $\sum_\mu A(\mu\vert\nu)\mu$:
\begin{eqnarray}
\hspace*{-10mm}
&&\hspace*{-15mm}
\int\! \rmd\x~  q_{\nu}(\x)\, \frac{  \sum_\mu  \Ind\left[    \Delta_{\mu}(     \x  )\!=\!0         \right] \tilde{A}(\mu \vert\nu)\mu  }{  Z_\nu(\x)        }\nonumber\\
\hspace*{-10mm}
&=& \int \!\rmd\x ~ \Big(\Ind\left[ \vert S_P(\x) \vert\!>\!1\right] \!+\!\Ind\left[ \vert S_P(\x) \vert\!=\!1\right] \Big) q_{\nu}(\x) \frac{   \sum_\mu \Ind\left[    \Delta_{\mu}(     \x  )\!=\!0         \right] \tilde{A}(\mu \vert\nu) \mu }{  Z_\nu(\x)        }
\nonumber\\
\hspace*{-10mm}
   &=&  \int \!\rmd\x ~ q_{\nu}(\x)\, \argmax_{\mu} \log P(   \x\vert\thetav_{\mu}      ) 
   \nonumber
   \\
   \hspace*{-10mm}&&
  \hspace*{10mm} +\int \!\rmd\x ~ \Ind\left[ \vert S_P(\x) \vert\!>\!1\right] q_{\nu}(\x)  \frac{ \sum_\mu   \Ind\left[    \Delta_{\mu}(     \x  )=0         \right] \tilde{A}(\mu \vert\nu) \mu }{  Z_\nu(\x)        }
    \end{eqnarray}
We note that  the second term is a contribution  of  sets  that can be characterized as $\{\x\,\vert\, P(   \x\vert\thetav_{\mu_1}      )\!=\! P(   \x\vert\thetav_{\mu_2}      ), ~\mu_1\!<\!\mu_2\}$, for some $(\mu_1,\mu_2)$.   If we assume that this term is zero\footnote{This is certainly true for model distributions $P(   \x\vert\thetav_{\mu}      )$ with non-overlapping  supports.}, then one of the consequences of  (\ref{eq:clust-recovery-cond-1}) is  equivalence  of the two averages  
\begin{eqnarray}
\sum_{\mu}A(\mu\vert\nu)\mu&=&\nu=\!\int\!  \rmd\x~q_\nu(   \x ) \argmax_{\mu}\log P(   \x\vert\thetav_{\mu}      )   \label{eq:clust-recovery-cond-2-comp-1},
    \end{eqnarray}
and 
\begin{eqnarray}
\hspace*{-10mm}
\int \! \rmd\x~q_\nu(   \x ) \argmax_{\mu}\log P(   \x\vert\thetav_{\mu}      )    
&=& \int \!\rmd\x ~q_\nu(   \x ) \argmin_{\mu} \log P^{-1}(   \x\vert\thetav_{\mu}      )   \nonumber\\
\hspace*{-10mm}
&=& \argmin_{\mu}\int\! \rmd\x ~ q_\nu(   \x ) \log P^{-1}(   \x\vert\thetav_{\mu}      )   \nonumber\\
\hspace*{-10mm}
&=& \argmin_{\mu}  D( q_\nu\vert\vert P_\mu)~=~\nu
\label{eq:clust-recovery-cond-2-comp-2},
    \end{eqnarray}
where $D( q_\nu\vert\vert P_\mu)$ is  the  Kullback-Leibler distance between the distributions $q_\nu(   \x )$ and $P(   \x\vert\thetav_{\mu}      )$.  Thus if (\ref{eq:clust-recovery-cond-1}) holds,  then the results (\ref{eq:clust-recovery-cond-2-comp-1},\ref{eq:clust-recovery-cond-2-comp-2}) show that  the $\max$ and expectation operators commute.  Using this property  in the average energy (\ref{eq:e-RS-T-0}) gives
\begin{eqnarray}
e(\infty)&=&- \sum_{\nu} \tilde{A}(\nu) \!\int\!     q_{\nu}(\x)  \max_{\mu}\log P(   \x\vert\thetav_{\mu}      ) \rmd\x \nonumber\\
&=&\sum_{\nu} \tilde{A}(\nu) \min_{\mu} \!\int\!     q_{\nu}(\x)   \log P^{-1}(   \x\vert  \thetav_{\mu}      ) )\rmd\x   \nonumber\\
&=&\sum_{\nu} \tilde{A}(\nu) \min_{\mu}   D( q_\nu\vert\vert P_\mu) +\sum_{\nu} \tilde{A}(\nu), H(q_\nu)%\label{eq:aver-f-energy-MF-lower-bound},
\end{eqnarray}
and in the distribution  (\ref{eq:Q-RS-T-0}) it leads to the equation
\begin{eqnarray}
Q_{\mu}(\x ) &=&\sum_{\nu} \tilde{A}(\nu)\, q_{\nu}(\x)\,\delta_{\mu; \argmax_{\tilde{\mu}}\log P(   \x\vert\thetav_{\tilde{\mu}}      )}  \nonumber \\      
 \thetav_{\mu} &=& \argmax_{\thetav} \int Q_{\mu}(\x )\log  P(   \x\vert  \thetav   ) \rmd\x\label{eq:Q-RS-T-0-simple}.
    \end{eqnarray}
We note that the above average energy and the MF (\ref{eq:F-MF})  average energy are both bounded from below by the average entropy $\sum_{\nu} \tilde{A}(\nu) H(q_\nu)$. This bound is saturated when all $D( q_\nu\vert\vert P_\mu)$ terms vanish, i.e. when the model matches the data exactly.

 \section{Implementation and application of the RS theory}

\subsection{Population dynamics algorithm\label{sssection:T-0-population-dynamics}} 
Equation (\ref{eq:Q-RS-T-0}) for the order parameter $Q_{\mu}(\x )$ can be solved numerically by a  population dynamics  algorithm~\cite{Mezard2009} which can be derived as follows.  Firstly, we re-arrange the equation for $Q_{\mu}(\x )$: 
\begin{eqnarray}
\hspace*{-20mm}
Q_{\mu}(\x ) &=&\sum_{\nu} \tilde{A}(\nu)\, q_{\nu}(\x)\,
    \frac{    \Ind\left[    \Delta_{\mu}(     \x  )=0         \right] \tilde{A}(\mu \vert\nu) 
     }{   \sum_{\mu^{\prime}} \Ind\left[     \Delta_{\mu^{\prime}}(     \x  )=0             \right]  \tilde{A}(\mu^{\prime}\vert\nu)         }\nonumber\\
     \hspace*{-20mm}
  &=&   \sum_{\nu} \tilde{A}(\nu)\, q_{\nu}(\x)\,\Big(
  \!\Ind\left[ \vert S_P(\x) \vert\!>\!1\right] \!+\!\Ind\left[ \vert S_P(\x) \vert\!=\!1\right]\! \Big)
    \frac{    \Ind\left[    \Delta_{\mu}(     \x  )\!=\!0         \right] \tilde{A}(\mu \vert\nu) 
     }{   \sum_{\mu^{\prime}} \!\Ind\left[     \Delta_{\mu^{\prime}}(     \x  )\!=\!0             \right]  \tilde{A}(\mu^{\prime}\vert\nu)         }\nonumber\\
    \hspace*{-20mm}
 &=&   \sum_{\nu} \tilde{A}(\nu)\, q_{\nu}(\x)\,\Ind\left[ \vert S_P(\x) \vert\!=\!1\right] \Ind\left[    \Delta_{\mu}(     \x  )\!=\!0         \right]+\cdots
 \nonumber
 \\
 \hspace*{-20mm}
 &&~~~~\cdots+\sum_{\nu} \tilde{A}(\nu)\, q_{\nu}(\x)\,\Ind\left[ \vert S_P(\x) \vert\!>\!1\right] \frac{    \Ind\left[    \Delta_{\mu}(     \x  )\!=\!0         \right] \tilde{A}(\mu \vert\nu) 
     }{   \sum_{\mu^{\prime}} \Ind\left[     \Delta_{\mu^{\prime}}(     \x  )\!=\!0             \right]  \tilde{A}(\mu^{\prime}\vert\nu)         }     
     \label{eq:Q-popul-dynam-comp-1-new}. 
    \end{eqnarray}
Secondly,  we note that the data distribution  $\sum_{\nu} \tilde{A}(\nu)\,    q_{\nu}(\x)$ can be replaced by a large sample $\xmatrix$, i.e. by  the data itself, via  the empirical distribution $N^{-1}\sum_{i\leq N}\delta(\x\!-\!\x_i)$, which can be also written as $N^{-1}\sum_{\nu\leq L}\sum_{i_v\leq N_{\nu}}\delta(\x\!-\!\x_{i_\nu})$. Here $N_\nu$, which satisfies $\lim_{N\rightarrow\infty}N(\nu)/N=\tilde{A}(\nu)$, is the number of data-points sampled from $q_{\nu}(\x)$.  Upon using both of these representations of $\sum_{\nu} \tilde{A}(\nu)\,    q_{\nu}(\x)$ in equation (\ref{eq:Q-popul-dynam-comp-1-new}) we obtain 
\begin{eqnarray}
\hspace*{-15mm}
Q_{\mu}(\x ) &=&  \frac{1}{N}\sum_{i=1}^N\delta(\x\!-\!\x_i)\Ind\left[ \vert S_P(\x_i) \vert\!=\!1\right] \Ind\left[    \Delta_{\mu}(     \x_i  )\!=\!0         \right]+\cdots
  \label{eq:Q-popul-dynam-comp-2-new}
\\
\hspace*{-15mm}
 &&\cdots+\! \frac{1}{N}\!\sum_{\nu=1}^L\!\sum_{i_v=1}^{N_{\nu}}\!\delta(\x\!-\!\x_{i_\nu})\Ind\!\left[ \vert S_P(\x_{i_\nu}) \vert\!>\!1\right] \frac{    \Ind\left[    \Delta_{\mu}(     \x_{i_\nu}  )\!=\!0         \right] \tilde{A}(\mu \vert\nu) 
     }{   \sum_{\mu^{\prime}} \Ind\left[     \Delta_{\mu^{\prime}}(     \x_{i_\nu}  )\!=\!0             \right]  \tilde{A}(\mu^{\prime}\vert\nu)         }. 
     \nonumber
    \end{eqnarray}
Finally,  it is very unlikely to find in $\xmatrix$, sampled from a distribution  of continuous random variables $\sum_{\nu} \tilde{A}(\nu)\,    q_{\nu}(\x)$, data points which satisfy   $\vert S_P(\x) \vert\!>\!1$, so the second term in ( \ref{eq:Q-popul-dynam-comp-2-new})  is almost surely zero for any  sample  $\xmatrix$  of finite size. Thus 
\begin{eqnarray}
Q_{\mu}(\x ) &=&  \frac{1}{N}\sum_{i=1}^N\delta(\x\!-\!\x_i)\Ind\left[ \vert S_P(\x_i) \vert\!=\!1\right] \Ind\left[    \Delta_{\mu}(     \x_i  )\!=\!0         \right]\nonumber\\
    &=&  \frac{1}{N}\sum_{i=1}^N\delta(\x\!-\!\x_i)\delta_{\mu; \argmax_{\tilde{\mu}}\log P(   \x_i\vert\thetav_{\tilde{\mu}}      )} \nonumber\\
   &=&   \frac{1}{N}\sum_{i=1}^N\delta_{\mu; \mu_i} \delta(\x-\x_i) \label{eq:Q-popul-dynam-comp-3-new}
    \end{eqnarray}
where $\mu_i=\argmax_{\tilde{\mu}}\log P(   \x_i\vert\thetav_{\tilde{\mu}}      )$. 
 Using the above  in equation (\ref{eq:Q-RS-T-0}),  we obtain for  $\mu\in\left[K\right]$ the following system of equations 
\begin{eqnarray}
Q_{\mu}(\x )  &=&   \frac{1}{N}\sum_{i=1}^N\delta_{\mu, \mu_i} \delta(\x-\x_i) \nonumber\\ 
        \thetav_{\mu}&=&\argmax_{\thetav} \!\int \!\rmd\x~Q_{\mu}(\x ) \log P(\x\vert\thetav)\, \nonumber\\
     \mu_i&=&\argmax_{\tilde{\mu}}\log P(   \x_i\vert\thetav_{\tilde{\mu}}      ) \label{eq:popul-dynam}    
    \end{eqnarray}
This set can be solved numerically as follows. We create a `population'  of random variables $\left\{\mu_i : i\in\left[N\right]\right\}$ where $\mu_i\!\in\!\left[K\right]$  are at first sampled uniformly. We use this population to compute the parameters $\thetav_{\mu}$; The latter are then used to compute a new population $\{\mu_i\}$. The last two steps are repeated until  one observes convergence of  the  energy $e(\infty)=-\sum_{\mu=1}^K\int\!\rmd\x ~Q_\mu(\x) \log P(\x\vert\thetav_\mu)$. Finally, we note  that using instead  equation (\ref{eq:Q-RS-T-0-simple}) as our starting point  would lead us to the same population  dynamics equations. Thus, for continuous data distributions  $\sum_{\nu} \tilde{A}(\nu)\,    q_{\nu}(\x)$ represented by a large finite sample, the equations (\ref{eq:Q-RS-T-0})  and (\ref{eq:Q-RS-T-0-simple})  are equal. 
 
 The population dynamics simplifies  significantly if we assume that the distribution $p(\x\vert\thetav)$ is  the multivariate Gaussian
 \begin{eqnarray}
\mathcal{N}(\x\vert\m,\Lmatrix^{-1})&=&\vert2\pi\Lmatrix^{-1}\vert^{-\frac{1}{2}}\rme^{-\frac{1}{2} (\x-\m)^T\Lmatrix(\x-\m  )}\label{def:Normal}
\end{eqnarray}
 with mean $\m$ and precision matrix  (inverse covariance matrix)  $\Lmatrix$.  The parameters  $ \thetav_{\mu}=(\m_{\mu}, \Lmatrix_\mu^{-1})$  we can be estimated   directly from the population via the equations
\begin{eqnarray}
\m_{\mu}\  &=&   \frac{1}{\sum_{j=1}^N\delta_{\mu; \mu_j}}\sum_{i=1}^N\delta_{\mu; \mu_i} \x_i \nonumber\\ 
  \Lmatrix_\mu^{-1}&=&    \frac{1}{\sum_{j=1}^N\delta_{\mu; \mu_j}}\sum_{i=1}^N\delta_{\mu; \mu_i} (\x_i\!-\!\m_\mu)  (\x_i\!-\!\m_\mu)^T    \label{eq:popul-dynam-Normal-1},
    \end{eqnarray}
where  $\mu_i$ is given by
\begin{eqnarray}
\hspace*{-10mm}
  \mu_i&=&\argmax_{\mu}\log \mathcal{N}(\x_i\vert\m_\mu,\Lmatrix_\mu^{-1})
   \label{eq:popul-dynam-Normal-2}\\ 
  \hspace*{-10mm}
  &=&\argmax_{\mu}-\frac{1}{2}\Tr\left\{ \Lmatrix_\mu (\x_i-\m_\mu  )  (\x_i-\m_\mu)^T\right\}+\frac{1}{2}\log \left\vert  \Lmatrix_\mu \right\vert -\frac{d}{2}\log2\pi.
 \nonumber
    \end{eqnarray}

\subsection{Population dynamics algorithm for finite $\beta$\label{sssection:T-finite-population-dynamics}}

Also equation (\ref{eq:Q-RS})  can be solved via population dynamics. However, to replace the distribution of data   $\sum_{\nu}\tilde{A}(\nu)\, q_{\nu}(\x)$ with its empirical version  $N^{-1}\sum_{i=1}^N\delta(\x\!-\! \x_i)$ we 
must  assume that  $\tilde{A}(\tilde{\mu}\vert\nu)=\tilde{A}(\tilde{\mu})$. For $\mu\in[K]$, this gives us  the following equations: 
\begin{eqnarray}
Q_{\mu}(\x )  &=&\frac{1}{N}\sum_{i=1}^N\delta(\x-\x_i)w_i(\mu) \nonumber\\
w_i(\mu)&=& \frac{
   \tilde{A}(\mu) \,\rme^{    \beta \log P(   \x_i\vert\thetav_{\mu}      )      }     
    }{   \sum_{\tilde{\mu}}\tilde{A}(\tilde{\mu})\,\rme^{    \beta \log P(   \x_i\vert\thetav_{\tilde{\mu}}      )      }       }\nonumber\\
        \thetav_{\mu}&=&\argmax_{\thetav} \!\int\! \rmd\x~Q_{\mu}(\x )\log P(\x\vert\thetav).\label{eq:popul-dynam-finite-T}
    \end{eqnarray}
    They  can be solved by creating a population $\left\{(w_i(1),\ldots, w_i(K)) : i\in\left[N\right]\right\}$ and using the above equations to update this population until  convergence of the free energy 
\begin{eqnarray}
f(\beta)&=&-\frac{1}{\beta N}\sum_{i=1}^N   \log\Big[    \sum_{\mu=1  }^K   \tilde{A}(\mu)\,   \rme^{ \beta \log P(\x_i\vert\thetav_{\mu}) }  \Big]\label{eq:f-RS-popul-dynam}.
\end{eqnarray}
%\Red{For $\beta=1$,  the algorithm estimates the average $f(1)=\lim_{N\rightarrow\infty}\langle f_N(K, \xmatrix)\rangle_\xmatrix$ of (\ref{def:f-beta-1}). }

Finally,  we note that  both population dynamics algorithms derived in this subsection  look somewhat similar to  the  
Expectation-Maximisation (EM) algorithm,  see e.g.~\cite{Bishop2006}. Comparing the Gaussian EM, used for maximum likelihood inference  of Gaussian mixtures, with   (\ref{eq:popul-dynam-Normal-1}) shows that the main difference is that EM uses the average $\langle \delta_{\mu; \mu_i}     \rangle_{\mathrm{EM}}$, over some `EM-measure',  instead of the delta function $\delta_{\mu; \mu_i}$.  Gaussian EM is hence an `annealed' version of the population dynamics (\ref{eq:popul-dynam-Normal-1}), but exactly how to relate  the two algorithms  in a more formal manner is not yet clear. 
 
\subsection{Numerical experiments\label{section:numerical-experiments}}  
In the mean-field (MF) theory  of Bayesian clustering  in~\cite{Mozeika2018},  the average entropy (\ref{eq:F-MF}) (derived via a different route) was the  central object. It was mainly used  for the Gaussian data model $P(\x\vert\thetav_{\!\mu})\equiv \mathcal{N}\big(\x\vert\m_{\mu},\Lmatrix_{\mu}^{-1}\big)$, where it becomes the MF entropy   
\begin{eqnarray}
F(\tilde{A})&=&  \frac{1}{2}\sum_{\mu=1}^K\tilde{A}(\mu) \log \Big((2\pi\rme)^{d}\big\vert \Lmatrix_{\mu}^{-1}(\tilde{A})\big\vert  \Big)   \label{eq:F-MF-Norm-1},
\end{eqnarray}
where  $\Lmatrix_{\mu}^{-1}(\tilde{A})$ is the covariance  matrix
\begin{eqnarray}
\Lmatrix_{\mu}^{-1}(\tilde{A})&=&\sum_{\nu=1}^L\! \tilde{A}(\nu\vert\mu)\big\langle\!(\x\!-\!\m_{\mu}(\tilde{A}))(\x\!-\!\m_{\mu}(\tilde{A}))^{\!T}\big\rangle_\nu, 
~~~\label{eq:Cov-MF}
\end{eqnarray}
and   $\m_{\mu}(\tilde{A})=\sum_{\nu=1}^L \tilde{A}(\nu\vert\mu)\langle\x\rangle_\nu$ is the mean. Here we use  $\langle\cdots\rangle_\nu$ for the averages generated by $q_\nu(\x)$. We note that  (\ref{eq:F-MF-Norm-1}) is  also equal to 
\begin{eqnarray}
F(\tilde{A})&=&\!\sum_{\mu, \nu}\!\tilde{A}(\nu, \mu) D(q_{\nu} \vert\vert \mathcal{N}_\mu (\tilde{A}))+\sum_{\nu=1}^L\!\tilde{A}(\nu)H(q_\nu),~~   \label{eq:F-MF-Norm-2}
\end{eqnarray}
where $\mathcal{N}_\mu(\tilde{A})\equiv  \mathcal{N}\big(\x\vert\m_{\mu}(\tilde{A}),\Lmatrix_{\mu}^{-1}(\tilde{A})\big)$.
In addition,  for the Gaussian model, the Laplace method,  quite often used  in statistics to approximate likelihoods~\cite{Guihenneuc2005},  applied to the log-likelihood (\ref{def:F-hat}) for $N\rightarrow\infty$ gives the entropy 
\begin{eqnarray}
\hat{F}_N(\cmatrix,\,\xmatrix)&=& \frac{1}{2}\sum_{\mu=1}^K\frac{ M_\mu(\cmatrix)}{N} \log\Big((2\pi\rme)^{d}\big\vert \Lmatrix_{\mu}^{-1}(\cmatrix,\,\xmatrix)\big\vert  \Big),  ~\label{eq:F-hat-Norm}
\end{eqnarray}
where  $\Lmatrix_{\mu}^{-1}(\cmatrix,\,\xmatrix)$ is the empirical covariance  of data in the cluster $\mu$ and $M_\mu(\cmatrix)=\sum_{i\leq N}c_{i\mu}$ is its size.  This expression can be minimized for clustering, either by gradient descent~\cite{Mozeika2018} or any other  algorithm.  The MF (\ref{eq:F-MF-Norm-1}) makes non-trivial predictions about $\hat{F}_N(\cmatrix,\,\xmatrix)$, such as on structure of its local minima, etc., and correctly estimated $\hat{F}_N\equiv\min_{\cmatrix}\hat{F}_N(\cmatrix,\,\xmatrix)$ for Gaussian data.  However, it  systematicaly  overestimates $\hat{F}_N$ when $K>L$  and when the  separations  between clusters are small \cite{Mozeika2018}.  

\begin{figure}[t]
 %\vspace*{-7mm}
 \setlength{\unitlength}{0.67mm}
% \begin{center}{
 %\hspace*{50mm}
 \begin{picture}(200,112)
%top
 \put(0,77){\includegraphics[width=40\unitlength,height=35\unitlength]{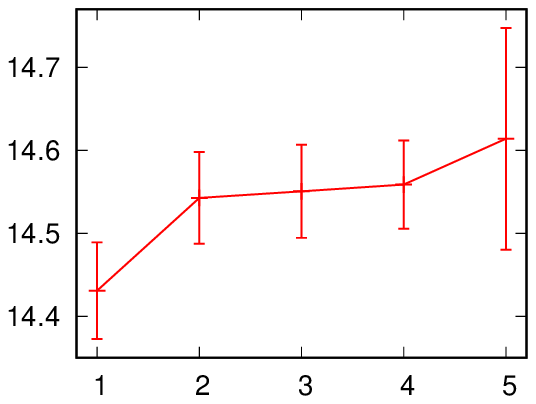}}
  \put(40,77){\includegraphics[width=40\unitlength,height=35\unitlength]{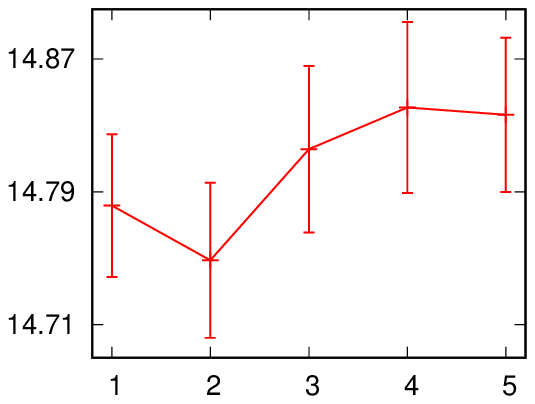}} \put(80,77){\includegraphics[width=40\unitlength,height=35\unitlength]{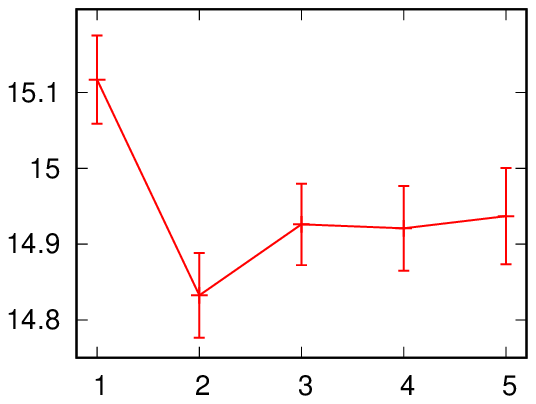}}
\put(120,77){\includegraphics[width=40\unitlength,height=35\unitlength]{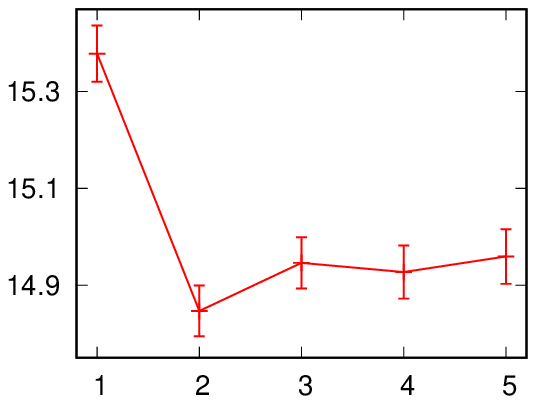}}\put(160,77){\includegraphics[width=40\unitlength,height=35\unitlength]{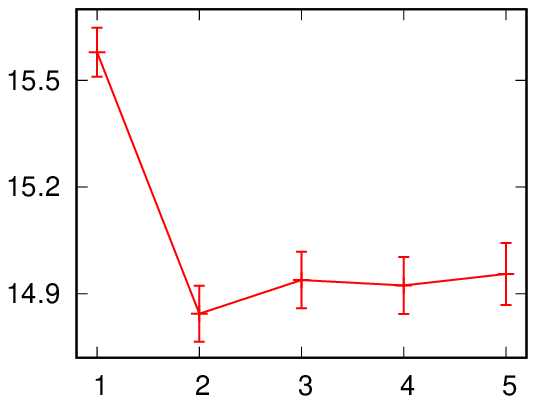}} 
 \put(-3 ,81){\rotatebox{90}{
\small{$\hat{F}_N+\log(K)$}}}

%middle  
\put(5,35){\includegraphics[width=190\unitlength, height=40\unitlength]{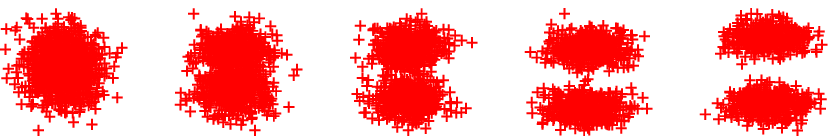}} 
 \put(-3 ,55){\rotatebox{90}{
{$\rho_N$}
} }  
%purity for the separation 1.58
 \put(35 ,67){\tiny{$0.783$}}%0.78315
\put(35 ,55){\tiny{$0.778$}}%0.778225
 \put(35 ,42){\tiny{$0.54$}} % 0.5396
  \put(39 ,61){$\vert$}
  \put(39 ,49){$\vert$}
 
 %purity for the separation 3.16
 \put(75 ,67){\tiny{$0.945$}}%0.94535
\put(75 ,55){\tiny{$0.944$}}%0.944025
 \put(75 ,42){\tiny{$0.941$}} % 0.94105
  \put(79 ,61){$\vert$}
  \put(79,49){$\vert$}
 
 %purity for the separation 4.74
 \put(115 ,67){\tiny{$0.993$}}%0.9927
\put(115 ,55){\tiny{$0.991$}}%0.9911
 \put(115 ,42){\tiny{$0.99$}} % 0.9896
  \put(119 ,61){$\vert$}
  \put(119 ,49){$\vert$}
 
 %purity for the separation 6.32
 \put(157 ,67){\tiny{$1.0$}}%0.99955
\put(155 ,55){\tiny{$0.999$}}%0.9992
 \put(155 ,42){\tiny{$0.999$}} % 0.9989
  \put(159 ,61){$\vert$}
  \put(159 ,49){$\vert$}
 
 %purity for the separation 7.91
 \put(192 ,67){\tiny{$1.0$}}%1.0
\put(192 ,55){\tiny{$1.0$}}%0.99995
 \put(192 ,42){\tiny{$1.0$}} % 0.99985
   \put(194 ,61){$\vert$}
  \put(194 ,49){$\vert$}
 
 %bottom
 \put(0,0){\includegraphics[width=40\unitlength,height=35\unitlength]{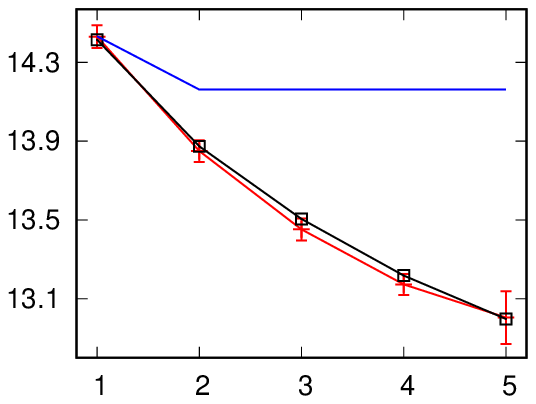}} \put(40,0){\includegraphics[width=40\unitlength,height=35\unitlength]{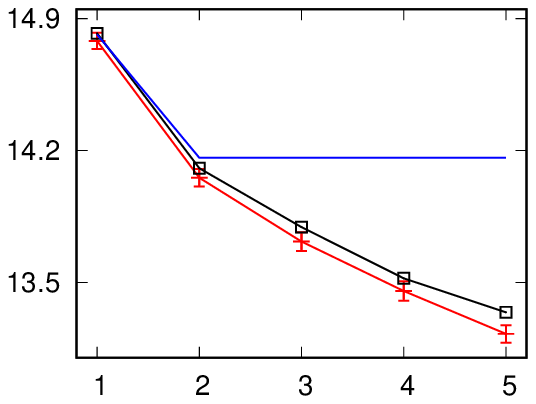}} \put(80,0){\includegraphics[width=40\unitlength,height=35\unitlength]{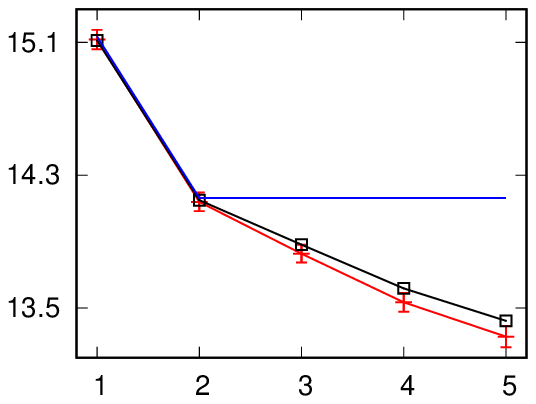}}
\put(120,0){\includegraphics[width=40\unitlength,height=35\unitlength]{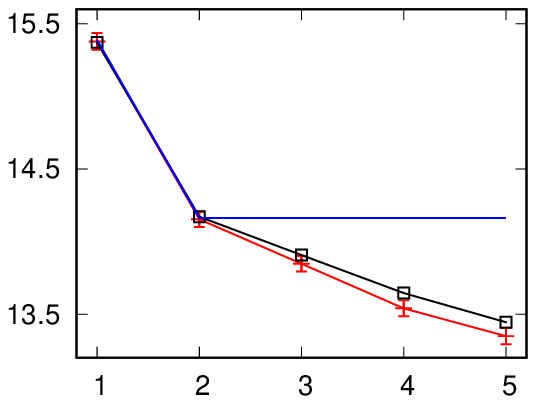}}\put(160,0){\includegraphics[width=40\unitlength,height=35\unitlength]{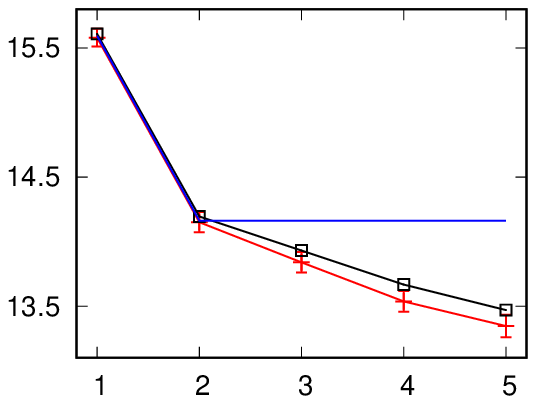}}
    
  \put(-3 ,15){\rotatebox{90}{
\small{$F$}
} }  
      \put(100 ,-5){\small{$K$}} 
      
\end{picture}
% }\end{center}
\vspace*{0mm}
 \caption{Bayesian clustering of data  generated  from Gaussian distributions $\mathcal{N}(\m_1,\I)$ and  $\mathcal{N}(\m_2,\I)$, with separation $\Delta=\vert\vert\m_1-\m_2\vert\vert$.  The sample,  split equally between the distributions,   is of size $N=2000$, and the data dimension  is $d=10$.  The data was generated for $\Delta/\sqrt{d}\in\left\{  \frac{1}{2},1,\frac{3}{2},2,\frac{5}{2}\right\}$, from left to right. Middle:  Data projected into two dimensions for  $\Delta$ increasing from left to right. The quality of the clustering, measured by the `purity'  $\rho_N$, obtained by the population dynamics clustering algorithm  is increasing with $\Delta$.  $\rho_N$ was measured for $10$ random samples of data, but only the minimum, median and maximum values  of $\rho_N$ (numbers connected by lines) are shown.  The size of each sample, split equally between the distributions, was  $N=20000$   and  the clustering algorithm assumed the number of clusters to be $K=2$. Top: $\hat{F}_N+\log(K)$ (red crosses connected by lines),  with the log-likelihood  $\hat{F}_N\equiv\min_{\cmatrix}\hat{F}_N(\cmatrix ,\xmatrix)$  computed  by  a gradient descent algorithm,  shown as a function of the assumed number of clusters $K$.  Symbols, connected by lines and with error bars, denote the average and $\pm$ one standard deviation, measured over $10$ random samples of data.  Bottom:  The log-likelihood $\hat{F}_N$ (red crosses connected by lines) is  compared  with the results of the mean-field theory (blue line) and population dynamics (connected black squares). For $K\geq2$ only the mean-field  lower bound $\frac{d}{2}\log(2\pi\rme)$  is plotted. }
 \label{figure:10dL2} 
 \end{figure}

We expect the present replica theory, related to the MF theory via inequality  $e(\infty)\leq F(\tilde{A})$, to be more accurate.  To test this expectation,  we generated samples from two isotropic Gaussian distributions $\mathcal{N}(\m_1,\I)$ and  $\mathcal{N}(\m_2,\I)$.  Each sample $\xmatrix$,  split equally between the distributions,   is of  size $N\!=\!2000$ and dimension  $d\!=\!10$.  We note that for any given $N$ and  $d$, there exists an $\epsilon>0$ such that  most of the $\x_i$ in sample $\xmatrix$ lie inside the two spheres centred at  $\m_1$ and $\m_2$ and both of radius $\sqrt{d(1\!+\!\epsilon)}$\footnote{ The probability of being outside a sphere is bounded from above by $N \rme^{-d I(\epsilon)}$, where  $I(\epsilon)=\left( \log(1\!+\!\epsilon)^{-1} \!+\!\epsilon\right)\!/2$ (see \ref{app:sphericity}). A much tighter bound,  given by $N\Gamma\left(d/2, d(1+\epsilon)/2\right)/\Gamma(d/2)$, uses that for $\x$ sampled from $\mathcal{N}(\m,\I)$ the squared Euclidean distance  $\vert\vert\x-\m\vert\vert^2$ follows  the $\chi^2$ distribution.}. The latter suggests that the Euclidean distance $\Delta=\vert\vert\m_1-\m_2\vert\vert$, measured relative to the natural scale $\sqrt{d}$, can be use as a measure of the degree of separation~\cite{Dasgupta1999} between the `clusters' centred at  $\m_1$ and $\m_2$ (see Figure  \ref{figure:10dL2}).

We used  gradient descent to find the low entropy states  of  (\ref{eq:F-hat-Norm}) for  our   data.  For each sample  $\xmatrix$ we ran the algorithm from $10$ different random initial states  $\cmatrix\,(0)$, and computed  $\hat{F}_N(\cmatrix\,(\infty),\,\xmatrix)$.  The latter was used to estimate $\hat{F}_N\equiv\min_{\cmatrix}\hat{F}_N(\cmatrix ,\xmatrix)$.  For this data, the log-likelihood function  $\hat{F}_N+\log(K)$  has a minimum at $K=2$, i.e. when the number of assumed  clusters $K$ equals  the number of true clusters $L$,  so it can be used reliably to infer true number of clusters. However, this inference method no longer works  when the separation $\Delta$ is too small  (see Figure  \ref{figure:10dL2}), but the `quality' of clustering, as measured by the purity $\rho_N(\cmatrix, \tilde{\cmatrix})=\frac{1}{N}\sum_{\mu=1}^K \max_{\nu}  \sum_{i=1}^N c_{i\mu}\tilde{c}_{i\nu}$ which compares~\cite{Manning2010}  the clustering  obtained by algorithm $\cmatrix$ with the true clustering $\tilde{\cmatrix}$, for $K=2$, i.e. for the true number of clusters, is still reasonable\footnote{We note that $0<\rho_N\leq 1$ with  $\rho_N=1$ corresponding  to a perfect recovery of true clusters and with $\rho_N\approx 1/L$ corresponding to  a random (unbiased) assignment into clusters.} as can be seen in Figure  \ref{figure:10dL2}.

The predictions of the MF theory for $\hat{F}_N$, $\min_{\tilde{A}} F(\tilde{A})$, is $F_1\!=\!\frac{1}{2}d \log (2\pi\rme)\!+\!\frac{1}{2} \log[1\!+\!(\Delta/2)^2 ]$ for $K\!=\!1$, and $F_2\!=\!\frac{1}{2} d\log (2\pi\rme)$ for  $K\!=\!2$. Thus $F_1\geq F_2$, as required.  Furthermore, if  $\log(2)\geq \frac{1}{2} \log[ 1\!+\!(\Delta/2)^2 ]$, which happens when $\Delta\leq 2\sqrt{3}$, then $F_2\!+\!\log(K)\!\geq \!F_1$, so the MF theory is unable to recover the true number of clusters when the separation $\Delta$ is small. The numerical results for  $\hat{F}_N+\log(K)$ are in qualitative agreement with the predicted values, but the MF predictions for  $\hat{F}_N$ are indeed found to be inaccurate when the separation $\Delta$ is small, and  wrong,   $F_K\geq F_2$ by equation (\ref{eq:F-MF-Norm-2}), when $K>2$. See Figure \ref{figure:10dL2}. 
 
To test the predictions of our replica theory we solve the Gaussian population dynamics equations  (\ref{eq:popul-dynam-Normal-1}) and  (\ref{eq:popul-dynam-Normal-2}) for the data with the same statistical properties as  in the above gradient descent experiments,  but with  a population size $N=20,000$.  We find that the average energy 
\begin{eqnarray}
e(\infty)=-\sum_{\mu\leq K}\int\!\rmd\x~ Q_\mu(\x) \log \mathcal{N}(\x\vert\m_\mu,\Lmatrix_\mu^{-1}), 
\end{eqnarray}
as computed by the population dynamics algorithm,  is in good agreement with the value of $\hat{F}_N$ obtained by gradient descent minimization (see Figure \ref{figure:10dL2}). The residual differences observed between $e(\infty)$ and  $\hat{F}_N$ are finite size effects. Furthermore,  we note that the numerical complexity of the population dynamics algorithm  is consistent with the  lower bound  that is \emph{linear} in $N$ (on average), as  follows from the complexity analysis in~\cite{Mozeika2018}.

\begin{figure}[t]
 %\vspace*{-7mm}
 \setlength{\unitlength}{0.67mm}
% \begin{center}{
 %\hspace*{50mm}
 \begin{picture}(200,107)
%top
\put(0,41){\includegraphics[width=67\unitlength, height=65\unitlength]{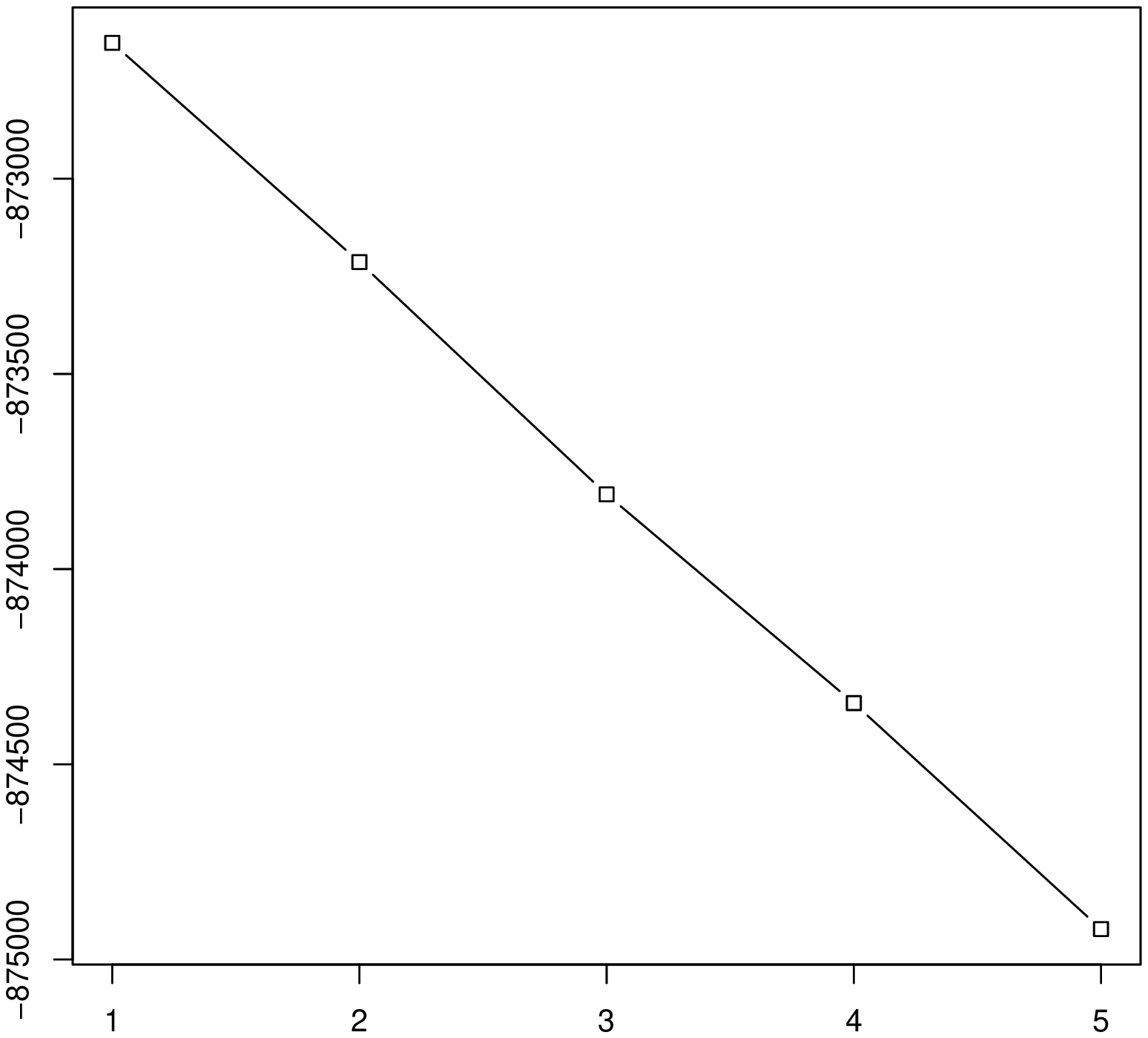}} 
\put(67,41){\includegraphics[width=67\unitlength, height=65\unitlength]{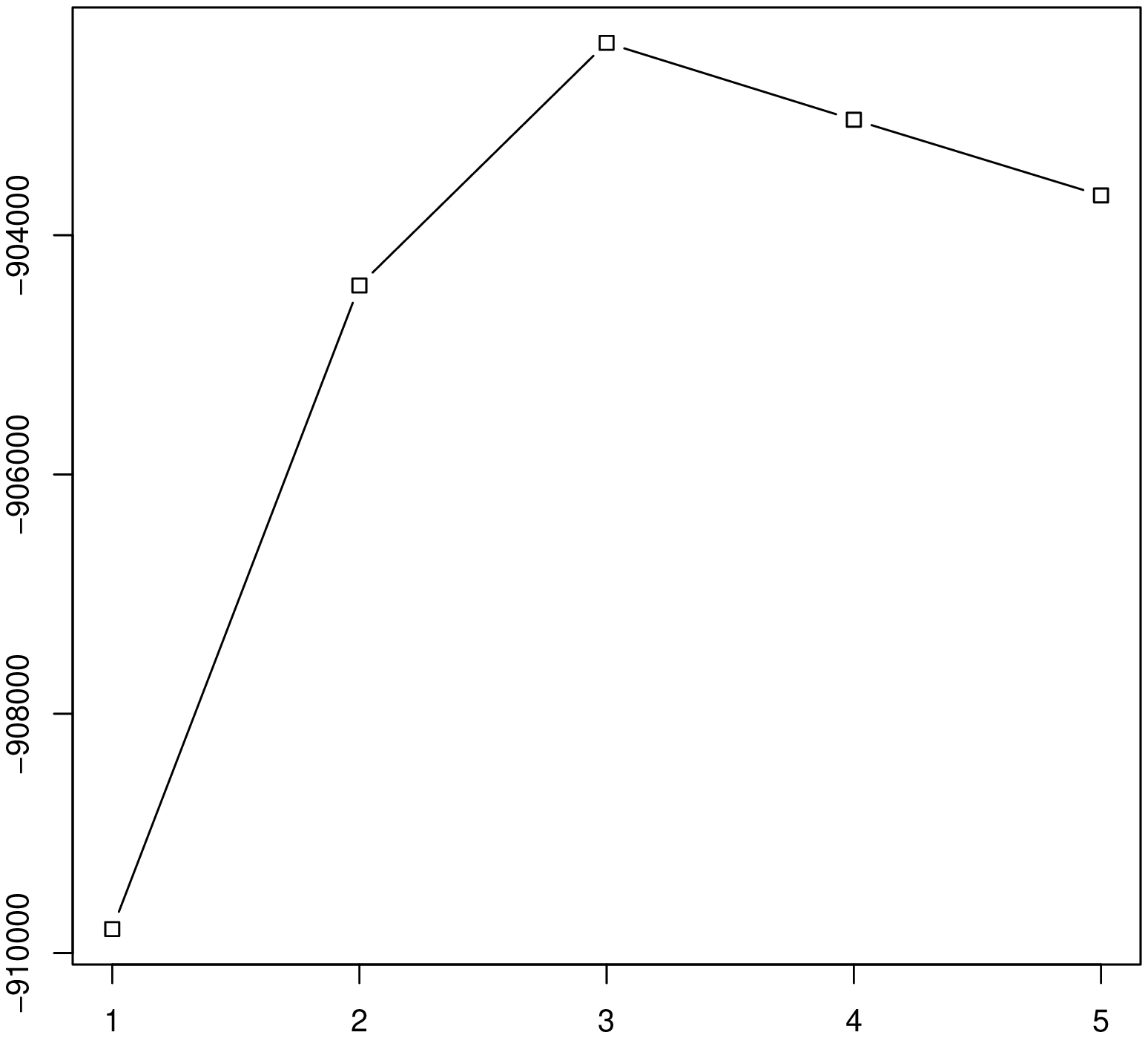}}
\put(134,41){\includegraphics[width=67\unitlength, height=65\unitlength]{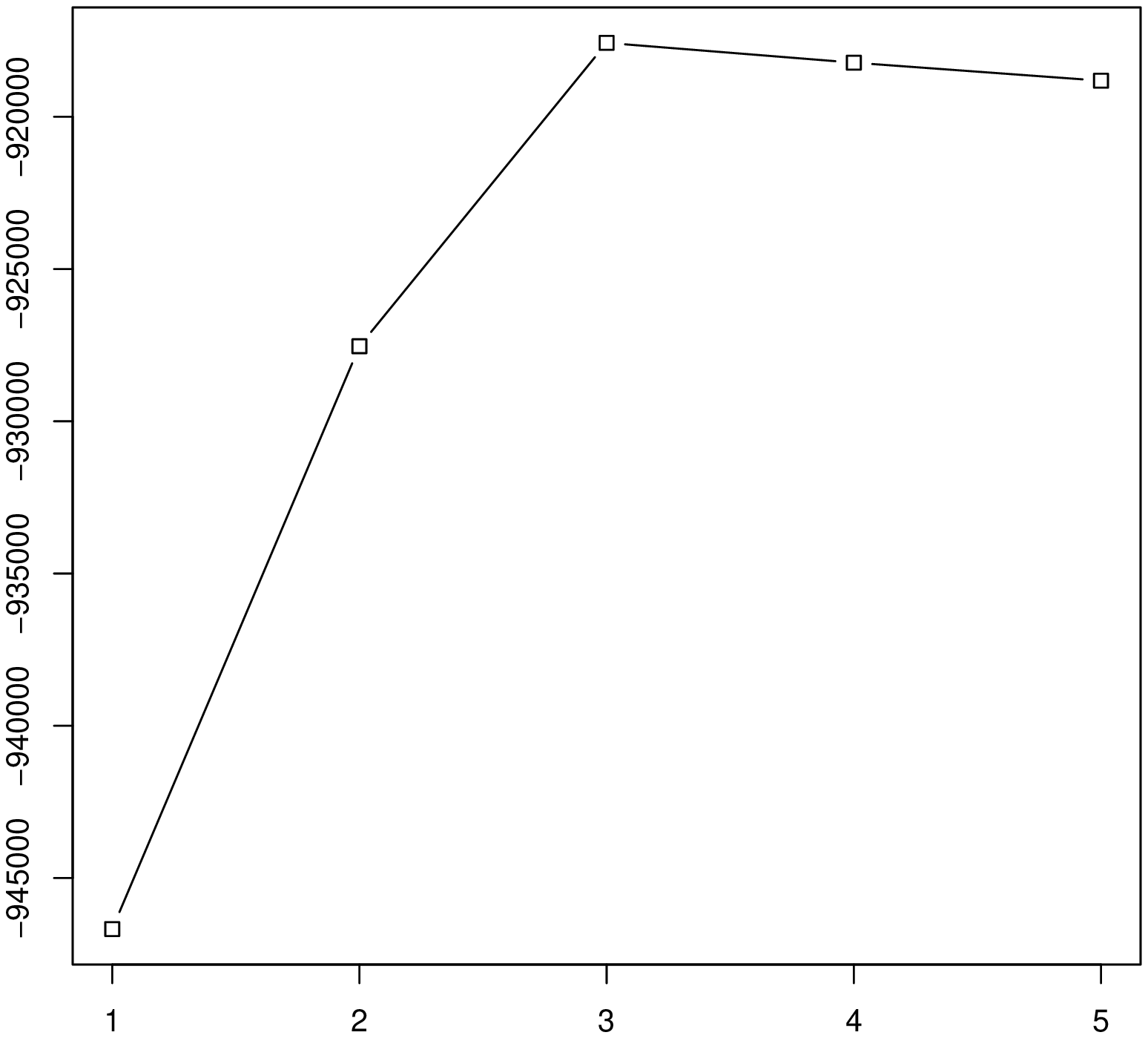}}
 \put(-3 ,71){\rotatebox{90}{\small{$BIC$}}}
 %bottom
\put(1,0){\includegraphics[width=67\unitlength, height=41\unitlength]{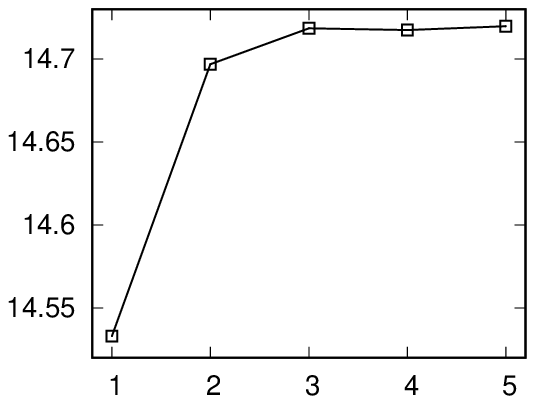}} 
\put(67.9,0){\includegraphics[width=67\unitlength, height=41\unitlength]{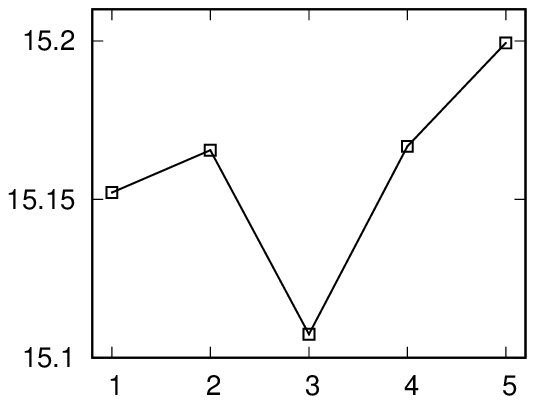}} 
\put(135,0){\includegraphics[width=67\unitlength, height=41\unitlength]{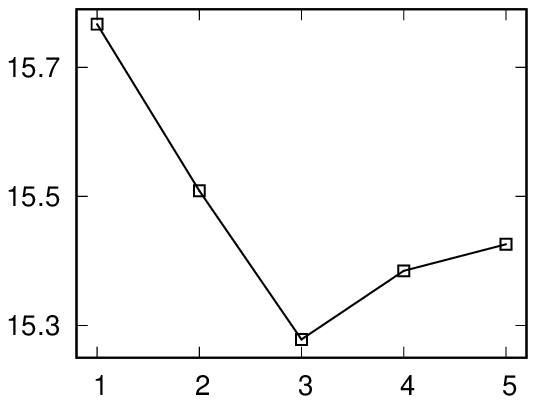}} 
 \put(-3 ,7){\rotatebox{90}{\small{$\hat{F}_N+\log(K)$}}}
     \put(100 ,-5){\small{$K$}} 
\end{picture}
% }\end{center}
\vspace*{0mm}
 \caption{Inferring the number of clusters in data  generated  from Gaussian distributions $\mathcal{N}(\m_\mu,\I)$ with separation $\Delta=\vert\vert\m_\mu-\m_\nu\vert\vert$, where $(\mu,\nu)\in[3]$.  The sample,  split equally between the distributions,   is of size $N=3\times10^4$, and the data dimension  is $d=10$.  The data was generated for $\Delta/\sqrt{d}\in\left\{  \frac{1}{2},1,\frac{3}{2},2,\frac{5}{2}\right\}$, but results shown here (from left to right) are only for  $\Delta/\sqrt{d}\in\left\{  \frac{1}{2},1,\frac{3}{2}\right\}$.   Top: BIC $\equiv2\mathcal{L}_N-n_\mathcal{N}\log(N)$, where $\mathcal{L}_N$ is  the  log-likelihood of GMM  estimated by EM algorithm and $n_\mathcal{N}$ is the number of parameters, as a function of $K$.   Bottom: $\hat{F}_N+\log(K)$, where $\hat{F}_N\equiv\min_{\cmatrix}\hat{F}_N(\cmatrix ,\xmatrix)$ is  the log-likelihood  function  computed by the population dynamics algorithm, as a function of $K$.}
 \label{figure:10dL3} 
 \end{figure}

\begin{figure}[t]
 %\vspace*{-7mm}
 \setlength{\unitlength}{0.67mm}
% \begin{center}{
 %\hspace*{50mm}
 \begin{picture}(200,107)
%top
\put(0,41){\includegraphics[width=67\unitlength, height=65\unitlength]{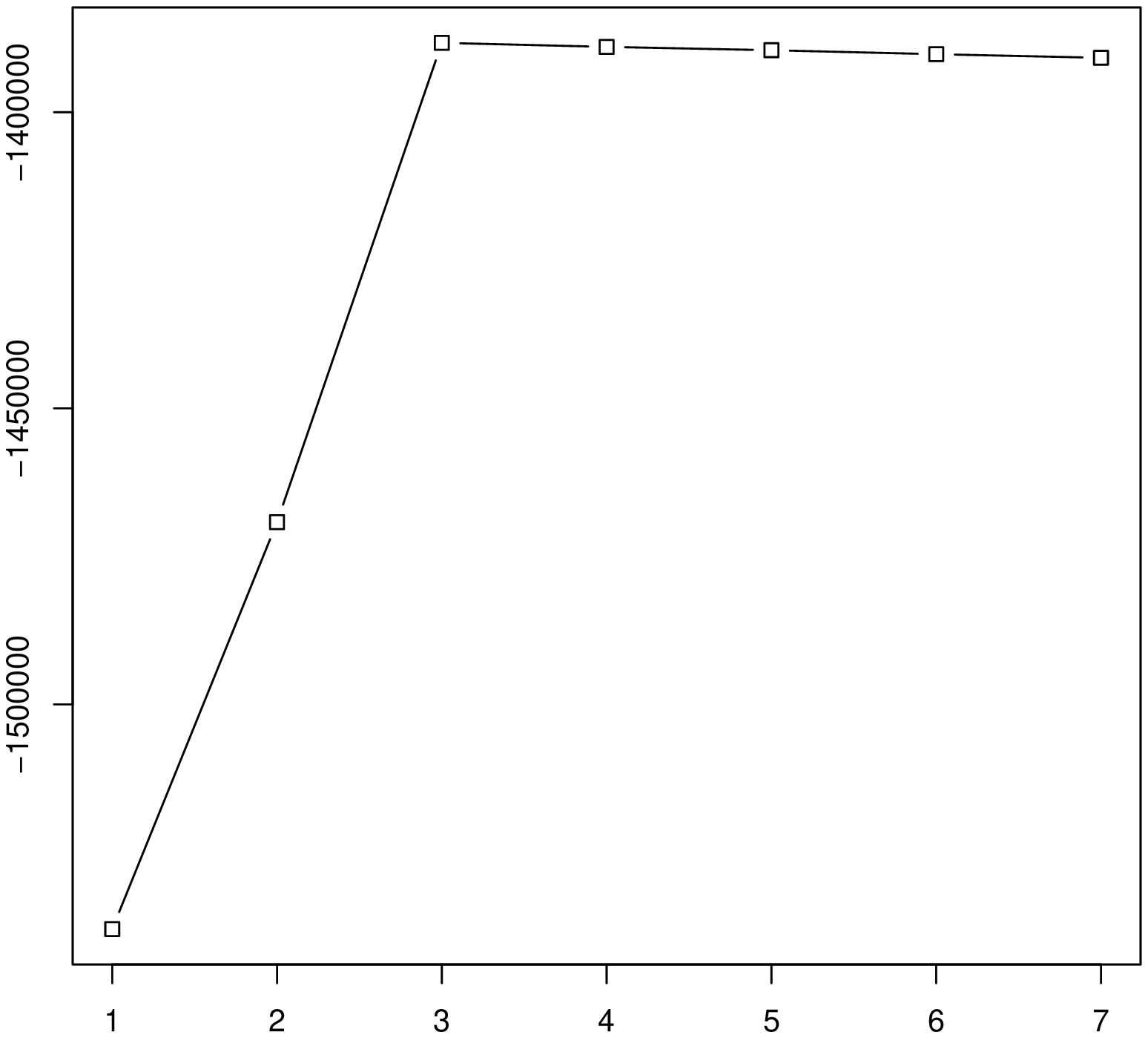}} 
\put(67,41){\includegraphics[width=67\unitlength, height=65\unitlength]{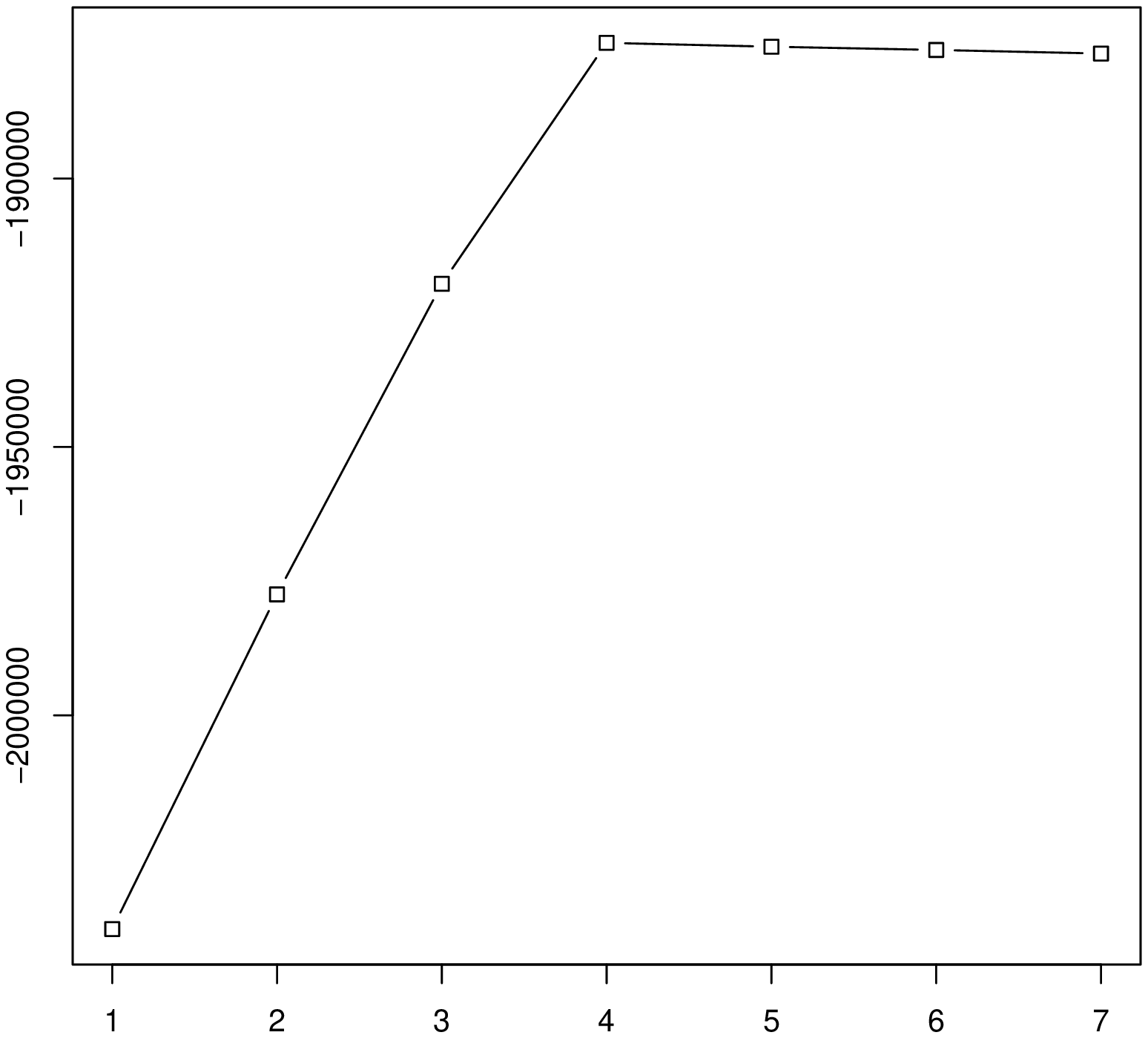}}
\put(134,41){\includegraphics[width=67\unitlength, height=65\unitlength]{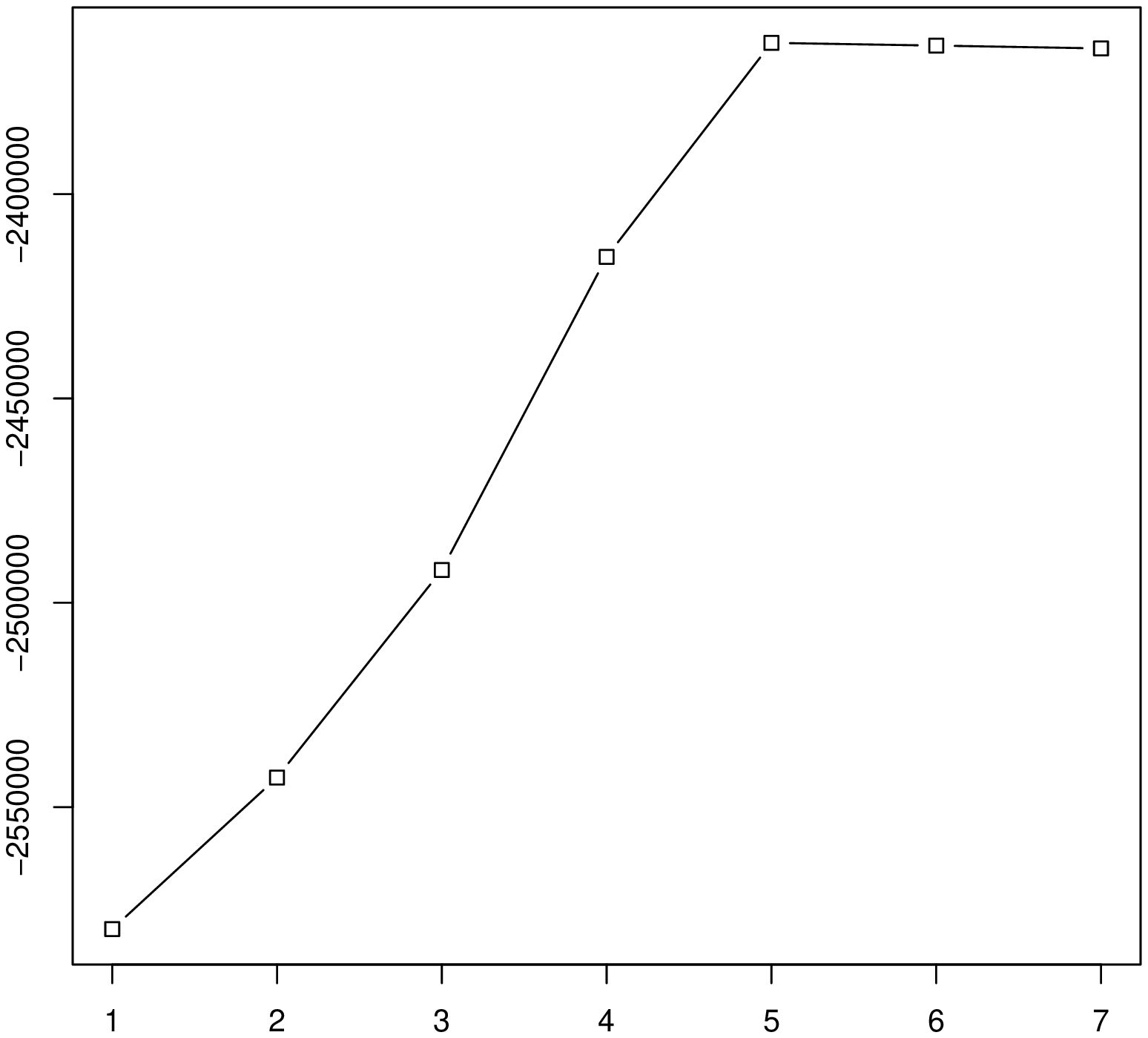}}
 \put(-3 ,71){\rotatebox{90}{\small{$BIC$}}}
 %bottom
\put(1,0){\includegraphics[width=67\unitlength, height=41\unitlength]{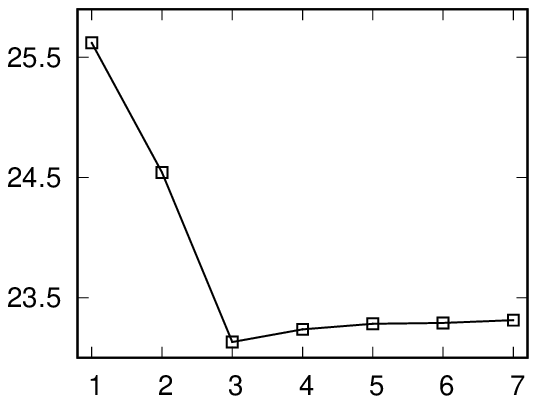}} 
\put(67.9,0){\includegraphics[width=67\unitlength, height=41\unitlength]{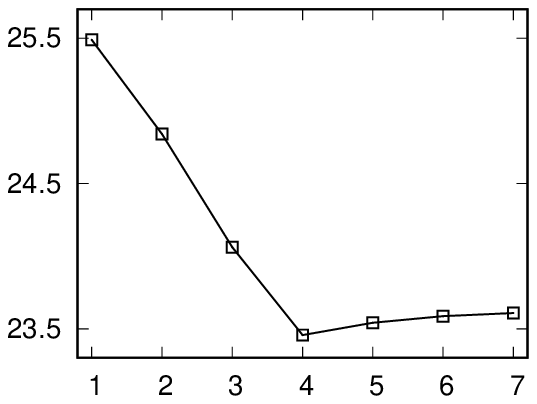}} 
\put(135,0){\includegraphics[width=67\unitlength, height=41\unitlength]{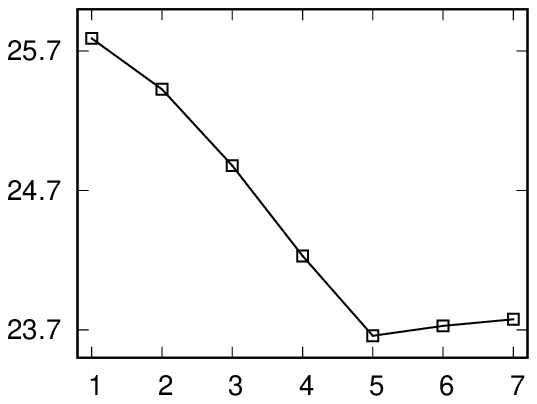}} 
 \put(-3 ,7){\rotatebox{90}{\small{$\hat{F}_N+\log(K)$}}}
     \put(100 ,-5){\small{$K$}} 
\end{picture}
% }\end{center}
\vspace*{0mm}
 \caption{Inferring the number of clusters in  data  generated  from Gaussian distributions $\mathcal{N}(\nullv,\Cov_\mu)$ with (from left to right) $\mu\in[3]$, $\mu\in[4]$ and $\mu\in[5]$.  The samples of dimension  $d=10$,  split equally between the distributions,   were, respectively, of the size $N=3\times10^4$, $N=4\times10^4$ and $N=5\times10^4$.  The covariance matrices $\Cov_\mu$ were sampled from the Wishart distribution with $d+1$ degrees of freedom and precision matrix $\I$.   Top: BIC $\equiv2\mathcal{L}_N-n_\mathcal{N}\log(N)$, where $\mathcal{L}_N$ is  the  log-likelihood of GMM  estimated by EM algorithm and $n_\mathcal{N}$ is the number of parameters, as a function of $K$.  Bottom: $\hat{F}_N+\log(K)$, with $\hat{F}_N$ computed by the population dynamics algorithm, as a function of $K$.}
 \label{figure:10dL3-5} 
 \end{figure}

\begin{figure}[t]
 %\vspace*{-7mm}
 \setlength{\unitlength}{0.67mm}
 \begin{center}{
 %\hspace*{50mm}
 \begin{picture}(200,71)
 \put(0,0){\includegraphics[width=100\unitlength]{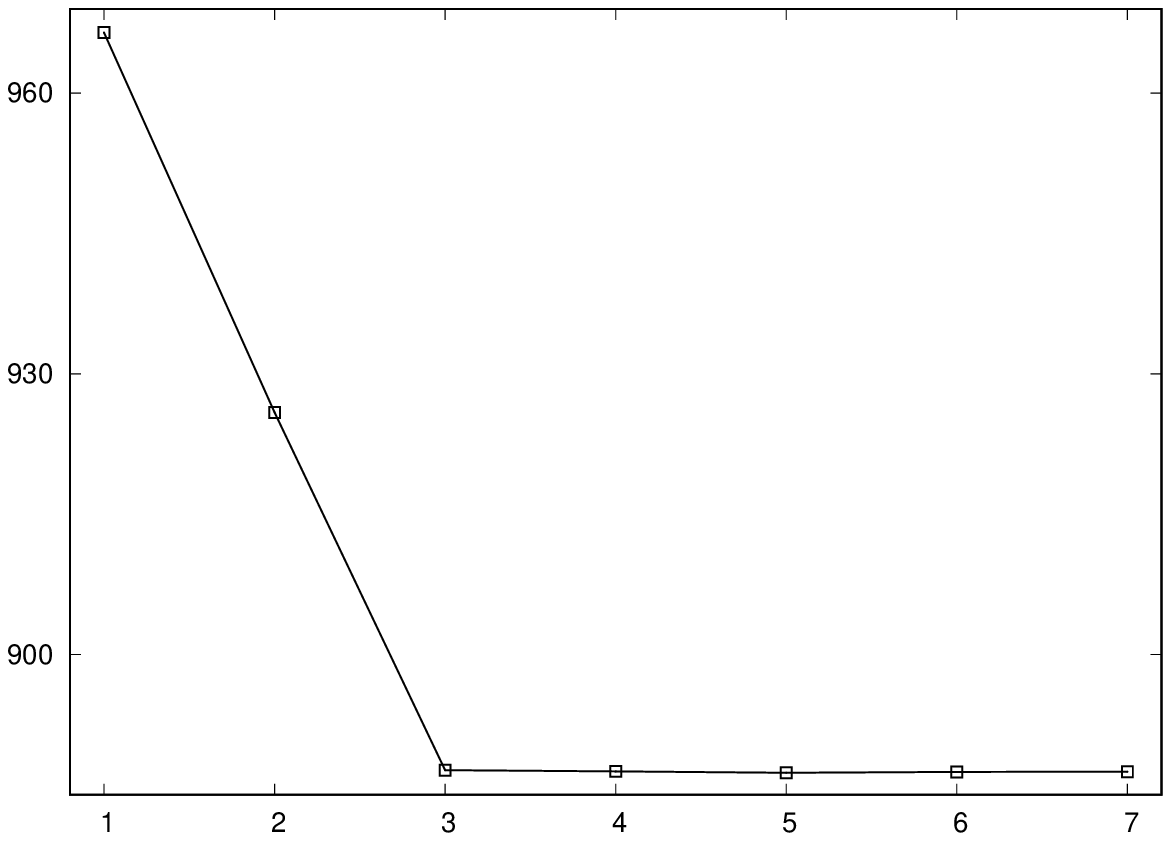}}%, height=61.8\unitlength
   \put(-5 ,25){\rotatebox{90}{\small{$ \hat{F}_N+\log(K)$}}}
 \put(50 ,-5){\small{$K$}} 
 \put(100,-9){\includegraphics[width=100\unitlength, height=87\unitlength]{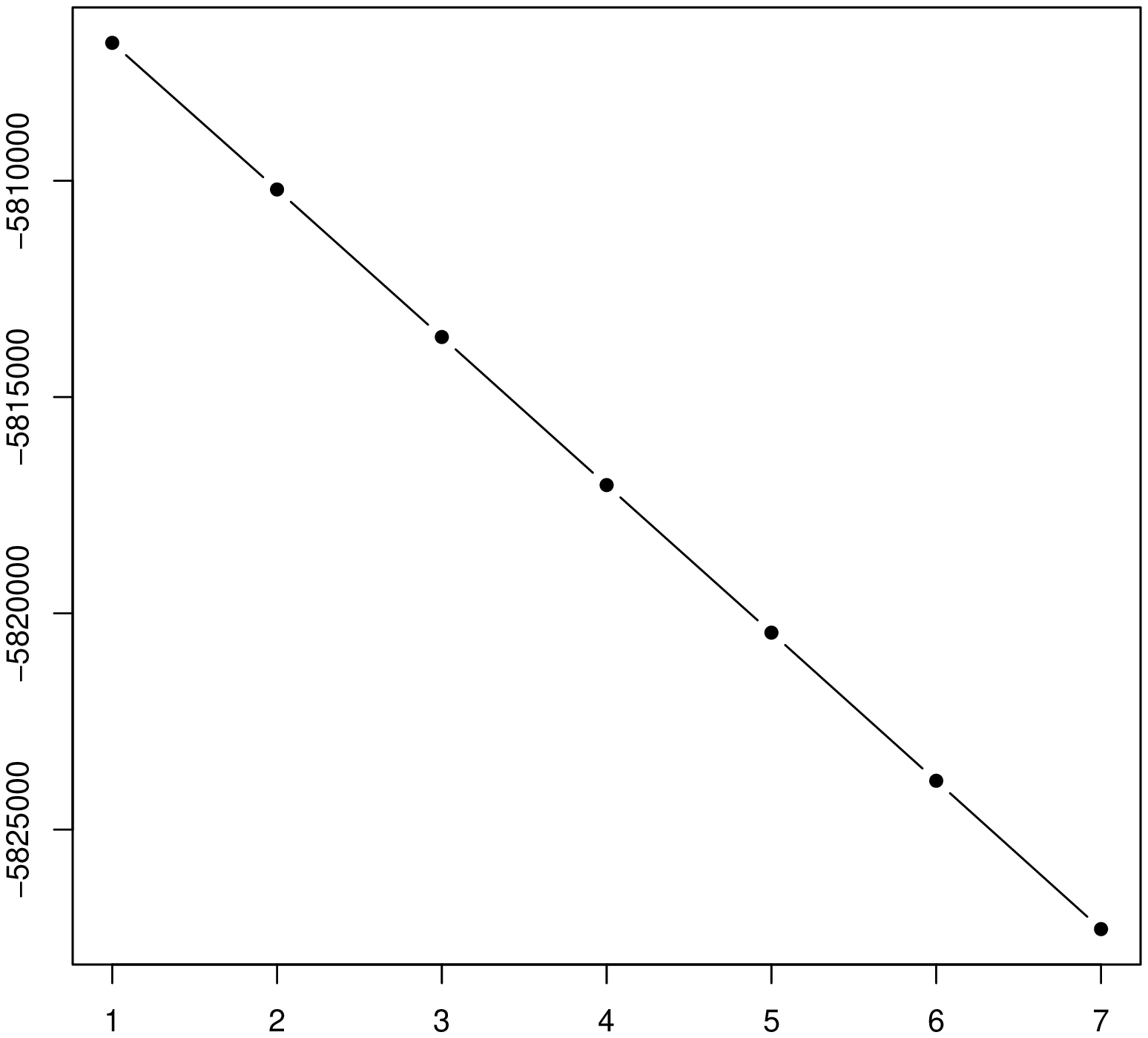}}
   \put(99,35){\rotatebox{90}{\small{$BIC$}}}
  \put(153 ,-5){\small{$K$}} 
  \end{picture}
 }\end{center}
% \vspace*{-7mm}
 \caption{Inferring number of clusters in the  data  generated  from Gaussian distributions $\mathcal{N}(\m_\mu,\Cov_\mu)$ with separation $\Delta/\sqrt{d}=\vert\vert\m_\mu-\m_\nu\vert\vert=5.2$, where $(\mu,\nu)\in[3]$.  The sample,  split equally between the distributions,   is of size $N=3\times10^3$, and of dimension  $d=500$. The (diagonal) covariance matrices $\Cov_\mu$  were sampled from $\chi^2$ distribution with $3$ degrees of freedom. The maximum and minimum diagonal entries in these matrices is, respectively,  $17.064$ and $0.017$, so $\Delta/\sqrt{d}=5.2$ ensures that clusters in the sample   are well separated (see \ref{app:sphericity}).  Left: $\hat{F}_N+\log(K)$, with$\hat{F}_N\equiv\min_{\cmatrix}\hat{F}_N(\cmatrix ,\xmatrix)$ computed by the population dynamics algorithm, as a function of $K$. Right:  BIC $\equiv2\mathcal{L}_N-n_\mathcal{N}\log(N)$, where $\mathcal{L}_N$ is  the  log-likelihood of GMM  estimated by EM algorithm and $n_\mathcal{N}$ is the number of parameters, as a function of $K$. }
 \label{figure:500dN1000L3} 
 \end{figure}

 Finally, we compare the Gaussian variant of the population dynamics clustering algorithm with a popular software package \cite{Scrucca2016}  which uses EM algorithm  to estimate the maximum $\mathcal{L}_N(\xmatrix)$ of the log-likelihood
 \begin{eqnarray}
\mathcal{\ell}_N(\xmatrix)&=& \sum_{i=1}^N \log\left( \sum_{\mu=1}^K w(\mu)\, \mathcal{N}(\x_i\vert\m_\mu,\Cov_\mu)\right)~\label{def:GMM}
\end{eqnarray}
with respect to  the 
parameters of the Gaussian mixture model (GMM)  $\sum_{\mu\leq K}  w(\mu)\, \mathcal{N}(\x_i\vert\m_\mu,\Cov_\mu)$, which are 
the 
means $\m_\mu$,  the covariances  $\Cov_\mu$ and  the weights  $w(\mu)\geq0$, where $\sum_{\mu\leq K}  w(\mu)=1$. To this end we consider inferring number of clusters in the samples of a Gaussian data with more than $L=2$ clusters, non-identity covariance matrices and a relatively large number of dimensions (see Figures  \ref{figure:10dL3},  \ref{figure:10dL3-5} and  \ref{figure:500dN1000L3}). The software package uses the Bayesian Information Criterion (BIC)  $2\mathcal{L}_N-n_\mathcal{N}\log(N)$, where  $n_\mathcal{N}$ is the number of parameters used in GMM,  and the population dynamics algorithm uses  $\hat{F}_N+\log(K)$, with the log-likelihood $\hat{F}_N\equiv\min_{\cmatrix}\hat{F}_N(\cmatrix ,\xmatrix)$ estimated by the average energy   $e(\infty)$, to infer the number of clusters in the data.
 \begin{figure}[t]
 %\vspace*{-7mm}
 \setlength{\unitlength}{0.67mm}
 \begin{center}{
 %\hspace*{50mm}
 \begin{picture}(200,71)
 \put(0,0){\includegraphics[width=100\unitlength]{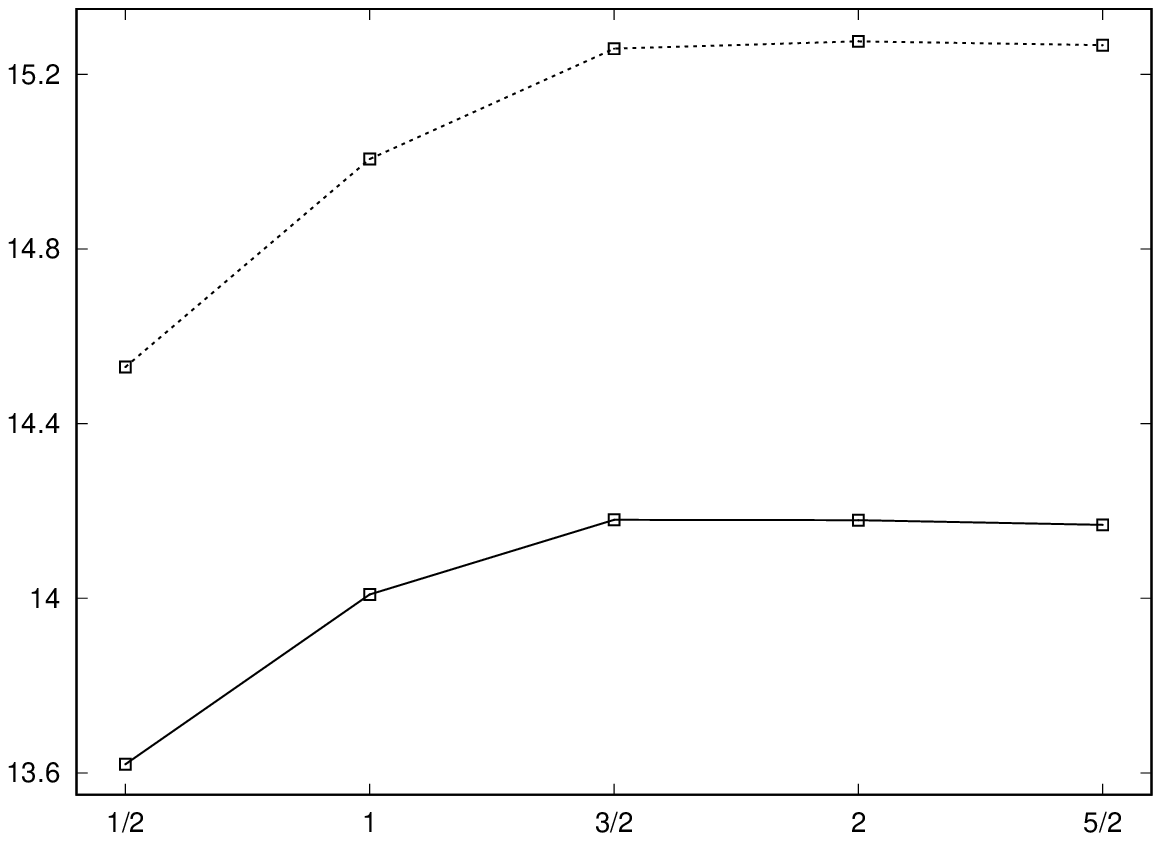}}%, height=61.8\unitlength
 %  \put(-5 ,25){\rotatebox{90}{\small{$ \min_{K}\hat{F}_N$}}}
 \put(50 ,-5){\small{$\Delta/\sqrt{d}$}} 
 \put(100,0){\includegraphics[width=100\unitlength]{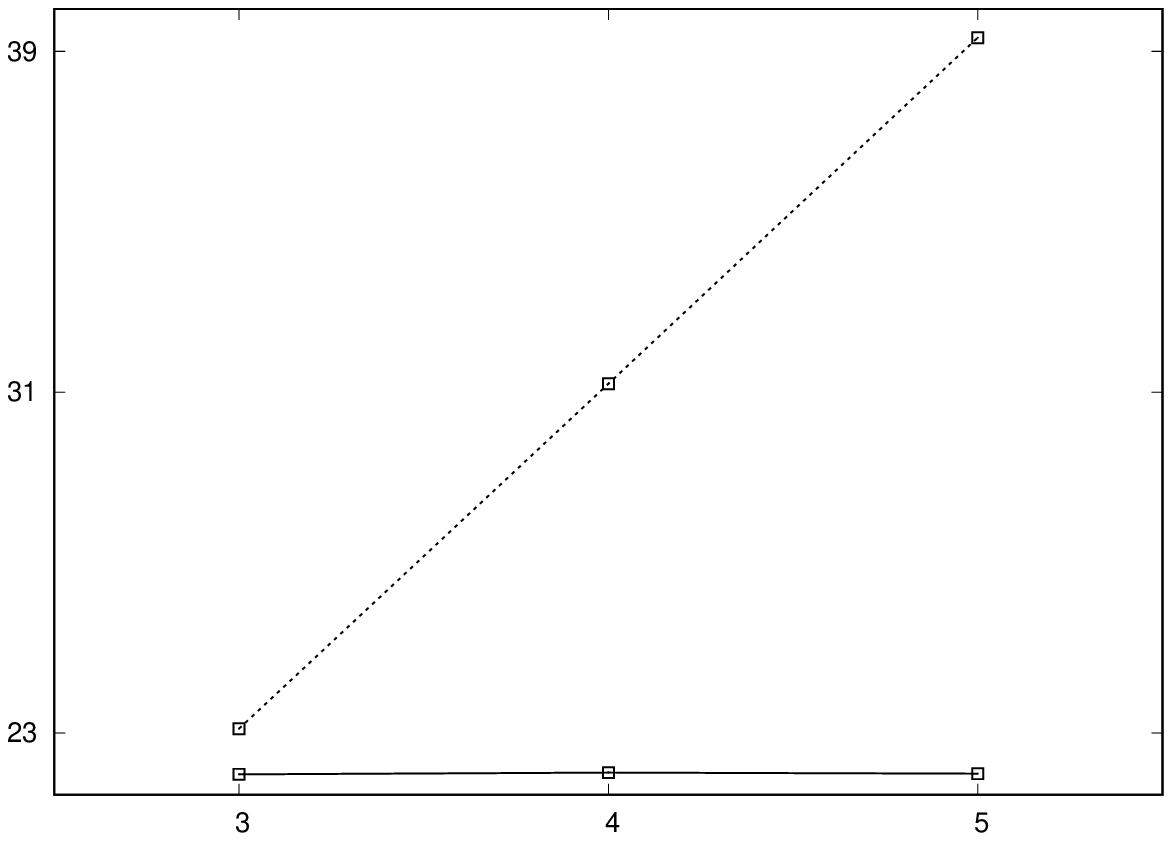}}
  \put(150 ,-5){\small{$K$}} 
  \end{picture}
 }\end{center}
% \vspace*{-7mm}
 \caption{The log-likelihood  densities $-\mathcal{L}_N/N$ (top dotted line)  and $\hat{F}_N$ (bottom solid line) plotted as functions  of, respectively, the cluster separation $\Delta/\sqrt{d}$ (computed at the inferred number of clusters)  and inferred number of clusters $K$ for the data described in Figures \ref{figure:10dL3} and \ref{figure:10dL3-5}.}
 \label{figure:EM-popul-dynam} 
 \end{figure}

For uncorrelated  data  we observe in  Figure \ref{figure:10dL3} that inference success in both methods  is strongly affected by the degree of  separation $\Delta$ of the clusters in the data, as  measured by the Euclidean distance between the means of Gaussians.  For small $\Delta$ the recovery of the true  number $L=3$  of clusters is not possible.   A simple MF argument, similar to the one used for $L=2$,  predicts that this inference failure latter will happen when $\Delta\leq 2\sqrt{3}$, i.e. exactly as for  $L=2$. However,  both algorithms are found to `work'  below this MF threshold (see  Figure \ref{figure:10dL3}) suggesting that the MF argument gives an upper bound. For correlated data, even when the separation parameter $\Delta$ is zero, the true number of clusters can still be recovered correctly by both algorithms (see Figure \ref{figure:10dL3-5}).  In all numerical experiments described in Figures  \ref{figure:10dL3}  and  \ref{figure:10dL3-5}  the log-likelihood density $-\mathcal{L}_N/N$, estimated by EM algorithm, is an upper bound for the log-likelihood density $\hat{F}_N$ computed by Gaussian population dynamics.   This points, at least in the regime of finite dimension $d$ and sample size $N\rightarrow\infty$,  to a possible relation between these likelihood functions.

In the high dimensional regime $d\rightarrow\infty$ and $N\rightarrow\infty$, with $d/N$ finite,  both algorithms fail to find the correct number of clusters (see Figure  \ref{figure:500dN1000L3}),  but they fail differently. The algorithm which uses Gaussian population dynamics, which was derived assuming finite $d$ and $N\rightarrow\infty$,  predicts more than $L=3$ clusters in the data, and the algorithm which uses EM predicts only one cluster. However, the population dynamics `almost'  predicts the correct number $L=3$  of clusters:  the changes in the  log-likelihood function $\hat{F}_N+\log(K)$ in the $K>3$ regime are   much smaller than in the  $K\leq3$ regime,   as can be seen in Figure  \ref{figure:500dN1000L3}.  This behaviour is also observed for similarly generated  data with the same sample size  but  with higher dimensions (not shown here), suggesting that taking into account the effect of the dimension $d$ properly in the present theoretical framework could lead to  improvements in  inference.  

\section{Discussion\label{section:discuss}}

 In this paper we use statistical mechanics to study model-based Bayesian clustering.   The partitions of data are microscopic states, the negative log-likelihood  of the data is the energy of these states, and  the data act as disorder in the model.  The optimal (MAP) partition corresponds to the  minimal energy state, i.e. the  ground state of this system.  The latter can be obtained from the free energy  via a low `temperature' limit, so to investigate MAP inference we evaluate the free energy.  We assume that  in a very large system,  i.e. for a large sample size, the free energy (density) is self-averaging.  This allows us to focus on the disorder-averaged free energy,  using the replica method. Following the prescription of the replica  method we first compute the average for an integer $n$ number of replicas, then we take the large system limit followed by the limit $n\rightarrow0$.  The latter is facilitated by assuming replica symmetry (RS)  in the  order parameter equation.   The main order parameter in the theory is the (average) distribution of data in each cluster $\mu\in[K]$.  
 
 In the low temperature limit,  the equations of the RS theory allow us to study the low energy states of the system. In this limit the average free energy and average energy are identical.   We show that the true partitions of the data are recovered exactly when the assumed number of clusters $K$ and the true number of clusters $L$ are equal, and the model distributions $P(\x\vert\thetav_\mu)$ have non overlapping supports for different  $\thetav_\mu$.  The high temperature limit  of the RS theory  recovers the  mean-field theory of~\cite{Mozeika2018}.  In this latter limit, the average energy, which equals the MF entropy~\cite{Mozeika2018}, is dominated by the  prior.  The MF entropy is an upper bound for  the low temperature average energy, and  can be optimised by selecting the prior.  Our order parameter equation can be solved numerically using a population dynamics algorithm.  Using this algorithm for the  Gaussian data  very accurately reproduces the results obtained  by gradient descent, minimising the negative log-likelihood of data,  algorithm  even  in the regime of a  small separations between clusters  and when $K>L$ where the MF theory gives incorrect predictions~\cite{Mozeika2018}.  The zero temperature population dynamics algorithm can be used for MAP inference.  %\Red{,  and its finite temperature version can be used for the full Bayesian inference of the most probable number of clusters.} 
 
 There are several interesting directions into which to extend the present work.  Many current studies use the so-called Rand index~\cite{Rand1971}, or the `purity'~\cite{Manning2010},  for measuring the dissimilarity between the true and inferred clusterings of data, but it would be also interesting  to estimate the probability that the  inferred clustering is `wrong'. Another direction is to consider  the high dimensional regime where $N\rightarrow\infty$ and  $d\rightarrow\infty$,  with $d/N$ finite.  We envisage that here the task of separating clusters may be `easier' than in the lower dimensional $d/N\rightarrow0$ regime, due to the  `blessing of dimensionality' phenomenon~\cite{Gorban2018}, according to which most data sampled from  high-dimensional Gaussian distributions reside in  the `thin' shell  of a  sphere (see \ref{app:sphericity}).  Both the early study \cite{Barkai1993}  and the more recent study \cite{Shalabi2018} on Bayesian  discriminant analysis indicate that  the classification of data, a supervised inference problem closely related to clustering,  becomes significantly easier   in the high-dimensional regime.  Alternatively,  the high dimensional regime could also cause  overfitting, and one may want to quantify this phenomena by using a more general information-theoretic measure of  overfitting~\cite{Coolen2017}.

\section*{Acknowledgements}
This work was supported by the Medical Research Council of the  United Kingdom (grant MR/L01257X/1).

\appendix 

\section{Disorder average\label{app:disorder-average}}
In this Appendix we study the average 
\begin{eqnarray}
\hspace*{-10mm} 
&&\hspace*{-15mm} \Big\langle \Big\langle  \rme^{-\beta N   \sum_{\alpha=1}^n\hat{F}_N(\cmatrix^\alpha, \xmatrix)}     \Big\rangle_{\left\{\cmatrix^\alpha\right\}}   \Big\rangle_{\xmatrix}   \nonumber\\
\hspace*{-10mm} 
&=&   \int\!\rmd\x_1\cdots \rmd\x_N \sum_{\cmatrix}q(\cmatrix\vert L) \Big\{\prod_{\nu=1}^L \prod_{i=1}^N     q_{\nu}^{c_{i\nu}}(\x_{i})\Big\}
 \Big\langle  \rme^{-\beta N   \sum_{\alpha=1}^n\hat{F}_N(\cmatrix^\alpha, \xmatrix)}        \Big\rangle_{\left\{\cmatrix^\alpha\right\}} \nonumber\\
\hspace*{-10mm} 
&=&   \int\!\rmd\x_1\cdots \rmd\x_N   \Big\langle\Big\{\prod_{\nu=1}^L \prod_{i=1}^N     q_{\nu}^{c_{i\nu}}(\x_{i})\Big\} \rme^{-\beta N   \sum_{\alpha=1}^n\hat{F}_N(\cmatrix^\alpha, \,\xmatrix)}   \Big \rangle_{\left\{\cmatrix^\alpha\right\};\cmatrix}    \label{eq:disorder-aver-comp-2}, 
\end{eqnarray}
where the average $\langle\cdots  \rangle_{\left\{\cmatrix^\alpha\right\};\cmatrix} $ now refers to the distribution  $\left\{\prod_{\alpha=1}^n P(\cmatrix^\alpha\vert K)\right\}q(\cmatrix\vert L)$.  If we define the density 
\begin{eqnarray}
Q_{\mu}(\x\vert\cmatrix^\alpha,\xmatrix)&=&\frac{1}{N}\sum_{i=1}^N c_{i\mu}^\alpha \delta(\x-\x_i)    \label{def:Q-replica-order-parameter}, 
\end{eqnarray}
then we may write
\begin{eqnarray}
\hspace*{-10mm}
-N \sum_{\alpha=1}^n\!\hat{F}_N(\cmatrix^\alpha, \,\xmatrix)&=&\sum_{\alpha=1}^n\sum_{\mu=1}^K\log\big\langle\rme^{\sum_{i=1}^N c_{i\mu}^\alpha\!\log P(\x_{i}\vert\thetav_{\mu})    } \big\rangle_{\thetav_{\mu}}   \nonumber\\
\hspace*{-10mm}
&=&\sum_{\alpha=1}^n\sum_{\mu=1}^K\!\log\! \big\langle \rme^{N\!\int\!\rmd\x~ Q_{\mu}(\x\vert\cmatrix^\alpha,\xmatrix)  \log P(\x\vert\thetav_{\mu})    } \big\rangle_{\thetav_{\mu}} ~~
\end{eqnarray}
and  for (\ref{eq:disorder-aver-comp-2}) we obtain 
\begin{eqnarray}
\hspace*{-10mm} 
&&\hspace*{-20mm} 
\int\!\rmd\x_1\cdots  \rmd\x_N~   \Big\langle\!  \Big\{\prod_{\nu=1}^L \prod_{i=1}^N     q_{\nu}^{c_{i\nu}}(\x_{i})\Big\} \rme^{   \beta\!\sum_{\alpha=1}^n\sum_{\mu=1}^K\log\big\langle \rme^{N\int Q_{\mu}(\x\vert\cmatrix^\alpha,\xmatrix)  \log P(\x\vert\thetav_{\mu})\rmd\x    } \big\rangle_{\thetav_{\mu}}       }   \Big\rangle_{\!\!\left\{\cmatrix^\alpha\right\}; \cmatrix}  \nonumber\\
\hspace*{-10mm} 
&=& \int\!\rmd\x_1\cdots \rmd\x_N~  \Big\langle \Big\{\prod_{\nu=1}^L \prod_{i=1}^N     q_{\nu}^{c_{i\nu}}(\x_{i})\Big\}
 \nonumber\\
\hspace*{-10mm} 
&& \hspace*{5mm}\times\prod_{\alpha=1}^n\prod_{\mu=1}^K \Big\{\prod_{\x}\int\! \rmd Q_{\mu}^\alpha(\x)~ \delta\left[ Q_{\mu}^\alpha(\x ) -  Q_{\mu}(\x\vert\cmatrix^\alpha,\xmatrix) \right]\Big\}\nonumber\\
\hspace*{-10mm} 
&& \hspace*{5mm} \times\rme^{   \beta\!\sum_{\alpha=1}^n\sum_{\mu=1}^K\log\left\langle \rme^{N\int Q_{\mu}^\alpha(\x )  \log P(\x\vert\thetav_{\mu})\rmd\x    } \right\rangle_{\thetav_{\mu}}       }   \Big \rangle_{\!\!\left\{\cmatrix^\alpha\right\}; \cmatrix}   \nonumber\\
\hspace*{-10mm} 
&=& \int \left\{\rmd \Q\,  \rmd\hat{\Q}  \right\}\rme^{\rmi N  \sum_{\alpha=1}^n\sum_{\mu=1}^K  \int \hat{Q}_{\mu}^\alpha(\x )  Q_{\mu}^\alpha(\x ) \rmd \x} \nonumber\\
\hspace*{-10mm} 
&& \hspace*{5mm}  \times\rme^{ \beta \sum_{\alpha=1}^n\sum_{\mu=1}^K\log \langle \rme^{N\int Q_{\mu}^\alpha(\x )  \log P(\x\vert\thetav_{\mu})\rmd\x    } \rangle_{\thetav_{\mu}}  }\nonumber\\
\hspace*{-10mm} 
 && \hspace*{5mm}  \times  \Big\langle \prod_{i=1}^N  \int\! \rmd\x_i     \Big\{\prod_{\nu=1}^L     q_{\nu}^{c_{i\nu}}(\x_{i})\Big\}  \rme^{-\rmi \sum_{\alpha=1}^n\sum_{\mu=1}^K  c_{i\mu}^\alpha \hat{Q}_{\mu}^\alpha(\x_i ) }    \Big \rangle_{\!\!\left\{\cmatrix^\alpha\right\};\cmatrix}    \label{eq:disorder-aver-comp-3}.
\end{eqnarray}
Using the properties of $\{c_{i\nu}\}$, the last line in the above expression can be rewritten as
\begin{eqnarray}
&&\hspace*{-20mm} 
 \Big\langle \prod_{i=1}^N  \int\! \rmd\x_i     \Big\{\prod_{\nu=1}^L     q_{\nu}^{c_{i\nu}}(\x_{i})\Big\}  \rme^{-\rmi \sum_{\alpha=1}^n\sum_{\mu=1}^K  c_{i\mu}^\alpha \hat{Q}_{\mu}^\alpha(\x_i ) }    \Big \rangle_{\!\!\left\{\cmatrix^\alpha\right\};\cmatrix} 
 \nonumber
 \\&=&
 \Big\langle \prod_{i=1}^N \Big\{   \sum_{\nu=1}^L c_{i\nu}  \!  \int\! \rmd\x ~      q_{\nu}(\x)    \rme^{-\rmi \sum_{\alpha=1}^n\sum_{\mu=1}^K  c_{i\mu}^\alpha \hat{Q}_{\mu}^\alpha(\x ) } \Big\}   \Big \rangle_{\!\!\left\{\cmatrix^\alpha\right\};\cmatrix}  \nonumber\\
&=& \Big\langle \rme^{  \sum_{i=1}^N \log   \sum_{\nu=1}^L c_{i\nu}    \int \rmd\x ~      q_{\nu}(\x)    \exp\big[-\rmi \sum_{\alpha=1}^n\sum_{\mu=1}^K  c_{i\mu}^\alpha \hat{Q}_{\mu}^\alpha(\x ) \big]  } \Big \rangle_{\!\!\left\{\cmatrix^\alpha\right\};\cmatrix} \label{eq:disorder-aver-comp-4}.
\end{eqnarray}
Since $c_{i\nu},c^\alpha_{i\nu}\in\{0,1\}$, 
subject to $\sum_{\nu=1}^L c_{i\nu}=\sum_{\mu=1}^K c_{i\mu}^\alpha=1$, it follows that 
 the vectors $\bc=(c_1,\ldots, c_L)$, $\bc_i=(c_{i1},\ldots, c_{iL})$, 
$\bc^\alpha=(c^\alpha_{1},\ldots, c^\alpha_{K})$  and $\bc_i^\alpha=(c^\alpha_{i1},\ldots, c^\alpha_{iK})$, will satisfy the identities $\bc\cdot\bc_i=\delta_{\bc,\bc_i}$ and $\bc^\alpha\!\cdot\bc_i^\alpha=\delta_{\bc^\alpha,\bc^\alpha_i}$.
Inserting $\sum_{\cv}\cv_i\cdot\cv=1$ and $\sum_{\cv^\alpha}\cv_i^\alpha\cdot\cv^\alpha=1$  into the exponential function in the average  (\ref{eq:disorder-aver-comp-4}) now gives, with 
$\muv=(\mu_1,\ldots,\mu_n)\in\{1,\ldots,K\}^n$:
\begin{eqnarray}
\hspace*{-10mm}
&&
\hspace*{-20mm}  \sum_{i=1}^N \log   \sum_{\nu=1}^L c_{i\nu}    \int\!\rmd\x~     q_{\nu}(\x)    \rme^{-\rmi \sum_{\alpha=1}^n\sum_{\mu=1}^K  c_{i\mu}^\alpha \hat{Q}_{\mu}^\alpha(\x ) }     \nonumber\\
\hspace*{-10mm}
&=&\sum_{\cv}    \sum_{\{\cv^\alpha\}}   \sum_{i=1}^N\cv\cdot\cv_i  \prod_{\alpha=1}^n \cv^\alpha\!\cdot\cv_i^\alpha
\log   \sum_{\nu=1}^L c_{\nu}    \int\!\rmd\x~     q_{\nu}(\x)    \rme^{-\rmi \sum_{\alpha=1}^n\sum_{\mu=1}^K  c_{\mu}^\alpha \hat{Q}_{\mu}^\alpha(\x ) }      \nonumber\\
\hspace*{-10mm}
&=& \sum_{\nu, \muv} \sum_{i=1}^N c_{i\nu} \left\{\prod_{\alpha=1}^n   c_{i\mu_\alpha}^\alpha \right\}            \sum_{\cv}    \sum_{\{\cv^\alpha\}}    c_{\nu}   \left\{\prod_{\alpha=1}^n  c_{\mu_\alpha}^\alpha\right\}            \nonumber\\
\hspace*{-10mm}
&&~~~~~~~~~~~~~\times\log   \sum_{\nu^\prime=1}^L c_{\nu^\prime}    \int\! \rmd\x ~      q_{\nu^\prime}(\x)    \rme^{-\rmi \sum_{\alpha=1}^n\sum_{\mu^\prime_{\alpha}=1}^K  c_{\mu^\prime_{\alpha}}^\alpha \hat{Q}_{\mu^\prime_{\alpha}}^\alpha(\x ) }    \nonumber\\
\hspace*{-10mm}
&=& \sum_{\nu, \muv} \sum_{i=1}^N c_{i\nu}\! \left\{\prod_{\alpha=1}^n   c_{i\mu_\alpha}^\alpha \!\right\}\! \log\!     \int \!  \rmd\x  ~    q_{\nu}(\x)\,    \rme^{-\rmi \sum_{\alpha=1}^n \hat{Q}_{\mu_{\alpha}}^\alpha(\x ) }  \label{eq:disorder-aver-comp-5},
\end{eqnarray}
 where we used the identities  $\sum_{\bc^\alpha}c^\alpha_\mu=1$ for all $(\alpha,\mu)$, and $\sum_{\bc}c_\nu\log[\sum_{\nu^\prime}c_{\nu^\prime}\phi_{\nu^\prime}]=\log \phi_\nu$ for all $\nu$. Let us now define the  density
\begin{eqnarray}
A(\nu,\muv\vert\cmatrix,\{\cmatrix^\alpha\})&=& \frac{1}{N}\!\sum_{i=1}^N c_{i\nu}\! \left\{\prod_{\alpha=1}^n   c_{i\mu_\alpha}^\alpha \!\right\}  \label{def:A-replica-order-parameter}, 
\end{eqnarray}
where $NA(\nu,\muv\vert\cmatrix,\{\cmatrix^\alpha\})$ is the number of data-points that are sampled  from the distribution $q_\nu(\x)$ and  assigned to clusters $\mu_1, \ldots,\mu_n$ for the $n$ replicas, respectively.  Using  this definition  and (\ref{eq:disorder-aver-comp-5})  in equation (\ref{eq:disorder-aver-comp-4}) converts the latter expression into
\begin{eqnarray}
&&
\hspace*{-20mm}
\Big\langle \rme^{ N\!\sum_{\nu, \muv } A(\nu,\muv\vert\cmatrix,\left\{\cmatrix^\alpha\right\})\! \log\!     \int \!  \rmd\x ~    q_{\nu}(\x)\exp\big[-\rmi \sum_{\alpha=1}^n \hat{Q}_{\mu_{\alpha}}^\alpha(\x ) \big]    } \Big \rangle_{\!\!\left\{\cmatrix^\alpha\right\};\cmatrix}  \nonumber\\
&=& \Big\langle \prod_{\nu, \muv  } \int\!\rmd A(\nu,\muv  )~ \delta\left[ A(\nu,\muv)-A(\nu,\muv\vert\cmatrix,\left\{\cmatrix^\alpha\right\})\right]\Big\rangle_{\!\!\left\{\cmatrix^\alpha\right\};\cmatrix}\nonumber\\
&&~~~\times\rme^{ N\!\sum_{\nu, \muv } A(\nu,\muv )\! \log\!     \int \! \rmd\x ~   q_{\nu}(\x)\exp[-\rmi \sum_{\alpha=1}^n \hat{Q}_{\mu_{\alpha}}^\alpha(\x )]     }\nonumber\\
&=&  \int\!\{\rmd A\,\rmd\hat{A}\}\, \rme^{N    \tilde{\Psi}[\{\hat{\Q}\};\{A,\hat{A}\}]  }\label{eq:disorder-aver-comp-6},%
\end{eqnarray}
where
\begin{eqnarray}
\hspace*{-5mm}
\tilde{\Psi}[\{\hat{\Q}\};\{A,\hat{A}\}]&=&\sum_{\nu, \muv  } A(\nu,\muv)\Big[ \rmi \hat{A}(\nu,\muv) 
+\log\!     \int \! \rmd\x ~    q_{\nu}(\x)\,    \rme^{-\rmi \sum_{\alpha=1}^n \hat{Q}_{\mu_{\alpha}}^\alpha(\x ) }   \Big]\nonumber\\
\hspace*{-5mm}
&&+\frac{1}{N}\log\left\langle\rme^{-\rmi N\sum_{\nu, \muv  }\hat{A}(\nu,\muv)  A(\nu,\muv\vert\cmatrix,\left\{\cmatrix^\alpha\right\})} \right \rangle_{\!\!\left\{\cmatrix^\alpha\right\};\cmatrix}.
\label{eq:disorder-aver-comp-7}
\end{eqnarray}
Finally, using (\ref{eq:disorder-aver-comp-6}) in the average (\ref{eq:disorder-aver-comp-3}) gives us the integral (\ref{eq:disorder-aver-comp-8}), as claimed. 

\section{Derivation of RS equations\label{app:replica-limit}}
The RS assumption implies that  $Q_{\mu_\alpha}^\alpha\!(\x )=Q_{\mu_\alpha}\!(\x )$, from which one deduces $\thetav_{\!\mu}^\alpha=\thetav_{\!\mu_\alpha}$ via  (\ref{eq:SP-comp-2}). Insertion of these forms into  the right-hand side of (\ref{eq:Q-R}), using  (\ref{eq:A-R}), 
leads to%
\begin{eqnarray}
&&\hspace*{-15mm}
\sum_{\nu, \muv }\delta_{\mu;\mu_\alpha}   A(\nu,\muv)   \frac{     q_{\nu}(\x)\,  
    \rme^{\sum_{\gamma=1}^n \beta \log P(\x\vert\thetav_{\mu_\gamma}) }
 }{   \int \!     q_{\nu}(\tilde{\x})\,   
      \rme^{\sum_{\gamma=1}^n \beta \log P( \tilde{\x}\vert\thetav_{\mu_\gamma}) } \rmd\tilde{\x}
    }  \nonumber\\
 &=&\sum_{\nu, \muv }\delta_{\mu;\mu_\alpha}\frac{\tilde{A}(\nu)  
 \int \! \rmd\x ~    q_{\nu}(\x)     \Big[ \prod_{\gamma=1}^n\tilde{A}(\mu_\gamma\vert\nu)\, \rme^{ \beta \log P(\x\vert\thetav_{\mu_\gamma}) }  \Big]  
    }{ \sum_{\tilde{\nu}} \tilde{A}(\tilde{\nu})
    \int \!  \rmd\x  ~    q_{\tilde{\nu}}(\x)   \Big[ \prod_{\gamma=1}^n\sum_{\tilde{\mu}_\gamma}\tilde{A}(\tilde{\mu}_\gamma\vert\tilde{\nu})\, \rme^{\beta \log P(\x\vert\thetav_{\tilde{\mu}_\gamma}) }   \Big]
    }\nonumber\\
&&~~~~~~~~~~\times\frac{     q_{\nu}(\x)\,  
    \rme^{\sum_{\gamma=1}^n \beta \log P(\x\vert\thetav_{\mu_\gamma}) }
 }{   \int \!     q_{\nu}(\tilde{\x})\,   
      \rme^{\sum_{\gamma=1}^n \beta \log P( \tilde{\x}\vert\thetav_{\mu_\gamma}) } \rmd\tilde{\x}
    }  \nonumber\\ 
     &=&
     \sum_{\nu}\tilde{A}(\nu) \frac{
     q_{\nu}(\x)\,\tilde{A}(\mu\vert\nu)\,\rme^{    \beta \log P(   \x\vert\thetav_{\mu}      )      } \Big[
     \sum_{\tilde{\mu}}\tilde{A}(\tilde{\mu}\vert\nu)\,\rme^{    \beta \log P(   \x\vert\thetav_{\tilde{\mu}}      )      }  \Big]^{n-1}      
    }{ \sum_{\tilde{\nu}} \tilde{A}(\tilde{\nu})
    \int \!   \rmd\x  ~   q_{\tilde{\nu}}(\x)   \Big[\sum_{\tilde{\mu}}\tilde{A}(\tilde{\mu}\vert\tilde{\nu})\, \rme^{\beta \log P(\x\vert\thetav_{\tilde{\mu}}) }   \Big]^n
    }.
    \nonumber\\[-1mm]
    &&
    \end{eqnarray}
We can now take the replica limit $n\rightarrow0$, and obtain (\ref{eq:Q-RS}).  
Using the RS assumption  in  (\ref{eq:A-R}) gives us the following expression for the marginal 
$A(\nu)=\sum_{\muv}  A(\nu, \muv)$:
\begin{eqnarray}
\hspace*{-0mm}
A(\nu)&=&  \frac{\tilde{A}(\nu)  
 \int \!  \rmd\x   ~    q_{\nu}(\x)     \Big[ \sum_\mu\tilde{A}(\mu\vert\nu)\, \rme^{ \beta \log P(\x\vert\thetav_{\mu}) }  \Big]^n 
    }{ \sum_{\tilde{\nu}} \tilde{A}(\tilde{\nu})
    \int \! \rmd\x  ~    q_{\tilde{\nu}}(\x)   \Big[\sum_{\tilde{\mu}}\tilde{A}(\tilde{\mu}\vert\tilde{\nu})\, \rme^{\beta \log P(\x\vert\thetav_{\tilde{\mu}}) }   \Big]^n }
    \end{eqnarray}
Hence $\lim_{n\to 0}A(\nu)= \tilde{A}(\nu)$. 
The RS equation for the conditional $A(\muv\vert\nu)$ becomes
\begin{eqnarray}
A(\muv\vert\nu)&=&  \frac{  
 \int \! \rmd\x ~    q_{\nu}(\x)     \Big[\prod_{\alpha=1}^n\tilde{A}(\mu_\alpha\vert\nu)\, \rme^{ \beta \log P(\x\vert\thetav_{\mu_\alpha}) }  \Big]
    }{ \sum_{\tilde{\nu}} \tilde{A}(\tilde{\nu})
    \int \!  \rmd\x ~   q_{\tilde{\nu}}(\x)   \Big[ \sum_{\tilde{\mu}}\tilde{A}(\tilde{\mu}\vert\tilde{\nu})\, \rme^{\beta \log P(\x\vert\thetav_{\tilde{\mu}}) }   \Big]^n  
    }
    \end{eqnarray}
  Its conditional marginal is
\begin{eqnarray}
\hspace*{-15mm}
A(\mu\vert\nu)&=&
 \frac{  
 \int \! \rmd\x ~    q_{\nu}(\x)   \tilde{A}(\mu\vert\nu)\, \rme^{ \beta \log P(\x\vert\thetav_{\mu}) }   \Big[\sum_{\tilde{\mu}}\tilde{A}(\tilde{\mu}\vert\nu)\, \rme^{ \beta \log P(\x\vert\thetav_{\tilde{\mu}}) }  \Big]^{n-1}
    }{ \sum_{\tilde{\nu}} \tilde{A}(\tilde{\nu})
    \int \!  \rmd\x ~   q_{\tilde{\nu}}(\x)   \Big[ \sum_{\tilde{\mu}}\tilde{A}(\tilde{\mu}\vert\tilde{\nu})\, \rme^{\beta \log P(\x\vert\thetav_{\tilde{\mu}}) }   \Big]^n  
    },
\end{eqnarray}
which for $n\to 0$ becomes (\ref{eq:A-RS}):
\begin{eqnarray}
A(\mu\vert\nu)&=&
 \int \! \rmd\x ~    q_{\nu}(\x)  \frac{ \tilde{A}(\mu\vert\nu)\, \rme^{ \beta \log P(\x\vert\thetav_{\mu}) } }{ \sum_{\tilde{\mu}}\tilde{A}(\tilde{\mu}\vert\nu)\, \rme^{ \beta \log P(\x\vert\thetav_{\tilde{\mu}}) } }
   \end{eqnarray}
Finally, inserting $Q_{\mu_\alpha}^\alpha\!(\x )=Q_{\mu_\alpha}\!(\x )$ and  $\thetav_{\!\mu}^\alpha=\thetav_{\!\mu_\alpha}$ into the nontrivial part of the average free energy (\ref{eq:f-R}) and taking the limit $n\to 0$ gives equation  (\ref{eq:f-RS}):
\begin{eqnarray}
\hspace*{-15mm}
f(\beta)-\phi(\beta)&=&-\lim_{n\rightarrow0}\frac{1}{\beta n}    \log  \Big\{  \sum_{\nu} \tilde{A}(\nu)\int \! \rmd\x~    q_{\nu}(\x)\Big[   
\sum_{\mu=1  }^K   \tilde{A}(\mu\vert\nu)\,   \rme^{ \beta \log P(\x\vert\thetav_{\mu}) }  \Big]^n      \Big\}\nonumber\\
 \hspace*{-15mm}
 &=&-\frac{1}{\beta }  \sum_{\nu} \tilde{A}(\nu) \int \!  \rmd\x ~    q_{\nu}(\x) \log\Big[     \sum_{\mu=1  }^K   \tilde{A}(\mu\vert\nu)\,   \rme^{ \beta \log P(\x\vert\thetav_{\mu}) }  \Big]
\end{eqnarray}

\section{Physical meaning of observables\label{app:physical-meaning}}

Let us consider the following two averages: 
\begin{eqnarray}
 Q_\mu(\x) &=&\left\langle \left\langle  Q_\mu(\x\vert\cmatrix,\xmatrix)\right\rangle_{\cmatrix\vert\xmatrix}  \right\rangle_{\xmatrix}, \label{def:Q-aver}
\\
 A(\nu,\mu) &=&\left\langle \left\langle  A(\nu,\mu\vert\cmatrix,\xmatrix)\right\rangle_{\cmatrix\vert\xmatrix}  \right\rangle_{\xmatrix},    \label{def:A-aver}
\end{eqnarray}
in which 
 $\left\langle\cdots\right\rangle_{\cmatrix\vert\xmatrix} $ is generated by  the Gibbs-Boltzmann distribution  (\ref{def:P(C|X)}) and  the disorder average $\left\langle\cdots\right\rangle_{\xmatrix}$ by the distribution (\ref{eq:q(X)}). 
Using the replica identity
\begin{eqnarray}
   \frac{\sum_{\cmatrix} W(\cmatrix) F(\cmatrix)}{\sum_{\cmatrix} W(\cmatrix)}&=&
   \lim_{n\rightarrow0}  \sum_{\cmatrix} W(\cmatrix)  F(\cmatrix) \Big\{\sum_{\tilde{\cmatrix}} W(\tilde{\cmatrix})\Big\}^{n-1}  
   \nonumber
   \\
   &=& \lim_{n\rightarrow0}  \sum_{\cmatrix^1}\ldots \sum_{\cmatrix^n} F(\cmatrix^1)  \prod_{\alpha=1}^n W(\cmatrix^\alpha)  \label{eq:replica-identity}
\end{eqnarray}
we may write for any test function $g(\x)$ 
\begin{eqnarray}
\hspace*{-21mm}
\int \!\rmd\x ~Q_\mu(\x)  &=&  \Big\langle\sum_{\cmatrix^1} \!\cdots\!  \sum_{\cmatrix^n}
 \int \! \rmd\x ~Q_\mu(\x\vert\cmatrix^1\!,\xmatrix)g(\x)  
 \prod_{\alpha=1}^n\left[
 P(\cmatrix^\alpha\vert K) \rme^{-\beta N \hat{F}_N(\cmatrix^\alpha, \xmatrix)}\right]\Big\rangle_{\!\xmatrix}\nonumber\\
 \hspace*{-21mm}
 &=&\Big\langle\Big\langle \rme^{-\beta N \sum_{\alpha=1}^n \hat{F}_N(\cmatrix^\alpha, \xmatrix)} \int \!\rmd\x~Q_\mu(\x\vert\cmatrix^1\!,\xmatrix)  g(\x) \Big\rangle_{\!\xmatrix}\Big\rangle_{\!\{\cmatrix^\alpha\}}.
 \label{def:Q-aver-comp-2}
\end{eqnarray}
Following the same steps we used in computing the disorder average in (\ref{eq:disorder-aver-comp-1}) we obtain 
\begin{eqnarray}
 &&\hspace*{-20mm}  \Big\langle\Big\langle \rme^{-\beta N \sum_{\alpha=1}^n \hat{F}_N(\cmatrix^\alpha, \xmatrix)} \int\! Q_\mu(\x\vert\cmatrix^1,\xmatrix)   g(\x)\, \rmd\x \Big\rangle_{\!\xmatrix}\Big\rangle_{\!\{\cmatrix^\alpha\}} \nonumber\\
 &=& \int\{\rmd \Q\,\rmd\hat{\Q}\,\rmd A\,\rmd\hat{A}\}\, \rme^{N    \Psi[\{\Q, \hat{\Q}\};\{A,\hat{A}\}]  } \int\!\rmd\x~ Q^1_\mu\!(\x)   g(\x)\label{def:Q-aver-comp-3},
\end{eqnarray}
and for $n\rightarrow0$, using $\int\{\rmd \Q\,\rmd\hat{\Q}\,\rmd A\,\rmd\hat{A}\}\, \rme^{N    \Psi[\{\Q, \hat{\Q}\};\{A,\hat{A}\}]  } \int\! Q^1_\mu\!(\x)  \, \rmd\x =1$, this leads us for $N\to\infty$ to the desired asymptotic result  
\begin{eqnarray}
\hspace*{-20mm}
\lim_{N\to\infty} \int \!\rmd\x~Q_\mu(\x) g(\x)
 &=& \lim_{N\to\infty} \frac{\int\{\rmd \Q\,\rmd\hat{\Q}\,\rmd A\,\rmd\hat{A}\}\, \rme^{N    \Psi[\{\Q, \hat{\Q}\};\{A,\hat{A}\}]  } \int\! \rmd\x~Q^1_\mu\!(\x)   g(\x)}{ \int\{\rmd \Q\,\rmd\hat{\Q}\,\rmd A\,\rmd\hat{A}\}\, \rme^{N    \Psi[\{\Q, \hat{\Q}\};\{A,\hat{A}\}]  } }\nonumber\\
 \hspace*{-20mm}
  &=& \int\! \rmd\x~Q^1_\mu\!(\x)   g(\x)
  \label{def:Q-aver-comp-4},
\end{eqnarray}
where the distribution $Q^1_\mu\!(\x)$ is the solution of equation (\ref{eq:Q-R}). Thus, assuming that the replica symmetry assumption is correct, the physical meaning of the distribution in the our RS equation  (\ref{eq:Q-RS}) is given by   (\ref{eq:Q-phys-meaning}). 
Similarly we can work out   
\begin{eqnarray}
\hspace*{-10mm}
 A(\nu,\mu) &=&\left\langle \left\langle  A(\nu,\mu\vert\cmatrix,\xmatrix)\right\rangle_{\cmatrix\vert\xmatrix}  \right\rangle_{\xmatrix}     \nonumber\\
 \hspace*{-10mm}
 &=& \int \! \rmd\xmatrix~P(\xmatrix\vert L)   \sum_{\cmatrix}P_\beta(\cmatrix\vert \xmatrix)   A(\nu,\mu\vert\cmatrix,\xmatrix) \nonumber\\
 \hspace*{-10mm}
 &=&\sum_{\tilde{\cmatrix}} q(\tilde{\cmatrix}\vert L) \int \! \rmd\xmatrix~P(\xmatrix\vert\tilde{\cmatrix})   \sum_{\cmatrix}P_\beta(\cmatrix\vert \xmatrix)  A(\nu,\mu\vert\cmatrix,\xmatrix)  \nonumber\\[-1mm]
 \hspace*{-10mm}
&=&\sum_{\tilde{\cmatrix}} q(\tilde{\cmatrix}\vert L) \!\int \! \rmd\xmatrix ~P(\xmatrix\vert\tilde{\cmatrix})   \sum_{\cmatrix}P_\beta(\cmatrix\vert \xmatrix) \Big[ \frac{1}{N}\!\sum_{i=1}^N  c_{i\mu} \tilde{c}_{i\nu} \Big],~~
\label{eq:A-aver-comp-1}
\end{eqnarray}
where we used the definitions $\tilde{c}_{i\nu}= \Ind\left[ \x_i \sim q_\nu(\x)\right]$ and $P(\xmatrix\vert \cmatrix) =\prod_{\nu=1}^L \prod_{i=1}^N     q_{\nu}^{c_{i\nu}}(\x_{i})$. Substitution of the definition of $P_\beta(\cmatrix\vert \xmatrix) $ allows us to work out the average further:
\begin{eqnarray}
\hspace*{-15mm}
 A(\nu,\mu) &=&\sum_{\tilde{\cmatrix}} q(\tilde{\cmatrix}\vert L)\! \int \!\rmd\xmatrix~ P(\xmatrix\vert\tilde{\cmatrix}) \sum_{\cmatrix} \frac{P(\cmatrix\vert K)}{Z_\beta(\xmatrix)   }\rme^{-\beta N  \hat{F}_N(\cmatrix, \xmatrix)} \Big[\frac{1}{N}\!\sum_{i=1}^N  c_{i\mu} \tilde{c}_{i\nu}\Big] \nonumber\\
 \hspace*{-15mm}
 &=&\lim_{n\rightarrow0}\sum_{\tilde{\cmatrix}} q(\tilde{\cmatrix}\vert L) \!\int \!\rmd\xmatrix~ P(\xmatrix\vert\tilde{\cmatrix})  \sum_{\cmatrix} P(\cmatrix\vert K)\rme^{-\beta N  \hat{F}_N(\cmatrix, \xmatrix)} 
 \nonumber
 \\[-3mm]
 \hspace*{-15mm}&& \hspace*{60mm} \times Z^{n-1}_\beta\! (\xmatrix)\Big[\frac{1}{N}\!\sum_{i=1}^N  c_{i\mu} \tilde{c}_{i\nu}\Big]
 \nonumber
 \\
  \hspace*{-15mm}
 &=&\sum_{\cmatrix} q(\cmatrix\vert L) \sum_{\cmatrix^1} \cdots\sum_{\cmatrix^n} \Big[\prod_{\alpha=1}^nP(\cmatrix^\alpha\vert K)\Big]
  \int \! \rmd\xmatrix~P(\xmatrix\vert\cmatrix)
 \nonumber
 \\[-1mm]
  \hspace*{-15mm}
 &&\hspace*{42mm} \times \rme^{-\beta N  \sum_{\alpha=1}^n\hat{F}_N(\cmatrix^\alpha, \xmatrix)} \Big[\frac{1}{N}\!\sum_{i=1}^N  c_{i\nu} c^1_{i\mu}\Big] \nonumber\\
  \hspace*{-15mm}
 &=& \Big\langle\!\Big\langle\!\Big\langle\rme^{-\beta N  \sum_{\alpha=1}^n\hat{F}_N(\cmatrix^\alpha, \xmatrix)} \!  \sum_{\muv}\delta_{\mu;\mu_1}A(\nu,\muv\vert\cmatrix,\!\{\cmatrix^\alpha\}\!)
 \Big\rangle_{ \!\!\xmatrix\vert\cmatrix}\Big\rangle_{\!\!\cmatrix}\Big\rangle_{\!\!\{\cmatrix^\alpha\}}  \label{eq:A-aver-comp-3},
\end{eqnarray}
in which $A(\nu,\muv\vert\cmatrix,\!\{\cmatrix^\alpha\}\!)$ is  defined in  equation (\ref{def:A-replica-order-parameter}).  The above expression can now be used, following the same steps as for the $Q_\mu(\x)$ order parameter, to show  that for $ N\rightarrow\infty $ and  $n\rightarrow0$ the following will hold:
\begin{eqnarray}
\hspace*{-10mm}
\lim_{N\to\infty} A(\nu,\mu)
&=& \lim_{N\to\infty} \frac{\int\{\rmd \Q\,\rmd\hat{\Q}\,\rmd A\,\rmd\hat{A}\}\, \rme^{N    \Psi[\{\Q, \hat{\Q}\};\{A,\hat{A}\}]  }    \sum_{\muv}\delta_{\mu;\mu_1}A(\nu,\muv)        }{ \int\{\rmd \Q\,\rmd\hat{\Q}\,\rmd A\,\rmd\hat{A}\}\, \rme^{N    \Psi[\{\Q, \hat{\Q}\};\{A,\hat{A}\}]  } }\nonumber\\
 \hspace*{-10mm}
  &=&   \sum_{\muv}\delta_{\mu;\mu_1}A(\nu,\muv)\label{eq:A-aver-comp-4},
\end{eqnarray}
where $A(\nu,\muv)$ is the solution of equation (\ref{eq:Q-R}).  From this we deduce that  (\ref{eq:A-phys-meaning}) indeed gives the physical meaning of  the RS expression (\ref{eq:A-RS}). 

\section{Average energy\label{app:energy}}

In this Appendix we  compute the average energy
\begin{eqnarray}
e(\beta)&=& \Big\langle \Big\langle  \hat{F}_N(\cmatrix,\, \xmatrix)\Big\rangle_{\!\cmatrix\vert\xmatrix}  \Big\rangle_{\!\xmatrix}\nonumber\\
&=&   \lim_{n\rightarrow0}\Big\langle \!\sum_{\cmatrix} P(\cmatrix\vert K)\rme^{-\beta N  \hat{F}_N(\cmatrix, \xmatrix)} Z_\beta^{n-1}(\xmatrix)   \hat{F}_N(\cmatrix, \xmatrix)  \!  \Big\rangle_{\!\xmatrix},  
\label{def:e}
\end{eqnarray}
where iwe used the replica identity (\ref{eq:replica-identity}). Assuming initially that  $n\in\mathbb{N}$ allows us to compute the average over $\xmatrix$ in the above expression as follows
\begin{eqnarray}
\hspace*{-12mm}
 &&\hspace*{-12mm}
 \Big\langle\Big\langle \rme^{-\beta N \sum_{\alpha=1}^n \hat{F}_N(\cmatrix^\alpha, \xmatrix)} 
 \hat{F}_N(\cmatrix^1, \xmatrix)
\Big\rangle_{\!\xmatrix}\Big\rangle_{\!\{\cmatrix^\alpha\}}\nonumber\\
\hspace*{-12mm}
&=&\Big\langle\!\Big\langle \rme^{-\beta N \sum_{\alpha=1}^n \hat{F}_N(\cmatrix^\alpha, \xmatrix)}
\Big[\!-\!\frac{1}{N}\!\sum_{\mu=1}^K\log \left\langle \rme^{\sum_{i=1}^N\!c^1_{i\mu}\!\log P(\x_{i}\vert\thetav)    } \right\rangle_{\!\thetav}
\!\Big]\Big\rangle_{\!\xmatrix}\!\Big\rangle_{\!\{\cmatrix^\alpha\}}\nonumber\\
 \hspace*{-12mm}
 &=&-\Big\langle\!\Big\langle \rme^{-\beta N \sum_{\alpha=1}^n \hat{F}_N(\cmatrix^\alpha, \xmatrix)} \Big[\frac{1}{N}\!\sum_{\mu=1}^K\log \left\langle \rme^{N\!\int\!\rmd\x ~Q_\mu(\x\vert\cmatrix^1\!,\xmatrix)\!\log P(\x\vert\thetav)    } \right\rangle_{\!\thetav}\!\Big]
 \Big\rangle_{\!\xmatrix}\!\Big\rangle_{\!\{\cmatrix^\alpha\}}
 \hspace*{-3mm}
 \nonumber
 \\
 \hspace*{-12mm}
 &&
 \label{eq:e-comp-1}
\end{eqnarray}
and, with the short-hand $\Psi[\ldots]=\Psi[\{\Q, \hat{\Q}\};\{A,\hat{A}\}]$ and after taking the replica limit $n\rightarrow0$ within the RS ansatz, we then arrive at equation (\ref{eq:e-RS}):
\begin{eqnarray}
\hspace*{-15mm}
  &&
  \hspace*{-10mm}
 \lim_{n\to 0}\lim_{N\rightarrow\infty} \left\langle\left\langle \rme^{-\beta N \sum_{\alpha=1}^n \hat{F}_N(\cmatrix^\alpha, \xmatrix)} 
 \hat{F}_N(\cmatrix^1, \xmatrix)
\right\rangle_{\xmatrix}\right\rangle_{\{\cmatrix^\alpha\}}\nonumber\\
 \hspace*{-10mm}
 &=& -\lim_{N\rightarrow\infty}
 \frac{\int\{\rmd \Q\,\rmd\hat{\Q}\,\rmd A\,\rmd\hat{A}\}\, \rme^{N    \Psi[\ldots]  } 
    \Big[\frac{1}{N}\!\sum_{\mu=1}^K\log \left\langle \rme^{N\!\int\! \rmd\x    ~Q^1_\mu(\x)\!\log P(\x\vert\thetav)} \right\rangle_{\!\thetav}\Big]}{\int\{\rmd \Q\,\rmd\hat{\Q}\,\rmd A\,\rmd\hat{A}\}\, \rme^{N    \Psi[\ldots]  }  }   \nonumber\\      
\hspace*{-10mm}
  &=& -\lim_{N\rightarrow\infty} \frac{1}{N}\sum_{\mu=1}^K\log \left\langle \rme^{N\!\int\!\rmd\x  ~Q^1_\mu(\x)\!\log P(\x\vert\thetav)   } \right\rangle_{\thetav}     \nonumber\\ 
  \hspace*{-10mm}
  &=& -\sum_{\mu=1}^K\max_{\thetav} \int\!\rmd\x ~ Q^1_\mu(\x) \log P(\x\vert\thetav).
  \label{eq:e-comp-2}
\end{eqnarray}

\section{`Sphericity'  of Normally  distributed samples \label{app:sphericity}}
Here we show that almost all points  of any random sample  from the $d$-dimensional Normal distribution $\mathcal{N}(\x\vert\m,\Cov )$, with mean $\m$ and covariance $\Cov$, lie in the annulus $d(\lambda_{\rm max}\!-\!\epsilon)<\vert\vert\x\!-\!\m\vert\vert^2<d(\lambda_{\rm max}\!+\!\epsilon)$,  where $\vert\vert\cdots\vert\vert$ is the Euclidean norm and  $\lambda_{\rm max}$ is  the maximum eigenvalue of $\Cov$, for sufficiently large  $d$ and  $0\!<\!\epsilon\! \ll \!1$.
If $\x$ is sampled from $\mathcal{N}(\x\vert\m,\Cov )$, then  $\left\langle\vert\vert\x-\m\vert\vert^2\right\rangle=\Tr (\Cov)$. We want to  bound the following probability:
\begin{eqnarray}
\hspace*{-5mm}
&&\hspace*{-15mm}
\mathrm{Prob}(\vert\vert\x\!-\!\m\vert\vert^2    \notin   (   \Tr (\Cov)\!-\!d\epsilon,  \Tr (\Cov)\!+\!d\epsilon) )
\nonumber\\
\hspace*{-5mm}
&=&\mathrm{Prob}(\vert\vert\x\!-\!\m\vert\vert^2\!\leq\!\Tr (\Cov)\!-\!d\epsilon )
+\mathrm{Prob}(\vert\vert\x\!-\!\m\vert\vert^2\!
\geq \!\Tr (\Cov)\!+\!d\epsilon )
\label{eq:annulus}.
\end{eqnarray} 
Firstly, for sufficienty small positive $\alpha$   we can use the Markov inequality to obtain
\begin{eqnarray}
&& \hspace*{-15mm}
\mathrm{Prob}(\vert\vert\x\!-\!\m\vert\vert^2\!\geq\! \Tr (\Cov)\!+\!d\epsilon )
\nonumber\\
&=&\mathrm{Prob}(\rme^{\frac{\alpha}{2} \vert\vert\x-\m\vert\vert^2}\geq\rme^{\frac{\alpha}{2}  (\Tr (\Cov)+d\epsilon)      })
\leq      \Big\langle\rme^{\frac{\alpha}{2} \vert\vert\x-\m\vert\vert^2}\Big\rangle  \rme^{-\frac{\alpha}{2}(\Tr (\Cov)+d\epsilon)} \nonumber\\
&=&\rme^{-\frac{1}{2}(\log\vert\I-\alpha \Cov \vert +\alpha (\Tr (\Cov)+d\epsilon)    )}\label{eq:outside-o-ineq-comp-1}.
\end{eqnarray} 
The last line,  which assumes that $\Cov^{-1}\!\!-\!\alpha\I$ is positive definite, follows from  (\ref{def:Normal}).  
Denoting the eigenvalues of the covariance matrix $\Cov$ by $\lambda_1,\ldots,\lambda_d$, we can bound  $\log\vert\I-\alpha \Cov \vert=\sum_{\ell=1}^d\log(1-\alpha\lambda(\ell))$  from below by $d\log(1-\alpha\lambda_{\rm max})$, where $\lambda_{\rm max}=\max_{\ell}\lambda(\ell)$. Using this in (\ref{eq:outside-o-ineq-comp-1}) gives us the simpler inequality 
\begin{eqnarray}
\hspace*{-5mm}
\mathrm{Prob}(\vert\vert\x\!-\!\m\vert\vert^2\!\geq\! \Tr (\Cov)\!+\!d\epsilon )
&\leq & \rme^{-\frac{1}{2}(  d\log(1-\alpha\lambda_{\rm max})      +\alpha  (\Tr (\Cov)+d\epsilon)                 )         }\label{eq:outside-o-ineq-comp-2},
\end{eqnarray} 
The function $d\log(1-\alpha  \lambda_{\rm max}  )      +\alpha(\Tr (\Cov)+d\epsilon)   $ is found to have its maximum at $\alpha=(\Tr (\Cov )+d\epsilon-d\lambda_{\rm max})/(\lambda_{max}   (\Tr (\Cov)+d\epsilon))$, which allows us to optimise the upper bound in (\ref{eq:outside-o-ineq-comp-2}) and produce the inequality
\begin{eqnarray}
\mathrm{Prob}(\vert\vert\x\!-\!\m\vert\vert^2\!\geq\Tr (\Cov)\!+\!d\epsilon )
&\leq&\exp\Big[-\frac{d}{2}\Phi\Big(\frac{d\lambda_{\rm max}}{\Tr (\Cov)  +d\epsilon}\Big)\Big]
\label{eq:outside-o-ineq},
\end{eqnarray} 
 where $\Phi(x)=\log(x)+x^{-1}-1$. We note that $\Phi(x)\geq0$, by the inequality $\log(x)\geq1-\frac{1}{x}$.  Also, $\Phi(x)$ is monotonic increasing (decreasing) for $x>1$ ($x<1$), and is exactly zero when $x=1$.  
Secondly, we derive a similar bound for the  second probability in (\ref{eq:annulus}):
\begin{eqnarray}
&& \hspace*{-15mm}
\mathrm{Prob}(\vert\vert\x\!-\!\m\vert\vert^2\leq\Tr (\Cov)\!-\!d\epsilon )\nonumber\\
&=&\mathrm{Prob}(\rme^{-\frac{\alpha}{2} \vert\vert\x-\m\vert\vert^2}\geq\rme^{-\frac{\alpha}{2}  (\Tr (\Cov)-d\epsilon)      })\leq     \left\langle\rme^{-\frac{\alpha}{2} \vert\vert\x-\m\vert\vert^2}\right\rangle  \rme^{\frac{\alpha}{2}(\Tr (\Cov)-d\epsilon)} \nonumber\\
&=& \rme^{-\frac{1}{2}(\log\vert\I+\alpha \Cov \vert -\alpha (\Tr (\Cov)-d\epsilon)    )}\label{eq:inside-o-ineq-comp-1}.
\end{eqnarray} 
Now  $\log\vert\I+\alpha \Cov \vert=\sum_{\ell=1}^d\log(1+\alpha\lambda(\ell))$ is bounded from below by $d\log(1+\alpha\lambda_{\rm min})$, where $\lambda_{\rm min}=\min_{\ell}\lambda(\ell)$. Using this in (\ref{eq:inside-o-ineq-comp-1}) gives us the inequality 
\begin{eqnarray}
\hspace*{-5mm}
\mathrm{Prob}(\vert\vert\x\!-\!\m\vert\vert^2\leq\Tr (\Cov)\!-\!d\epsilon )
\leq   \rme^{-\frac{1}{2}(   d\log(1+\alpha\lambda_{\rm min})    -\alpha (\Tr (\Cov)    -d\epsilon)    )}\label{eq:inside-o-ineq-comp-2}.
\end{eqnarray} 
 We note that the quantity $d\log(1+\alpha\lambda_{\rm min})    -\alpha (\Tr (\Cov)    -d\epsilon)$ takes its maximum for $\alpha=(\Tr (\Cov)    -d\epsilon+d  \lambda_{min})/(\lambda_{min}(\Tr (\Cov)    -d\epsilon))$, which in (\ref{eq:inside-o-ineq-comp-2}), gives the new bound
\begin{eqnarray}
\mathrm{Prob}(\vert\vert\x\!-\!\m\vert\vert^2\!\leq\Tr (\Cov)\!-\!d\epsilon )&\leq &  \exp\Big[-\frac{d}{2}\Phi\Big(  \frac{d  \lambda_{\rm min}}{\Tr (\Cov)    -d\epsilon}      \Big)\Big]
\label{eq:inside-o-ineq}.
\end{eqnarray} 
 By using  the two inequalities (\ref{eq:outside-o-ineq},\ref{eq:inside-o-ineq})  in  (\ref{eq:annulus}),  we obtain the inequality 
\begin{eqnarray}
&&\hspace*{-15mm}
\mathrm{Prob}(\vert\vert\x\!-\!\m\vert\vert^2    \notin   (   \Tr (\Cov)\!-\!d\epsilon,  \Tr (\Cov)\!+\!d\epsilon) )
\nonumber\\
&\leq &   2 \exp\Big[-\frac{d}{2}   \min\Big\{ \Phi\Big(  \frac{d  \lambda_{\rm min}}{\Tr (\Cov)    \!-\!d\epsilon}  \Big),\Phi\Big(\frac{d\lambda_{\rm max}}{\Tr (\Cov)  \!+\!d\epsilon}\Big)  \Big\}       \Big]      
\label{eq:annulus-upper-bound-I}
\end{eqnarray} 
Moreover, since $\Tr (\Cov)\leq d\lambda_{max}$, we may also write
\begin{eqnarray}
&&\hspace*{-15mm}
\mathrm{Prob}(\vert\vert\x\!-\!\m\vert\vert^2    \notin   (   \Tr (\Cov)\!-\!d\epsilon,  \Tr (\Cov)\!+\!d\epsilon) )
\nonumber\\
&\leq &   2 \exp\Big[-\frac{d}{2}   \min\Big\{ \Phi\Big(  \frac{\lambda_{\rm min}}{\lambda_{\rm max} \!-\!\epsilon}  \Big),\Phi\Big(\frac{\lambda_{\rm max}}{\lambda_{\rm max} \!+\!\epsilon}\Big)  \Big\}       \Big]      
\label{eq:annulus-upper-bound-II}
\end{eqnarray} 
The remaining extrema are given by 
\begin{eqnarray}
\hspace*{-10mm}
 \epsilon\in(0, \epsilon_1 )\!:&~~& 
 \min\Big\{ \Phi\Big(  \frac{\lambda_{\rm min}}{\lambda_{\rm max} \!-\!\epsilon}  \Big),\Phi\Big(\frac{\lambda_{\rm max}}{\lambda_{\rm max} \!+\!\epsilon}\Big)  \Big\} = \Phi\Big(\frac{\lambda_{\rm max}}{\lambda_{\rm max} \!+\!\epsilon}\Big)
 \\
 \hspace*{-10mm}
\epsilon\in(\epsilon_1,\epsilon_2)\!:&~~& 
  \min\Big\{ \Phi\Big(  \frac{\lambda_{\rm min}}{\lambda_{\rm max} \!-\!\epsilon}  \Big),\Phi\Big(\frac{\lambda_{\rm max}}{\lambda_{\rm max} \!+\!\epsilon}\Big)  \Big\} = 
 \Phi\Big(\frac{\lambda_{\rm max}}{\lambda_{\rm max} \!-\!\epsilon}\Big) 
 \end{eqnarray}
with
\begin{eqnarray}
\epsilon_1=\frac{\lambda_{\rm max}(\lambda_{\rm max}\!-\!\lambda_{\rm min})}{\lambda_{\rm max}+\lambda_{\rm min}},~~~~~~
\epsilon_2= \lambda_{\rm max}\!-\!\lambda_{\rm min}
\end{eqnarray}
Furthermore, when $\lambda_{\rm max}=\lambda_{\rm min}=\lambda$, i.e.  $\Cov=\lambda\I$, one obtains
  \begin{eqnarray}
\epsilon\in(0,\lambda)\!:&~~&    \min\Big\{ \Phi\Big(  \frac{  \lambda }{ \lambda\!-\!\epsilon}\Big),\Phi\Big(\frac{\lambda }{\lambda \!+\!\epsilon}\Big) \Big\} =  \Phi\Big(\frac{\lambda  }{\lambda   +\epsilon}\Big)
   \end{eqnarray} 
   
 If, in contrast, we observe a sample $\x_1,\ldots,\x_N$ from $\mathcal{N}(\x\vert\m,\Cov)$, instead of a single vector $\x$, then the probability $\mathrm{Prob}(\cup_{i=1}^N\left\{   \vert\vert\x_i\!-\!\m\vert\vert^2\!    \notin   (   \Tr (\Cov)\!-\!d\epsilon,  \Tr (\Cov)\!+\!d\epsilon)      \right\})$ that at least one of the events  $  \vert\vert\x_i\!-\!\m\vert\vert^2 \!   \notin   (   \Tr (\Cov)\!-\!d\epsilon,  \Tr (\Cov)\!+\!d\epsilon) $ occurs,  can be bounded by combining  Boole's inequality with inequalities (\ref{eq:outside-o-ineq}) and (\ref{eq:annulus-upper-bound-I}):
\begin{eqnarray}
&&\hspace*{-15mm}  \mathrm{Prob}(\cup_{i=1}^N\left\{   \vert\vert\x_i\!-\!\m\vert\vert^2  \!  \notin   (   \Tr (\Cov)\!-\!d\epsilon,  \Tr (\Cov)\!+\!d\epsilon)      \right\})             \nonumber\\
&\leq& \sum_{i=1}^N    \mathrm{Prob}(\vert\vert\x_i\!-\!\m\vert\vert^2  \!  \notin   (   \Tr (\Cov)\!-\!d\epsilon,  \Tr (\Cov)\!+\!d\epsilon) )                 \nonumber\\
&\leq & 2 N   \exp\Big[\!-\!\frac{d}{2}   \min\Big\{ \Phi\Big(  \frac{d  \lambda_{\rm min}}{\Tr (\Cov)   \! -\!d\epsilon}      \Big),\Phi\Big(\frac{d\lambda_{\rm max}}{\Tr (\Cov)  \!+\!d\epsilon}\Big)  \Big\}                   
\end{eqnarray}
Repeating similar steps to those followed earlier then gives for $\lambda_{\rm max}>\lambda_{\rm min}$:
\begin{eqnarray}
\hspace*{-16mm}  \mathrm{Prob}(\cup_{i=1}^N\left\{   \vert\vert\x_i\!-\!\m\vert\vert^2  \!  \notin   (   \Tr (\Cov)\!-\!d\epsilon,  \Tr (\Cov)\!+\!d\epsilon)      \right\})      &\leq&
2 N   \exp\Big[\!-\!\frac{d}{2}      \Phi\Big(\frac{\lambda_{\rm max}}{\lambda_{\rm max}  \!+\!\epsilon}\Big)      \Big] 
\nonumber\\[-0mm] \hspace*{-16mm} &&  \label{eq:annulus-ineq-N-II}
\end{eqnarray} 
provided 
$\epsilon\in(0, \lambda_{\rm max}(\lambda_{\rm max}\!-\!\lambda_{\rm min})/(\lambda_{\rm max}\!+\!\lambda_{\rm min}))$, whereas for $\Cov=\lambda\I$ we have 
\begin{eqnarray}
\hspace*{-15mm}  \mathrm{Prob}(\cup_{i=1}^N\left\{   \vert\vert\x_i\!-\!\m\vert\vert^2  \!  \notin   (   \Tr (\Cov)\!-\!d\epsilon,  \Tr (\Cov)\!+\!d\epsilon)      \right\})      &\leq&
2 N   \exp\Big[\!-\!\frac{d}{2}      \Phi\Big(\frac{\lambda}{\lambda  \!+\!\epsilon}\Big)\Big],         \nonumber
\\
\hspace*{-16mm}&&         \label{eq:annulus-ineq-N-III}
\end{eqnarray} 
provided $\epsilon\in(0, \lambda)$. It is now clear that there is a function $d(\epsilon, \lambda_{\rm max}, N )>0$ such that  for  $d>d(\epsilon, \lambda_{\rm max}, N )$ almost all points  of a sample from  $\mathcal{N}(\x\vert\m,\Cov )$ lie in the annulus\footnote{
For small $d$, the bound in (\ref{eq:annulus-ineq-N-II}) is very loose,   so it makes more sense to consider the probability that $\cup_{i\leq N}\{   \vert\vert\x_i-\m\vert\vert^2   \geq d(\lambda_{max}+\epsilon)  \} $, i.e. that at least one $\x_i$  in the sample $\xmatrix$ lies outside the ball $\mathcal{B}_{\sqrt{d(\lambda_{max}+\epsilon)}}(\m)$, given by $\mathrm{Prob}(\cup_{i\leq N} \{   \x_i\notin    \mathcal{B}_{\sqrt{d(\lambda_{max}+\epsilon)}}(\m) \}           )  \leq  N   \exp[-\frac{d}{2}      \Phi(\frac{\lambda_{\rm max}}{\lambda_{\rm max}  +\epsilon})] $. } $ \sqrt{d(\lambda_{max}-\epsilon)}<\vert\vert\x-\m\vert\vert<\sqrt{d(\lambda_{max}+\epsilon)}$.

\section*{References}
%\bibliographystyle{unsrt}
%\bibliography{jpa_clust_refs}

\end{document}